\documentclass[aps,onecolumn,preprint,superscriptaddress,nofootinbib,floats,tightenlines,12pt]{revtex4}
\usepackage{amsmath,amssymb,color,mathrsfs, graphicx,verbatim,epsfig, bbm, wasysym, axodraw}
\usepackage[hyperfootnotes=false]{hyperref}
\usepackage{slashed}
\allowdisplaybreaks

\usepackage{slashed} % Han uses this
\usepackage{color, multirow} % Han uses thist

\newcommand{\ie} {{\it i.e.}}  % Han uses this

\begin{document}

\baselineskip=18pt

%%%%%%%%%%
%%%%%%%%%%    Title page
%%%%%%%%%%

\thispagestyle{empty}
\vspace{20pt}
\font\cmss=cmss10 \font\cmsss=cmss10 at 7pt

\begin{flushright}
\small 
\end{flushright}

\hfill
\vspace{20pt}

\begin{center}
{\Large \textbf
{
Combined analysis of double Higgs production \\[0.15cm] 
via gluon fusion at the HL-LHC \\[0.15cm]
in the effective field theory approach
}}
\end{center}

\vspace{15pt}
\begin{center}
{ Jeong Han Kim$^{\, a}$, Yasuhito Sakaki$^{\, b}$ and Minho Son$^{\, b}$}
\vspace{30pt}

$^{a}$ {\small \it Department of Physics and Astronomy, \\ University of Kansas, Lawrence, Kansas, 66045 USA
}
\vskip 3pt
$^{b}$ {\small \it Department of Physics, Korea Advanced Institute of Science and Technology, \\ 291 Daehak-ro, Yuseong-gu, Daejeon 34141, Republic of Korea
}

\end{center}

\vspace{20pt}
\begin{center}
\textbf{Abstract}
\end{center}
\vspace{5pt} {\small
We perform the combined analysis of the double Higgs production via gluon fusion in the $b\bar{b} \gamma\gamma$ and $b\bar{b}\tau^+\tau^-$ decay channels at the High-Luminosity LHC (HL-LHC).  To validate our analysis, we reproduce the ATLAS result of the $b\bar{b} \gamma\gamma$ process including all contributions from fakes. For the $b\bar{b}\tau^+\tau^-$ decay channel, we perform the similar analysis to the CMS one. As an improvement, we also perform the multivariate analysis employing the boosted decision tree algorithm. Then, we derive 68\% probability contours on anomalous Higgs couplings in the effective field theory (EFT) approach for various analyses. We find that the $b\bar{b}\tau^+\tau^-$ process outperforms the $b\bar{b}\gamma\gamma$ for the measurement of energy-growing operators, while adding the $b\bar{b}\tau^+\tau^-$ process is least beneficial for improving the precision of the Higgs self-coupling (mainly set by the $b\bar{b}\gamma\gamma$ process). We illustrate that the double Higgs production alone can be comparable to the single Higgs process in constraining the modification of the top Yukawa coupling in the positive direction. Focusing on the Higgs self-coupling as a special interest, we derive the precision as a function of various improvable parameters such as tag and mistag rates of tau leptons, heavy flavor jets, photon identification, diphoton mass resolution, and jet energy resolution to take into account future phenomenological studies. As an optimistic benchmark scenario, we illustrate that the 68\% and 95\% probability intervals of the Higgs self-coupling, $\lambda_3/\lambda_{3}^{SM}$, at the HL-LHC can reach $[0.2,\, 2.3]$ and $[-0.1,\, 3.5] \cup [4.0,\, 6.5]$, respectively, where the correlation among the EFT coefficients is taken into account.
}

\vfill\eject
\noindent

\tableofcontents
\newpage

%%%%%%%%%%%%%%%%%%%%%%%%%%%%%%%%%%%%%%%%
%%%%%%%%%%%%%%%%%%%%%%%%%%%%%%%%%%%%%%%%
%%%%%%%%%%%%%%%%%%%%%%%%%%%%%%%%%%%%%%%%
%%%%%%%%%%%%%%%%%%%%%%%%%%%%%%%%%%%%%%%%
%%%%%%%%%%%%%%%%%%%%%%%%%%%%%%%%%%%%%%%%
%%%%%%%%%%%%%%%%%%%%%%%%%%%%%%%%%%%%%%%%
%%%%%%%%%%%%%%%%%%%%%%%%%%%%%%%%%%%%%%%%
\section{Introduction}
\label{sec:intro}

The precision measurements of the interactions of the recently discovered Higgs boson~\cite{Aad:2012tfa,Chatrchyan:2012ufa} with the fermions and gauge bosons at the Large Hadron Collider (LHC) indicate that we live in a special theory that stays weakly coupled up to very high energy scales, namely the Standard Model (SM)~\cite{Englert:1964et,Higgs:1964pj,Guralnik:1964eu, Weinberg:1967tq}. 
On the other hand, the story of the purely Higgs sector is a bit pessimistic. The global picture of the Higgs potential in the bottom-up approach is currently unavailable~\footnote{The constraints on the Higgs self-coupling extracted from the LHC data at $\sqrt{s} = 8$ and $\sqrt{s} = 13$ TeV are found to be weak~\cite{Khachatryan:2016sey, CMS-PAS-HIG-17-008}.
}, and we may have to wait a long time until we achieve a high precision on the Higgs self-coupling according to recent studies at various future colliders~\cite{Baer:2013cma,Barr:2014sga,Azatov:2015oxa,He:2015spf,Contino:2016spe}.  Without the precision measurement of the Higgs self-coupling, it will be unlikely for us to complete our understanding of the origin of electroweak symmetry breaking and the thermal history of the Higgs potential through the evolution of the Universe. It is not hard to imagine a new physics scenario that manifests itself only via the modification of the Higgs self-coupling. For instance, a large deviation of the Higgs self-coupling from the SM prediction can be linked to the baryon-anti-baryon asymmetry based on the strong first-order electroweak phase transition~\cite{Trodden:1998ym, Cline:2006ts, Morrissey:2012db}.
 
In this work, we revisit the double Higgs production via gluon fusion known as the most prominent process~\footnote{There also have been attempts to constrain the Higgs self-coupling via single Higgs process~\cite{Degrassi:2016wml,Gorbahn:2016uoy,Bizon:2016wgr,Maltoni:2017ims} (see~\cite{DiVita:2017eyz} for a related discussion), the $Vhh$ ($V=W,Z$) process~\cite{Cao:2015oxx}, and double Higgs production in the vector boson fusion~\cite{Bishara:2016kjn}.} to access modifications of the Higgs self-coupling. We target the HL-LHC as it will likely be the earliest future collider to provide us a meaningful precision on the Higgs self-coupling. We explore three directions to exploit the benefit from the HL-LHC: i) we combine two decay channels, $b\bar{b}\gamma\gamma $ and $b\bar{b}\tau^+\tau^-$, of the double Higgs production in the effective field theory (EFT) approach just like combining various decay channels in the single Higgs coupling measurements is beneficial; ii) we apply a multivariate technique employing the boosted decision tree (BDT) algorithm to our analysis as a way to enhance the significance; iii) we parametrize the precision of the Higgs self-coupling as functions of various improvable variables such as tag and mistag rates of tau jets and heavy flavor jets, photon identification efficiency, and invariant mass resolution etc to take into account future phenomenological studies.
 
As the evidence of the new physics is not seen with the increasing reach of the new physics scale, the viewpoint of the SM as an effective field theory provided by the large mass gap between the electroweak scale and the new physics scale makes more sense as a way to parametrize the effects of the new physics. In the EFT approach, the new physics effects are encoded in the coefficients of the effective Lagrangian. The precision measurements of those coefficients will constrain the possible structure of the underlying new physics from which the EFT originated. When the new physics is not far from the cutoff scale of the EFT, it may manifest itself as a large deviation of the EFT coefficient.  
The total amplitude of the double Higgs production process in the EFT approach contains many diagrams due to new types of interactions in the EFT Lagrangian (see Fig.~\ref{fig:diagrams}). The amplitude is parametrized in terms of five anomalous couplings in the nonlinear basis, namely the top Yukawa coupling, the Higgs self-coupling, contact interactions between two Higgs bosons and two top quarks, a Higgs boson and two gluons, and two Higgs bosons and two gluons. From the viewpoint of the EFT, setting all the EFT coefficients to the SM values except $\lambda_3$ (Higgs self-coupling), as commonly done in literature, may not be justified unless the selection of only $\lambda_3$ is associated with a symmetry or a hidden fine-tuning is involved. To take into account the correlation between the Higgs self-coupling and other anomalous couplings, we keep all the EFT coefficients in our analysis. We will perform the marginalization over other parameters when we derive the precision of the Higgs self-coupling as our special interest. 

Previous studies on the double Higgs production at the HL-LHC have been performed in various final states including $b\bar{b}\gamma\gamma $~\cite{Kling:2016lay, Baur:2003gp, Baglio:2012np, Huang:2015tdv, Azatov:2015oxa,Cao:2015oaa,Cao:2016zob,Alves:2017ued,CMS-PAS-FTR-15-002, ATL-PHYS-PUB-2014-019, Barger:2013jfa}, $b\bar{b}\tau^+\tau^-$~\cite{Baur:2003gpa, Dolan:2012rv,CMS-PAS-FTR-15-002}, $b\bar{b} W^+W^-$~\cite{Papaefstathiou:2012qe, Huang:2017jws} and $b\bar{b} b\bar{b}$~\cite{deLima:2014dta, Wardrope:2014kya, Behr:2015oqq}. Among those, despite its lower signal rate, the $ b\bar{b}\gamma\gamma$ ($BR \simeq$ 0.264 \%) decay channel is the cleanest and most thoroughly studied in literature. Apparently, switching to the $b\bar{b}\tau^+\tau^-$ ($BR \simeq$ 7.31 \%) decay channel may look promising, as the cross section increases by a factor of 27.7 compared to the $b\bar{b}\gamma\gamma$ decay channel.
However, the signal rate is penalized by a series of factors, the first one being the sub-branching ratios of the $\tau^+\tau^-$ system, namely $42.3 \%$ and $45.5 \%$ for the fully hadronic and semileptonic final states, respectively.
The signal rate is further penalized by the tau tagging efficiency.  Finally, reconstructing the $h\rightarrow \tau^+ \tau^-$ system, for instance, against $Z\rightarrow \tau^+\tau^-$ is challenging due to the irreducible loss of information via invisible neutrinos~\cite{Barr:2011he}. As a result, the performance of the $b\bar{b}\tau^+\tau^-$ channel does not look better than the $b\bar{b}\gamma\gamma$ channel~\cite{CMS-PAS-FTR-15-002}. The $b\bar{b}\tau^+\tau^-$ may have the potential to be at best comparable to the $ b\bar{b}\gamma\gamma$. While most previous studies of the double Higgs production at the HL-LHC focused only on the variation of $\lambda_3$, the EFT approach was initiated in~\cite{Contino:2012xk} and~\cite{Azatov:2015oxa} (see~\cite{Cao:2015oaa,Cao:2016zob} for related studies) for the $b\bar{b}\gamma\gamma$ decay channel and~\cite{Goertz:2014qta} for the $b\bar{b}\tau^+\tau^-$ decay channel~\footnote{The result in~\cite{Goertz:2014qta} should be considered optimistic due to optimistic tau reconstruction efficiency and a smaller set of backgrounds. For instance, truth tau leptons with 70\% reconstruction efficiency was assumed with a negligible fake rate and only three irreducible backgrounds, $t\bar{t}$, $hZ$, and $ZZ$, were included.}. In our work, we for the first time perform the combined analysis of two decay channels, $b\bar{b}\gamma\gamma$ and $b\bar{b}\tau^+\tau^-$, in the EFT approach. We show that the $b\bar{b}\tau^+\tau^-$ process outperforms the $b\bar{b}\gamma\gamma$ process for the measurement of energy-growing operators, while adding the $b\bar{b}\tau^+\tau^-$ process is least beneficial for improving the precision of the Higgs self-coupling.

A multivariate analysis seems the right method beyond cut-based analysis to improve the performance when kinematic variables are correlated in a complicated way, since it can efficiently identify a signal region in a multidimensional parameter space. Some studies are found in~\cite{Wardrope:2014kya,Alves:2017ued}. We utilize the boosted decision tree (BDT) algorithm to estimate its impact on the sensitivity of the EFT coefficients~\footnote{However, the BDT analysis applied to the rare process like the double Higgs production for the $b\bar{b}\gamma\gamma$ and $b\bar{b}\tau^+\tau^-$ decay channels at the HL-LHC should be taken with a grain of salt. As the BDT analysis also imposes a BDT cut, the analysis may suffer from a low statistics. While this issue can be improved in a Monte Carlo-based analysis, a purely data-driven analysis will be difficult to be realized. We leave this issue for the future investigation.}. We find that the multivariate analysis improves the significance of the SM from 1.3 to 2.1 at the HL-LHC for the $b\bar{b}\gamma\gamma$ decay channel and from 0.9 (0.6) to 1.4 (0.8) for the fully hadronic (semileptonic) $b\bar{b}\tau^+\tau^-$ decay channel. We observe that the benefit of the multivariate analysis becomes more evident for the EFT coefficients

Improving the photon identification, tagging efficiencies of $\tau$ jets and heavy flavor jets (such as $b$ ,$\, c$ jets) as well as suppressing various fake rates, $j\rightarrow \gamma$, $j\rightarrow b$, and $c\rightarrow b$, are key to enhancing significance. Also, high-quality invariant mass resolutions of $m_{\gamma \gamma}$, $m_{b \bar{b}}$, and $m_{\tau \tau}$ are incontrovertible ingredients to disentangle a signal from the backgrounds. Since those improvements can be made via various independent phenomenological studies, it would be informative to express the precision as a function of those improvable factors to predict our capability in the future. A similar approach has been considered in the study at the 100 TeV $pp$ collider~\cite{Contino:2016spe}.

This paper is organized as follows. In Section~\ref{sec:crosssection}, we discuss the basis of anomalous Higgs couplings and the prametrization of the cross section in the EFT approach. In Section~\ref{sec:bbaa}, we perform the analysis of the $b\bar{b}\gamma\gamma$  decay channel. We first validate our cut-based analysis by reproducing the ATLAS result at the HL-LHC~\cite{ATL-PHYS-PUB-2014-019} (see also~\cite{ATL-PHYS-PUB-2017-001}). Then we carry out a simple multivariate analysis using the BDT technique. In Section~\ref{sec:bbtautau}, we perform a similar exercise for the $b\bar{b}\tau^+\tau^-$ decay channel. Here, we follow the similar CMS cut-based analysis~\cite{CMS-PAS-FTR-15-002}. In Section~\ref{sec:results}, we discuss the sensitivity of the EFT coefficients obtained by either individual or combined analysis. We present the constraint on the Higgs self-coupling taking into account the correlation among the EFT coefficients. We also discuss the impact of the individual improvable factors on the precision of the Higgs self-coupling. Finally, we present our projected precision on the Higgs self-coupling based on an optimistic benchmark scenario at the HL-LHC. In Section~\ref{sec:Summary}, we summarize our results and reach out to a conclusion. In Appendix~\ref{app:sec:simdetails}, we provide the detail of our background simulation.\\

%%%%%%%%%%%%%%%%%%%%%%%%%%%%%%%%%%%%%%%%
%%%%%%%%%%%%%%%%%%%%%%%%%%%%%%%%%%%%%%%%
%%%%%%%%%%%%%%%%%%%%%%%%%%%%%%%%%%%%%%%%
%%%%%%%%%%%%%%%%%%%%%%%%%%%%%%%%%%%%%%%%
%%%%%%%%%%%%%%%%%%%%%%%%%%%%%%%%%%%%%%%%
%%%%%%%%%%%%%%%%%%%%%%%%%%%%%%%%%%%%%%%%
%%%%%%%%%%%%%%%%%%%%%%%%%%%%%%%%%%%%%%%%
\section{Effective Field Theory Approach}
\label{sec:crosssection}
\subsection{Parametrizations of Higgs boson couplings}

%%%%%%%%%%%%%%%%%%%%%%%%%%%%%%%%%%%%%%%%%%%%%%%%%%%%%%%%%%%%
\begin{figure}[tp]
\begin{center}
\includegraphics[width=0.30\textwidth]{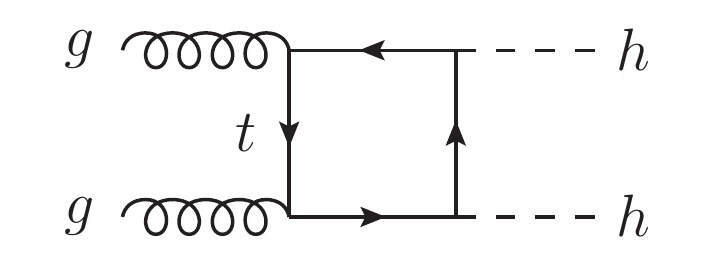}
\hspace{0.8em}
\includegraphics[width=0.335\textwidth]{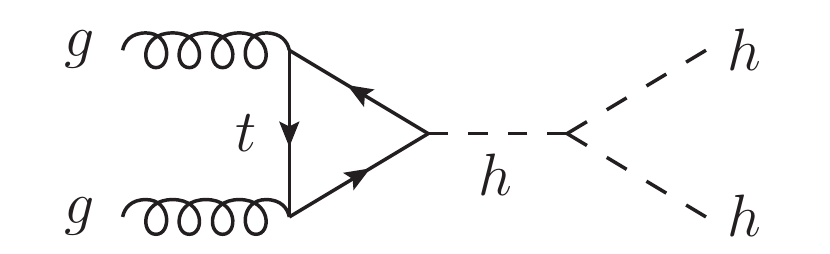}
\hspace{0.8em}
\includegraphics[width=0.28\textwidth]{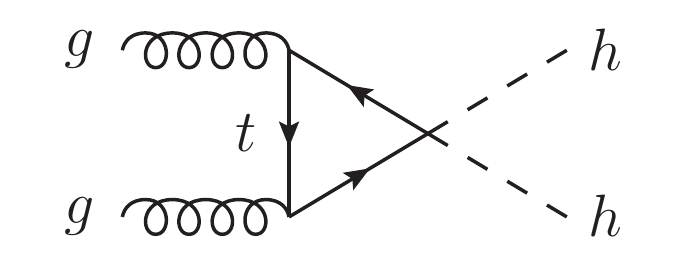}\\
\vspace{1.2em}
\includegraphics[width=0.30\textwidth]{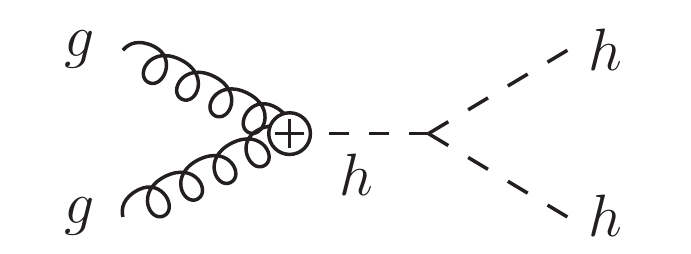}
\hspace{2.5em}
\includegraphics[width=0.233\textwidth]{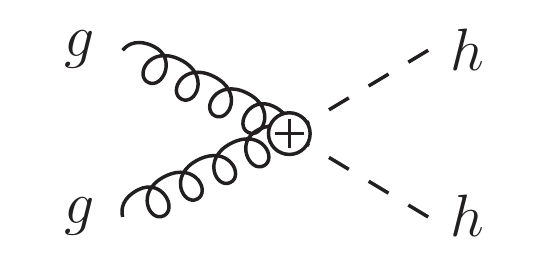}
\caption{Feyman diagrams contributing to double Higgs production via gluon fusion. The first three diagrams in the upper line are labeled as ${\cal A}_\square$, ${\cal A}_\triangle$, and ${\cal A}_{\triangle nl}$ and those in the bottom by ${\cal A}_3$ and ${\cal A}_4$ from the left in Section~\ref{sec:crosssection}.}
\label{fig:diagrams}
\end{center}
\end{figure}
%%%%%%%%%%%%%%%%%%%%%%%%%%%%%%%%%%%%%%%%%%%%%%%%%%%%%%%%%%%%
%

Embedding the Higgs boson into an effective field theory approach can be straightforwardly done once a guiding principle that the interactions of the Higgs boson with the other SM fields are determined. In the SM, the physical Higgs boson is a part of the doublet of the $SU(2)_L \times U(1)_Y$ gauge group. This linear representation of the Higgs doublet is well supported by the observation at the LHC which points toward the SM prediction. A new physics effect will be allowed to appear in the vicinity of the SM set by the experimental uncertainty. The corrections to the SM are encoded in coefficients of higher-dimensional operators ${\cal O}_i$ organized by their dimensions,
\begin{equation}
{\cal L} = {\cal L}_{SM} + \sum_i \frac{c^{(6)}_i}{\Lambda^2} {\cal O}_i^{(6)} + \sum_i \frac{c^{(8)}_i}{\Lambda^4} {\cal O}_i^{(8)} + \cdots~,
\end{equation}
where the $\Lambda$ is the cutoff scale where a new physics kicks in. The operators with an odd dimension have been neglected by assuming the accidental global symmetries in the SM such as the lepton number.
The leading contribution comes from the dimension-six operators in this framework. We list only the dimension-six operators which participate in the double Higgs production process, and they are written as, in the SILH basis~\cite{Giudice:2007fh},
\begin{equation}
\label{eq:linearL}
\begin{split}
\Delta {\cal L}_{d=6}  =
& \, \frac{\overline c_H}{2 v^2} \partial_\mu \big(H^\dagger H\big) \partial^\mu \big(H^\dagger H\big)
   + \frac{\overline c_u}{v^2} y_t \left(H^\dagger H\, \overline q_L H^c t_R + \textrm{h.c.}\right)  \\
& - \frac{\overline c_6}{v^2}\frac{m_h^2}{2 v^2} \big(H^\dagger H\big)^3 
   + \overline c_g \frac{g_s^2}{m_W^2} H^\dagger H G^a_{\mu\nu} G^{a\, \mu\nu}\, ,
\end{split}
\end{equation}
where $v = 1/(\sqrt{2} G_F)^{1/2} =  246$ GeV and $m_h = 125$ GeV. These dimension-six operators are formed by adding two additional powers of the Higgs doublet, $H^\dagger H$, to the renormalizable terms in the SM while keeping the same number of derivatives. An additional power of the Higgs doublet in the SILH basis is accompanied by the factor, $g_*/m_* \equiv 1/f$, where a new physics is characterized by one coupling $g_*$ and one scale $m_*$ which is associated with the scale of new states. The SILH power counting of the coefficients estimates 
\begin{equation}
\label{eq:SILH:linearL}
 \overline c_H,\ \overline c_u,\ \overline c_6 \sim \frac{v^2}{f^2},\quad \quad \overline c_g \left ( \frac{4\pi}{\alpha_2} \right ) \sim \frac{v^2}{f^2}\times \frac{\lambda^2}{g^2_*} ~,
\end{equation}
where $\lambda$ is a weak spurion coupling suppressing $\bar{c}_g$.

The typical size of the coefficients varies depending on an assumption on the structure of the UV completion. For instance, dropping the assumption on the Higgs boson as a pseudo-Goldstone boson in the SILH power counting will remove the suppressions in Eq.~(\ref{eq:linearL}) which breaks the shift symmetry. In this modified power counting, the degeneracy in the size of the coefficients in Eq.~(\ref{eq:SILH:linearL}) breaks, and a parametric separation between coefficients can be achieved~\cite{Azatov:2015oxa,DiVita:2017eyz}, 
\begin{equation}
  \overline c_H \sim \frac{v^2}{f^2} \quad < \quad \overline c_6 \sim \frac{v^2}{f^2} \times \frac{g^2_*}{\lambda_4}~,
\end{equation}
where $\lambda_4$ is the quartic coupling in the SM Higgs potential. 

The parametrization in the basis of the Higgs boson that belongs to the linear representation in Eq.~(\ref{eq:linearL}) is suitable to the small sizes of the coefficients in order for the EFT expansion to make sense.  In the situation where the size of the coefficients can be substantially large, the parametrization in the nonlinear basis in terms of the custodial singlet physical Higgs boson $h$ becomes more suitable. It is called a nonlinear basis in a sense that $SU(2)_L \times U(1)_Y$ gauge group is nonlinearly realized. The Higgs boson $h$ in the nonlinear basis is not necessarily the $SU(2)_L$ doublet, but it can be more generic. The following five parameters in the nonlinear basis are relevant for double Higgs production,
\begin{equation}
\label{eq:nonlinearL}
{\cal L}  = - m_t \,\overline t t \left(c_t \frac{h}{v} + c_{2t} \frac{h^2}{v^2}\right)
- \frac{c_3}{6} \left ( \frac{3m_h^2}{v} \right ) h^3
+ \frac{g_s^2}{4 \pi^2}\left(c_g \frac{h}{v} + c_{2g} \frac{h^2}{2v^2}\right)
G^a_{\mu\nu} G^{a\, \mu\nu}\, ,
\end{equation}
where we set top quark mass to $m_t = 173$ GeV.  As is evident in Eq.~(\ref{eq:nonlinearL}), the coefficients of terms with one Higgs and two Higgs bosons are not necessarily related in the nonlinear basis unlike the case of the linear basis. 
The coefficients in the nonlinear basis can be matched to those in the linear basis by resuming over all powers of $H$ and expanding operators in terms of the physical Higgs $h$. At the level of dimension-six operators, the relations are given by
\begin{equation}
\label{eq:dictionary}
c_t = 1 - \frac{\overline c_H}{2} - \overline c_u\,, \quad
c_{2t} = -\frac{1}{2}\left( \bar c_H + 3\bar c_u \right) \,,
\quad  c_3 = 1  - \frac{3}{2}\overline c_H + \overline c_6\,,
\quad  c_g = c_{2g} = \overline c_g \left( \frac{4\pi}{\alpha_2} \right)  , 
\end{equation}
where $\alpha_2 \equiv g^2/4\pi$. 

The ${\cal O}_H$ operator in the linear basis (the first term of Eq.~(\ref{eq:linearL})) leads to the universal modification of the processes involving the Higgs bosons via the wave function renormalization of the Higgs boson, and it is strongly constrained by the $h-Z$ coupling measurement. Then, a sizable deviation of the Higgs self-coupling in the nonlinear basis is translated into the sizable value of $\overline c_6$ according to the relation of Eq.~(\ref{eq:dictionary}). The validity of the EFT with a large $\overline c_6$ will be an important issue that has to be considered. As is indicated in Eq.~(\ref{eq:dictionary}), the modification of the top Yukawa coupling, $\overline c_u$, can be measured via three types of amplitudes in Fig.~\ref{fig:diagrams} involving $c_t$ and $c_{2t}$ couplings. It will be interesting to know how well $\overline c_u$ can be constrained by the double Higgs production process alone compared to the sensitivity extracted from the single Higgs process such as $t\bar{t}h$. We will illustrate in Section~\ref{sec:SenAnomalCouplings} that combining various decay channels of the double Higgs process can be comparable to the single Higgs process in constraining the modification of the top Yukawa coupling in the positive direction.

An important characteristic feature of the double Higgs production process with the set of coefficients in Eq.~(\ref{eq:nonlinearL}) is that the amplitudes in Fig.~\ref{fig:diagrams} exhibit the different scaling behaviors with the scale of the process, $\sqrt{\hat{s}}$, in the high energy limit. The one-loop amplitude due to the $t\bar{t}hh$ interaction with $c_{2t}$ has a high-energy behavior scaling like ${\cal A}_{\triangle nl} \sim c_{2t}\, {\rm log}^2 \left ( m^2_t/{\hat{s}} \right )$. The amplitude with $c_t$ is either suppressed by the $s$-channel Higgs boson exchange, ${\cal A}_\triangle \sim  c_t c_3\, \left(m_h^2/{\hat{s}}\right )\, {\rm log}^2 \left ( m^2_t/{\hat{s}} \right )$, or it does not show the energy-dependent scaling behavior, ${\cal A}_\square \sim c_t^2$. Similarly the first amplitude in the bottom of Fig.~\ref{fig:diagrams} scales like ${\cal A}_{3} \sim  c_g c_3$, where $\hat{s}$ from the $s$-channel Higgs boson exchange is canceled by the $\hat{s}$ in $hgg$ vertex. The last amplitude in Fig.~\ref{fig:diagrams} grows most rapidly with the energy, ${\cal A}_{4} \sim c_{2g}\, \left (\hat{s}/v^2 \right )$.

Since the Higgs self-coupling $c_3$ (or $\overline c_6$) only appears in ${\cal A}_\triangle$ and ${\cal A}_3$ which have the suppression by $m^2_h/\hat{s}$ factor, its sensitivity relies on the threshold region where the size of the backgrounds is largest. The modification of the top Yukawa coupling $\overline c_u$ might gain a better sensitivity via the amplitude ${\cal A}_{\triangle nl}$ which grows with the energy. The energy-dependent behavior suggests to us to exploit the exclusive analysis utilizing the differential distribution of $m_{hh} \equiv \sqrt{\hat{s}}$ to enhance the sensitivity on the couplings and break the degeneracy among various coefficients. This exercise has been done at 14 TeV and 100 TeV $pp$ colliders for the $b\bar{b}\gamma\gamma $ decay channel~\cite{Azatov:2015oxa,He:2015spf} and for the $b\bar{b}\tau^+\tau^-$ decay channel at 14 TeV~\cite{Goertz:2014qta}.

\subsection{Cross section of double Higgs production}
The square of the summed amplitudes in Fig.~\ref{fig:diagrams} has various terms. We parametrize the cross section as a function of the coefficients in the nonlinear basis as in~\cite{Azatov:2015oxa,DiVita:2017eyz},
\begin{equation}
\label{eq:tot_xsec}
\begin{split}
{\sigma \over \sigma_{SM}}  = \,  & A_1\, c_t^4 + A_2 \, c_{2t}^2  + A_3\,  c_t^2 c_3^2  + A_4 \, c_g^2 c_3^2  + A_5\,  c_{2g}^2  + A_6\, c_{2t} c_t^2 + A_7\,  c_t^3 c_3 \\[0.1cm]
& + A_8\,  c_{2t} c_t\, c_3 + A_9\, c_{2t} c_g c_3 + A_{10}\, c_{2t} c_{2g} + A_{11}\,  c_t^2 c_g c_3 + A_{12}\, c_t^2 c_{2g} \\[0.1cm]
& + A_{13}\, c_t c_3^2 c_g  + A_{14}\, c_t c_3 c_{2g} +A_{15}\, c_g c_3 c_{2g} \,.
\end{split}
\end{equation}
We choose the same $m_{hh}$ bin size and division for the $b\bar{b}\gamma\gamma$ channel as done in~\cite{Azatov:2015oxa} (see also~\cite{DiVita:2017eyz}). In this way, the fit coefficients of the cross section for six $m_{hh}$ bins in~\cite{Azatov:2015oxa} can be recycled. The cross section fit for the inclusive analysis can be constructed from the exclusive ones.

The similar type of exclusive analysis in the $b\bar{b}\tau^+\tau^-$ decay channel is more challenging. Since the decay of the $\tau^+\tau^-$ system always involves invisible neutrinos, it is difficult to extract the exact scale of the process, $\sqrt{\hat s}$ ($=m_{hh}$). Instead, one can at most reconstruct the transverse mass of the $b\bar{b}\tau^+\tau^-$ system, denoted as $m^{vis}_{hh}$, out of the available information in the final state~\footnote{A sophisticated matrix element method (or something analogous to it) may reproduce the overall invariant mass distribution of the $b\bar{b}\tau^+\tau^-$ system that matches to the parton level $m_{hh}$ distribution. However, the one-to-one matching between two quantities in an event-by-event basis is still difficult.}.  Then, it is technically difficult to get a set of fit coefficients of Eq.~(\ref{eq:tot_xsec}) with a great accuracy for many $m^{vis}_{hh}$ bins by the similar trick used in~\cite{Azatov:2015oxa} (using the events at the hadron level necessarily lose statistics due to a low efficiency, reducing the accuracy of the fit coefficients). In this work, we will perform only the inclusive analysis of $b\bar{b}\tau^+\tau^-$ channel without using $m^{vis}_{hh}$ distribution. More importantly, $\sqrt{\hat s}$ is a scale that controls the validity of the EFT. Only events below the cutoff scale, $\sqrt{\hat{s}} < \Lambda$, must be included in the analysis to derive the sensitivity on the EFT coefficients. Imposing a cut on any type of transverse mass, $m_{hh}^{vis}$, will inevitably introduce a contamination from the events above the cutoff scale when the correlation between $\sqrt{\hat{s}}$ and $m_{hh}^{vis}$ is poor~\cite{Falkowski:2016cxu}. At the HL-LHC, the number of the events of the SM beyond the TeV scale after analysis cuts is extremely tiny~\footnote{See Table~\ref{tab:bbaa:mhhBinAnalysis} for the differential distribution of $m_{hh}$ in the $b\bar{b}\gamma\gamma$ decay channel. The situation will be similar for the $b\bar{b}\tau^+\tau^-$ decay channel as its overall signal rate is similar to that of $b\bar{b}\gamma\gamma$.}. Although the signal rate away from the SM for operators growing with the energy might be enhanced, we expect that this issue does not cause a severe problem as long as the cutoff $\Lambda$ is not assumed to be sub-TeV scale.

%%%%%%%%%%%%%%%%%%%%%%%%%%%%%%%%%%%%%%%%
%%%%%%%%%%%%%%%%%%%%%%%%%%%%%%%%%%%%%%%%
%%%%%%%%%%%%%%%%%%%%%%%%%%%%%%%%%%%%%%%%
%%%%%%%%%%%%%%%%%%%%%%%%%%%%%%%%%%%%%%%%
%%%%%%%%%%%%%%%%%%%%%%%%%%%%%%%%%%%%%%%%
%%%%%%%%%%%%%%%%%%%%%%%%%%%%%%%%%%%%%%%%
%%%%%%%%%%%%%%%%%%%%%%%%%%%%%%%%%%%%%%%%
\section{Double Higgs production at HL-LHC}
\label{sec:DHiggsHL}

The signal events were generated using our internal C++ code linked to QCDLoop~\cite{Ellis:2007qk} which evaluates one-loop diagrams in Fig.~\ref{fig:diagrams} with anomalous couplings (see Eq.~(\ref{eq:nonlinearL})). The leading-order (LO) SM cross section is found to be $16.2\,$fb at $\sqrt{s} = 14\,$TeV using CTEQ6ll PDF's (LO PDF with LO $\alpha_s$), setting the factorization and renormalization scales to $Q = m_{hh}$, and $m_h = 125\,$GeV, $m_t=173\,$GeV. We then rescale the signal cross section by the next-to-next-to-leading order (NNLO) k-factor of $ 2.27$~\cite{deFlorian:2013jea}~\footnote{While the k-factor in~\cite{deFlorian:2013jea} was obtained in the heavy top mass limit (see~\cite{Dawson:1998py} for the first NLO calculation), the NNLL matched to the NNLO cross section at 14 TeV including the finite top quark mass effects to the NLO  has been calculated to be $39.64 \; \rm{fb}$ with uncertainties of $^{+ 4.4 \%}_{-6.0 \%}$ from the QCD scale, and $\pm 2.1\%$ and $\pm 2.2 \%$ from the PDF4LHC15\_nnlo\_mc parton distribution function and $\alpha_s$ respectively~\cite{deFlorian:2016spz,deFlorian:2015moa,Borowka:2016ehy}. The cross section used in our analysis is about 7\% smaller (conservative) but it is within the uncertainty. The dependence of the NLO, NNLO k-factors on EFT coefficients of the double Higgs production at $\sqrt{s} = 14$ TeV have been studied in~\cite{Grober:2015cwa,Borowka:2016ypz,deFlorian:2017qfk}.} as done in~\cite{Azatov:2015oxa}. The hard-scattering events have been showered and hadronized by \textsc{PYTHIA} v8.219~\cite{Sjostrand:2014zea}. 

All backgrounds samples were generated by \textsc{MadGraph}5\_aMC$@$NLO v2.3.3~\cite{Alwall:2014hca,Hirschi:2015iia} with the default factorization and renormalization scales, interfaced with \textsc{PYTHIA} v6.4 for the parton showering and hadronization. The full description on the background generation can be found in Appendix~\ref{app:sec:simdetails}.

We include some of the detector effects based on the ATLAS detector performances~\cite{ATLAS:2013004}. We smear the momenta and energies of the reconstructed jets, photons and leptons depending on their energies~\footnote{For smearing jets, we take noise (N), stochastic (S), and constant (C) parameters 13.15, 0.74, and 0.05, respectively~\cite{ATLAS:2013004}. Since photons and muons can be remarkably well reconstructed, we use the effective resolutions of $\sigma / E = 0.02$ and $\sigma / E = 0.05 \sqrt{E/1000}$ for photons and muons, respectively.}. On the other hand, we do not include the multiple interaction and pile-up in our simulation.
Throughout our paper, the significance is defined as $N_S/\sqrt{N_B}$, where $N_S$ ($N_B$) is the number of the signal (total backgrounds), assuming 3 ab$^{-1}$ at the HL-LHC.

%%%%%%%%%%%%%%%%%%%%%%%%%%%%%%%%%%%%%%%%
%%%%%%%%%%%%%%%%%%%%%%%%%%%%%%%%%%%%%%%%
\subsection{$b\bar{b}\gamma\gamma$ decay channel}
\label{sec:bbaa}

%%%%%%%%%%%%%%%%%%%%%%%%%%%%%%%%%%%%%%%%
%%%%%%%%%%%%%%%%%%%%%%%%%%%%%%%%%%%%%%%%
%\subsubsection{Object reconstructions}
%\label{sec:bbaa_OR}

The events first go through the photon/lepton isolation criteria. A photon (lepton) is declared to be isolated if it satisfies $p^{\Sigma}_T/p_T(\gamma) < 0.1$ ($p^{\Sigma}_T/p_T(l) < 0.15$), where $p^{\Sigma}_T$ is the sum of the transverse momenta of final state particles within the $\Delta R = 0.4$ isolation cone. We include only the isolated photons (leptons) with $p_T(\gamma) > 25$ GeV ($p_T(l) > 20$ GeV) in the $|\eta(\gamma)| < 2.5$ ($|\eta(\gamma)| < 2.4$) region. Events with more than two isolated photons are vetoed. 

The remaining particles are clustered by the \verb|FastJet| \cite{Cacciari:2011ma} implementation using the anti-$k_T$ algorithm \cite{Cacciari:2008gp} with a distance parameter of $R = 0.4$. Events with at least two jets which pass minimum cuts $p_T(j)>25$ GeV and $|\eta(j)| < 2.5$ are considered. Jets are further classified into three categories based on the heavy flavors that matches to them. Our heavy flavor tagging algorithm runs iteratively over jets to search for $b$ hadrons or $c$ hadrons inside them. If a $b$ hadron ($c$ hadron) is found inside, it is classified as a $b$ jet ($c$ jet). The remaining unmatched jets are called light jets. 

We will only consider the events with at least four reconstructed objects, \ie~$N_{b}+ N_{c} + N_{light-jet}  + N_\gamma + N_l \ge 4$.
The isolated photons and reconstructed jets are iteratively paired into two photon candidates and two $b$-jet candidates for the cut-based analysis. Here, the photon candidate can be either an isolated photon, a lepton or a jet faking a photon. Similarly the $b$-jet candidate can be either a $b$ jet, a $c$ jet faking a $b$ jet, or a light jet faking a $b$ jet. We then reconstruct two invariant masses $m_{\gamma\gamma}$ and $m_{b\bar{b}}$ out of those two photon candidates and two $b$ jet candidates.

An efficient $b$-tagging remains essential for obtaining a higher signal efficiency and lowering background contaminations. While a realistic $b$-tagging method incorporates all $\eta$ and $p_T$ dependences, we apply a flat $b$-tag rate of $\epsilon_b = 0.7$ and a mistag rate that a $c$ jet (light jet) is misidentified as a $b$ jet of $\epsilon_{c\rightarrow b} = 0.3$ ($\epsilon_{j\rightarrow b} = 0.015$). The effects of varying mistag rates will be discussed in Section~\ref{sec:varyingFakes}.
The fake rate of a jet passing a photon identification is parameterized by ATLAS~\cite{ATL-PHYS-PUB-2013-009},
\begin{equation}
\epsilon_{j\rightarrow \gamma}(p_T) = 9.3 \cdot 10^{-3} \times e^{-p_T/27.5 {\rm GeV}}~.
\end{equation}
The fake rate of a lepton passing a photon identification is set to $\epsilon_{l \rightarrow \gamma}= 0.02~(0.05)$  in the barrel $|\eta| < 1.37$ (end-cap calorimeter $1.37 < |\eta| < 2.37$) region, based on the estimate by ATLAS~\cite{ATL-PHYS-PUB-2014-019} for the HL-LHC. Finally, a photon reconstruction efficiency is implemented based on the parametrization by ATLAS~\cite{ATL-PHYS-PUB-2013-009}
\begin{equation}
\label{eq:photonRecoEff}
\epsilon_\gamma(p_T) = 0.76 - 1.98 \times e^{-p_T/16.1 {\rm GeV}} .
\end{equation} 
%

%%%%%%%%%%%%%%%%%%%%%%%%%%%%%%%%%%%%%%%%
%%%%%%%%%%%%%%%%%%%%%%%%%%%%%%%%%%%%%%%%
\subsubsection{Cut-based analysis}
\label{sec:bbaa:cutbased}
We validate our cut-based analysis by reproducing the result by ATLAS at the HL-LHC~\cite{ATL-PHYS-PUB-2014-019} and they are summarized in Table~\ref{tab:bbaaSummary}. 
The cuts in ATLAS~\cite{ATL-PHYS-PUB-2014-019} are listed below;
\begin{equation}
\begin{split}
&p_{T}(\gamma) > 30\ {\rm GeV}~, \quad p_{T>}(b) > 40\ {\rm GeV}~, \quad p_{T<}(b)>25\ {\rm GeV}~,\\
&|\eta(\gamma)| < 1.37,\quad 1.52 < |\eta(\gamma)| < 2.37,\quad |\eta(b)| < 2.5, \\
&0.4 < \Delta R(b,b) < 2.0~,\quad 0.4 < \Delta R (\gamma,\gamma) < 2.0~,\quad \Delta R (\gamma, b) > 0.4~,\\
&100 < m^{\rm reco}_{bb} < 150\ {\rm GeV}~, \quad 123 < m^{\rm reco}_{\gamma\gamma} < 128\ {\rm GeV}~,\\
&p_T(\gamma\gamma) > 110\ {\rm GeV}~, \quad p_T (bb) > 110\ {\rm GeV}~, \quad N_{jet} < 6,
\end{split}
\end{equation}
where 
 $p_{T>}(b)$ ($p_{T<}(b)$) are the transverse momentum of the harder (softer) $b$-jet candidate.
The $\eta$ denotes a pseudorapidity, and $\Delta R$ is a distance in the $\eta$-$\phi$ (pseudorapidity-azimuthal angle) plane between two objects. For the $\Delta R (\gamma, b)$, the minimum distance between photon and $b$-jet candidates is taken among four combinations. The  $m_{\gamma\gamma}$ and $m_{bb}$  ($p_T (\gamma\gamma)$ and $p_T (bb)$) are the invariant masses (transverse momenta) of two photon- and two $b$-jet systems. The $N_{jet}$ is the number of jets which satisfy $p_{T}(j)>25$ GeV and $|\eta(j)| < 2.5$.

The background samples were categorized in terms of jet flavors at the hadron level as in~\cite{ATL-PHYS-PUB-2014-019}. Among the backgrounds in the Table~\ref{tab:bbaaSummary}, the $b\bar{b}\gamma\gamma$ ($c\bar{c}\gamma\gamma$) category includes the events with at least two $b$-jet candidates ($c$-jet candidates). The $jj\gamma\gamma$ category in the Table~\ref{tab:bbaaSummary} is the collection of events with less than two $b$ or $c$-jet candidates which includes $jj\gamma\gamma$, $cj\gamma\gamma$, $bj\gamma\gamma$, and $bc\gamma\gamma$ at the hadron level. We find that the dominant contribution in the $jj\gamma\gamma$ category comes from $bc\gamma\gamma$ followed by $bj\gamma\gamma$ and $cj\gamma\gamma$. The dominant contribution to the $t\bar{t}\gamma$ background is caused by the leptons faking the photons. As is evident in Table~\ref{tab:bbaaSummary}, the signal rate and the backgrounds look consistent with the ATLAS simulation. The significance of the SM for the $b\bar{b} \gamma\gamma $ decay channel is estimated to be 1.3 which is consistent with the ATLAS simulation. The detail of the background simulation is given in Appendix~\ref{app:sec:simdetails}.

In addition to the analysis with the ATLAS cuts, we also provide the column for the analysis with cuts in~\cite{Azatov:2015oxa} except a few differences. In the analysis in~\cite{Azatov:2015oxa}, the fakes of $j,l\rightarrow \gamma$ were not included in the background estimation, and the $c$-jet candidate was not separately considered. Instead, the $c$-flavor jet was included in the definition of the light jet where the universal mistag rate, $\epsilon_{j\rightarrow b}=0.01$, was applied. The photon reconstruction efficiency in~\cite{Azatov:2015oxa} was assumed to be 80\%. The numbers (fourth column) in the Table~\ref{tab:bbaaSummary} were obtained by imposing the same set of cuts in~\cite{Azatov:2015oxa} but with the same tag/mistag rates and photon reconstruction efficiency along with the separate treatment of the $c$-flavor jet as in ATLAS~\cite{ATL-PHYS-PUB-2014-019}. Consequently, the total background size is significantly larger than 37.1 events in~\cite{Azatov:2015oxa} whereas the signal is found to be degraded~\footnote{The final signal rate (8.1) is smaller than 12.8 events in~\cite{Azatov:2015oxa}. We find that the discrepancy is caused by two reasons: $p_T$-dependent photon reconstruction efficiency in Eq.~(\ref{eq:photonRecoEff}) and the jet smearing that we applied~\cite{ATLAS:2013004} (further jet smearing in addition to those caused by the parton shower and hadronization was not applied in~\cite{Azatov:2015oxa}).
}.

%%%%%%%%%%%%%%%%%%%%%%%%%%%%%%%%%%%
\begin{table}[tbp]
\centering
\begin{tabular}{ccccc}  
\hline
 \quad Expected yields (3 ab$^{-1}$)  
 & \quad ATLAS~\cite{ATL-PHYS-PUB-2014-019}  \quad 
 & \quad With ATLAS cuts \quad & \quad With cuts in~\cite{Azatov:2015oxa} \quad & \quad MVA \quad \\
\hline \hline
 $h(b\bar{b})h(\gamma\gamma)$  & 8.4 & 8.0 & 8.1 & 8.7\\
\hline
 $b\bar{b} \gamma\gamma$  & 9.7 & 12.3 & 23 & 6.4\\
 $c\bar{c} \gamma\gamma$  & 7.0 &  7.4 & 14 & 2.4\\
 $b\bar{b} \gamma j           $  & 8.4 &  7.5 & 16 & 1.2\\
 $j j \gamma\gamma          $  & 7.4 &  4.1 & 8.7 & 1.7\\ 
$t\bar{t}\gamma                 $  & 3.2 &  1.5 & 4.4 & 1.5\\
 $t\bar{t}h(\gamma\gamma)       $& 6.1 & 5.5 & 6.8  & 3.7\\
 $Z(b\bar{b})h(\gamma\gamma)$ & 2.7 &  1.2 & 0.86 & 1.0\\
 $b\bar{b}h(\gamma\gamma)     $& 1.2 & 0.24 & 0.25 & 0.2\\
 \hline
 Total backgrounds & 45.7 & 39.8 & 73.4 & 18.0\\
\hline \hline
\end{tabular}
\caption{The expected number of events using 3 ab$^{-1}$ at the HL-LHC for the double Higgs production by ATLAS~\cite{ATL-PHYS-PUB-2014-019}, by our analysis with ATLAS cuts (as the validation) and with cuts in~\cite{Azatov:2015oxa} and by MVA (multivariate analysis). The $t\bar{t}(\ge 1$ lepton) and $b\bar{b}jj$ backgrounds are not displayed in the table due to their small sizes.}
\label{tab:bbaaSummary}
\end{table}
%%%%%%%%%%%%%%%%%%%%%%%%%%%%%%%%%%%

%%%%%%%%%%%%%%%%%%%%%%%%%%%%%%%%%%%%%%%%
%%%%%%%%%%%%%%%%%%%%%%%%%%%%%%%%%%%%%%%%
\subsubsection{Multivariate Analysis}
\label{MVA_bbaa}

In this section, we carry out a more sophisticated analysis approach, namely a multivariate analysis, to improve the performance  compared to the cut-based analysis. We proceed it by employing the BDT algorithm with the help of the {\tt TMVA-Toolkit} \cite{Speckmayer:2010zz} in the {\tt ROOT} framework \cite{Brun:1997pa}. The cut-based analysis cuts out the phase space by applying a series of windows (bounded by either both boundaries or one side). This approach might be able to reach the maximal efficiency via an optimization when the variables are not correlated. When a signal region has a more complicated boundary due to the correlations among variables, the cut-based analysis may not be the best option. The MVA technique is one way to achieve a better performance since it can efficiently identify the signal region in the multidimensional phase space~\footnote{Identifying the signal region accurately requires a large statistics, and this approach will be difficult to be a data driven in a process with a low signal rate. An alternative approach may be identifying the backgrounds better in a control region (where enough statistics is guaranteed) and extrapolating them to the signal region. By rejecting the extrapolated backgrounds in the signal region, one may be able to extract the signal rate. We implicitly assume that the final performance of either approach will be similar.}.

We prepare training samples for BDT analysis for each background.  The samples are required to pass following cuts:
\begin{align}
\begin{split}
& \text{number of isolated photon}\geq 2,\quad \text{number of $b$ jets}\geq 2,\\
& p_T (\gamma) > 30~\text{GeV},\quad p_T(b) > 25~\text{GeV},\quad \\
& |\eta (\gamma)| < 1.37,\quad 1.52 < |\eta (\gamma)| < 2.37,\quad |\eta (b)| < 2.5, \\
& \Delta R (\gamma,\gamma) > 0.4,\quad \Delta R (\gamma, b) > 0.4,\quad \Delta R (b, b) > 0.4, \\
& 120 < m^{\rm reco}_{\gamma\gamma}  < 130~\text{GeV},\quad 60 < m^{\rm reco}_{bb} < 160~\text{GeV}. 
\label{train_condition}
\end{split}
\end{align}
The variables entering into the BDT analysis, in addition to those in Eq.~(\ref{train_condition}), are $p_T(\gamma\gamma)$, $p_T(bb)$, $\Delta R (\gamma_i, b_j)$ ($i,j = 1,2$), $p_{T}(hh)$, $m^{\rm reco}_{hh}$, $N_{jet}$, $y (hh)$, $E_T^{\rm miss}$, and $p_T(j_1)$, where $p_T (j_1)$ is the transverse momentum of the hardest extra jet (if it exists) which is not assigned as a $b$ jet and $c$ jet. The $\Delta R (\gamma_i, b_j)$ is the distance between $i$th photon and $j$th $b$ jet, and $E_T^{\rm miss}$ is the missing transverse energy in the system. The detail of our BDT analysis is postponed to Appendix~\ref{app:MVA} where the impact of some individual variables is demonstrated.

%%%%%%%%%%%%%%%%%%%%%
\begin{figure}[!htb!]
\begin{center}
\includegraphics[width=0.32\linewidth]{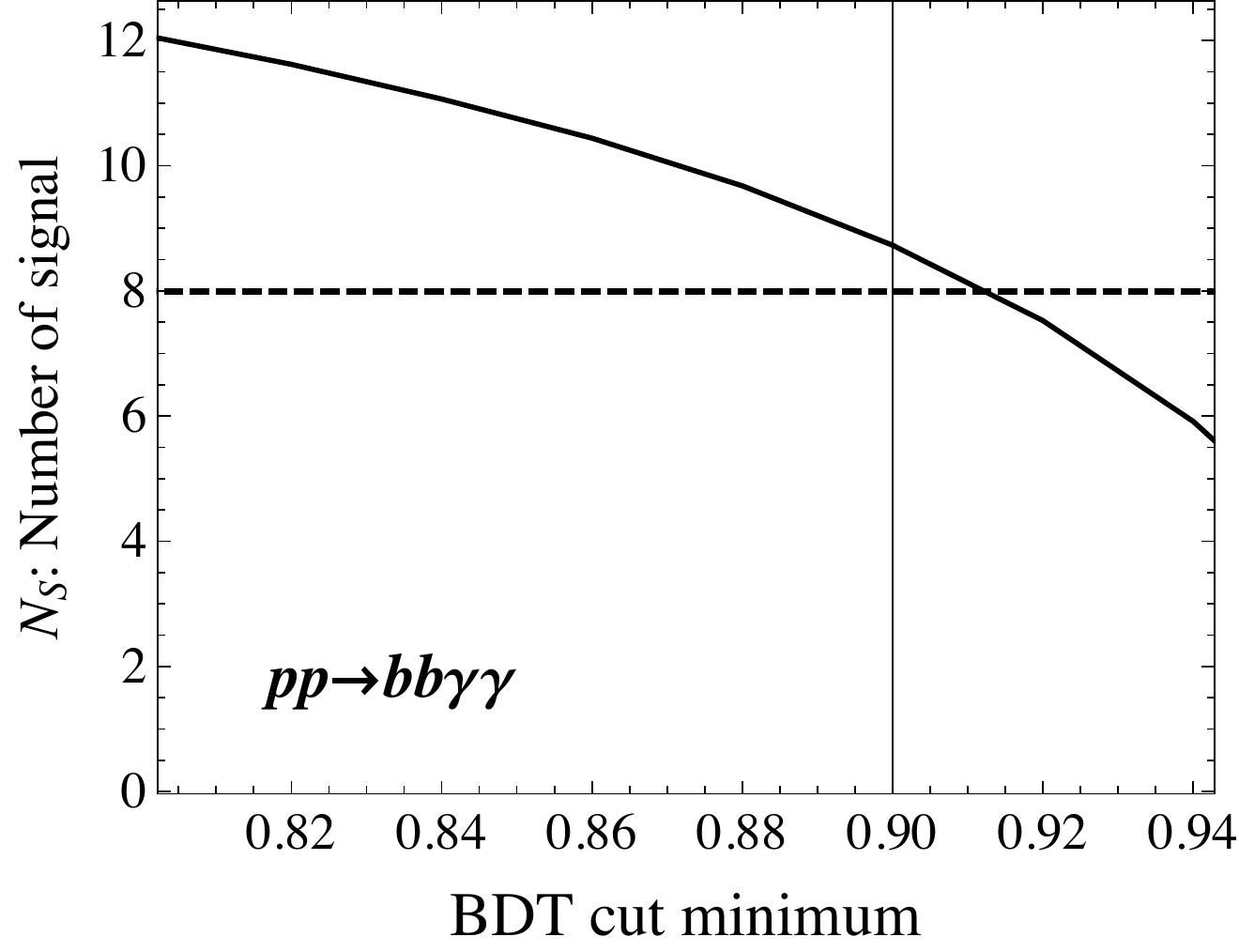}
\includegraphics[width=0.32\linewidth]{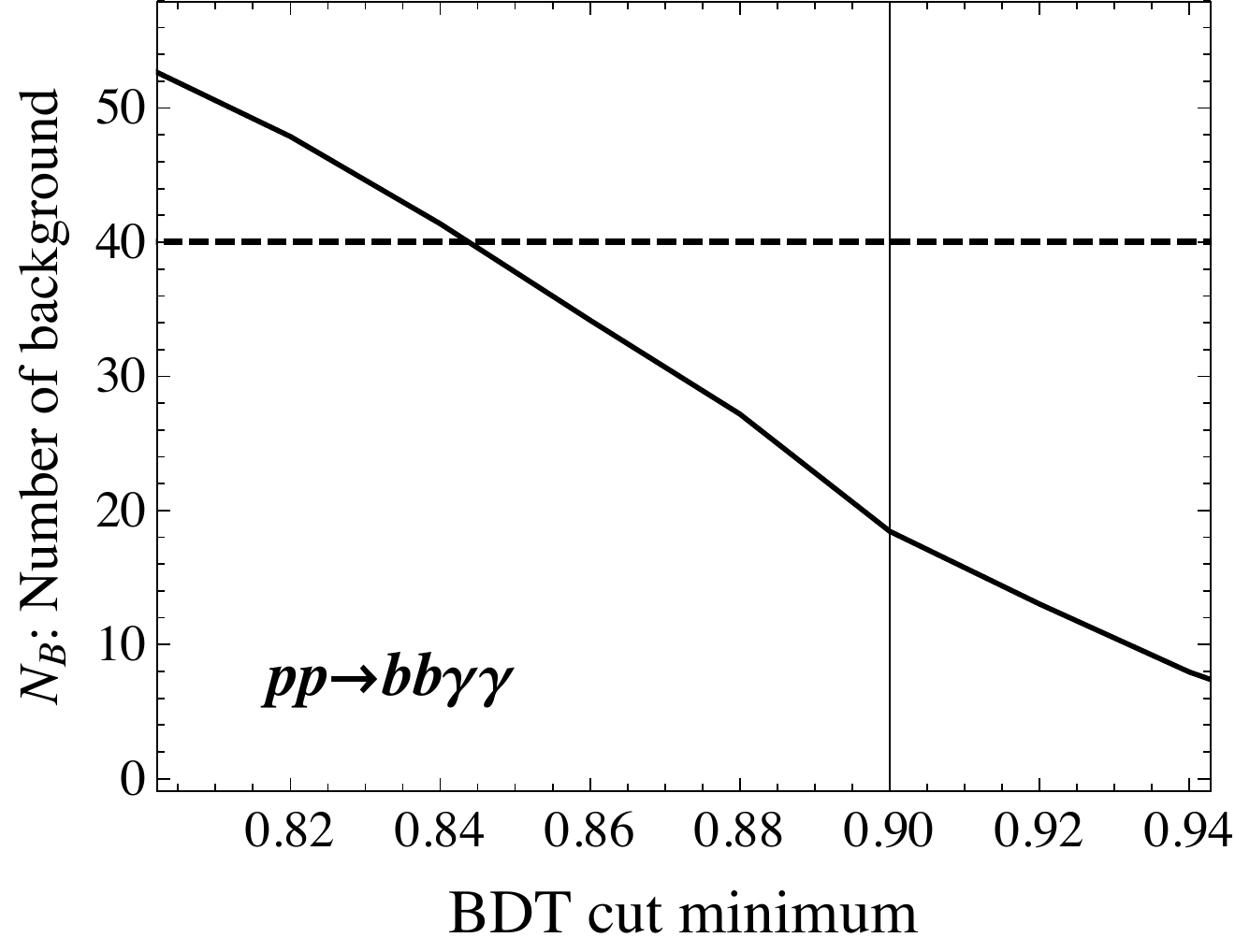}
\includegraphics[width=0.32\linewidth]{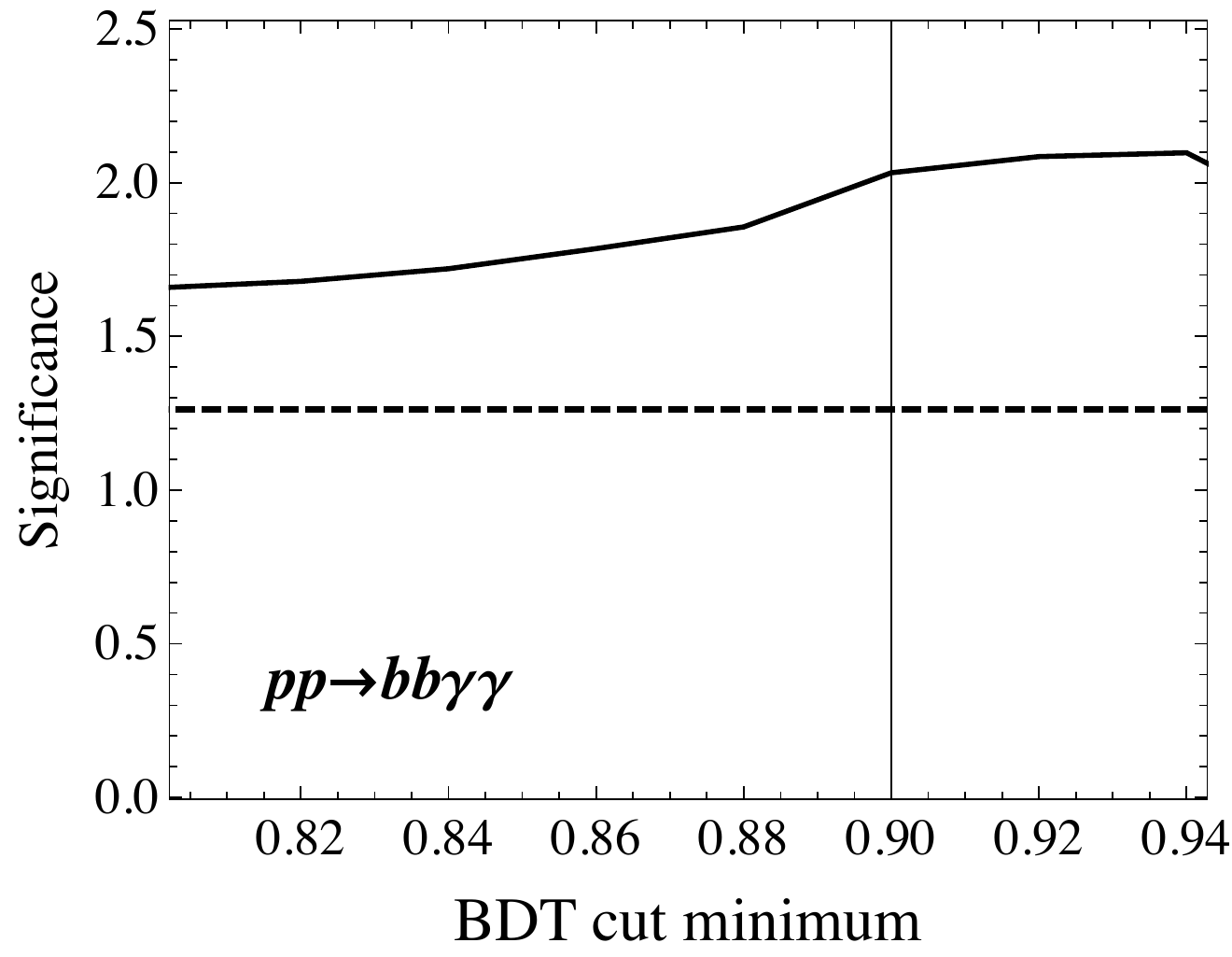}
\caption{
The number of signal ($N_S$), background ($N_B$), and significance after imposing some BDT cuts and the condition in Eq.~(\ref{train_condition}). Dashed lines correspond to the result by the cut-and-count analysis. The thin vertical line denotes our BDT cut minimum.
}
\label{fig:BDT_bbaa}
\end{center}
\end{figure}
%%%%%%%%%%%%%%%%%%%%%

As the significance plot as a function of BDT cut minimum in Fig.~\ref{fig:BDT_bbaa} shows roughly a plateau in the vicinity of the maximal significance, it is not straightforward to choose unambiguously a unique reference point in the BDT analysis. In our BDT analysis for the $b\bar{b} \gamma\gamma $ decay channel, we choose the BDT minimum $=0.90$ as a reference point to proceed deriving the sensitivities on various anomalous Higgs couplings in later sections. Increasing BDT minimum can further reduce the total backgrounds for only a small loss of the signal rate. In this situation, the significance becomes more robust to systematic uncertainties but, at the same time, it becomes more vulnerable to a larger statistical error caused by the limitation of the background simulation. While one ideally may select different BDT minima at different beyond the SM (BSM) points to take into account of the varying kinematics  point-by-point in the BSM phase space, we simply apply the universal BDT minimum chosen at the SM point to all the other BSM points. Using the reference BDT minimum, the significances are improved to 2.1.

%%%%%%%%%%%%%%%%%%%%%%%%%%%%%%%%%%%%
\begin{table}[tbp]
\centering
\begin{tabular}{c|cccccc}  
 $m_{hh}$ [GeV] 
 & \quad $250-400$ \quad & \quad $400-550$ \quad & \quad $550-700$ \quad & \quad $700-850$ \quad & \quad $850-1000$ \quad & \quad $1000-$ \quad \\
\hline \hline
$h(b\bar{b})h(\gamma\gamma)$   & 1.0  & 4.0  & 1.9 & 0.67  & 0.25  & 0.17 \\
\hline
%\multicolumn{5}{c}{Backgrounds}\\
 $b\bar{b} \gamma\gamma$  & 2.0 & 5.4 & 2.5 & 1.1 & 0.69 & 0.55 \\
 $c\bar{c} \gamma\gamma$  & 2.3 & 3.2 & 0.63 & 0.79 & 0.25 & 0.22 \\
 $b\bar{b} \gamma j$             & 1.4  & 4.4 & 0.95 & 0.5  & 0.0035  & 0.24 \\
 $j j \gamma\gamma$            & 1.0  & 2.2 & 0.46 & 0.22  & 0.063  & 0.11 \\ 
$t\bar{t}\gamma$                   &  0.44 & 0.63 & 0.36 & 0.063  & 0.015  & 0.02 \\
 $t\bar{t}h(\gamma\gamma) $ & 1.4 & 2.7  & 0.96  & 0.3 & 0.12 & 0.057 \\
 $Z(b\bar{b})h(\gamma\gamma)$ &  0.29 & 0.49 & 0.22 & 0.09  & 0.058  & 0.24 \\
 $b\bar{b}h(\gamma\gamma)$  & 0.087  & 0.12 & 0.025 & 0.0049 & 0.0023 & 0.00048 \\
\hline
\end{tabular}
\caption{The expected number of events, using 3 ab$^{-1}$, at the HL-LHC for the exclusive analyses of the double Higgs production with ATLAS cuts~\cite{ATL-PHYS-PUB-2014-019}.}
\label{tab:bbaa:mhhBinAnalysis}
\end{table}
%%%%%%%%%%%%%%%%%%%%%%%%%%%%%%%%%%%%%%%

%%%%%%%%%%%%%%%%%%%%%%%%%%%%%%%%%%%%%%%%
%%%%%%%%%%%%%%%%%%%%%%%%%%%%%%%%%%%%%%%%
%%%%%%%%%%%%%%%%%%%%%%%%%%%%%%%%%%%%%%%%
%%%%%%%%%%%%%%%%%%%%%%%%%%%%%%%%%%%%%%%%
%%%%%%%%%%%%%%%%%%%%%%%%%%%%%%%%%%%%%%%%
%%%%%%%%%%%%%%%%%%%%%%%%%%%%%%%%%%%%%%%%
%%%%%%%%%%%%%%%%%%%%%%%%%%%%%%%%%%%%%%%%
\subsection{$b\bar{b}\tau^+\tau^-$ decay channel}
\label{sec:bbtautau}

The $\tau^+\tau^-$ system has three branches, fully hadronic ($\tau_{h} \tau_{h}$), semileptonic ($\tau_{l} \tau_{h}$), and dileptonic ($\tau_{l} \tau_{l}$) modes, where $h$ ($l$) stands for hadronic (leptonic) tau decays. For the leptonic tau decay, we consider only muons, or $l = \mu$, due to its high reconstruction efficiency. We include two leading final states, $\tau_{h} \tau_{h}$ and $\tau_{\mu} \tau_{h}$, in our analysis.

The typical signature of the leptonic tau decay is simply an isolated lepton as it decays to a lepton and two neutrinos, $\tau \rightarrow l + \nu_\tau + \nu_l$. The hadronic tau decay, on the other hand, goes through $\tau \rightarrow \nu_\tau + X$, where X denotes collimated hadrons. The traditional tau-tagging algorithm is based on that most hadronic activity is concentrated in the vicinity of the hardest track inside a jet, and it typically gives a percent level mistag rate for $50-60\%$ tag rate. A recent more sophisticated $\tau$ identification technique achieves $\mathcal{O}(1\%-0.1 \%)$ level misidentification rate for $50-60\%$ of tagging efficiency~\cite{CMS-PAS-TAU-16-002}. We take a somewhat semi-realistic approach for the $\tau$-jet identification. Just like a nominal $b$-tagging algorithm searching for a $b$ hadrons inside a jet, our $\tau$-tagging works in a similar way by searching for the truth-level $\tau$ parton inside a $R=0.4$ anti-$k_T$ jet. The $\tau$-jet candidate is further multiplied by an appropriate tag rate. In this study, we will take 50\% tagging efficiency and 0.48\% misidentification rate as central values. The impact of different tau-tagging performances will be examined in Section~\ref{sec:varyingFakes}. 

The event preselection of the $b\bar{b}\tau^+\tau^-$ final state starts from the lepton isolation which uses the same criterion as that in the previous section. Only the events without an isolated lepton are selected for the fully hadronic $b\bar{b}\tau^+ \tau^-$ final state, whereas exactly one isolated lepton is required for the semileptonic $b\bar{b}\tau^+ \tau^-$ final state. The remaining particles are clustered into anti-$k_T$ jets with a distance parameter $R=0.4$, and only the jets with $p_T (j) > 30$ GeV and $|\eta(j)| < 2.4$ are considered. As in the case of $b\bar{b} \gamma\gamma$, our heavy flavor/tau tagging algorithm runs iteratively over jets to classify them as $b$ jets, $c$ jets, light jets, or $\tau$ jets which are then iteratively paired into two $\tau$-jet candidates (one $\tau$-jet candidate and a lepton) and two $b$-jet candidates for the fully hadronic (semileptonic) $b\bar{b}\tau^+ \tau^-$. Since the decay products of both leptonic and hadronic taus include invisible neutrinos, the missing transverse momentum ($\slashed{\vec{p}}_T$) is defined such that it balances the $p_T$ sum of all visible final state particles including those from the $b\bar{b}$ system.

%%%%%%%%%%%%%%%%%%%%%%%%%%%%%%%%%%%%%%%%
%%%%%%%%%%%%%%%%%%%%%%%%%%%%%%%%%%%%%%%%
\subsubsection{Cut-based analysis}

In our cut-based analysis, we follow the similar strategy to CMS~\cite{CMS-PAS-FTR-15-002} except that the reconstruction of the $h\rightarrow \tau^+\tau^-$ resonance is differently done. We impose the following sets of cuts for the fully hadronic $b\bar{b}\tau^+\tau^-$ final state:
\begin{equation}
\begin{split}
&
p_T (\tau) > 60\ {\rm GeV}~ \quad {\rm or ~}\quad
p_{T>} (\tau) > 90\ {\rm GeV}~, \quad p_{T<}(\tau) > 45\ {\rm GeV}~,\\
&
p_{T}(b) > 30\ {\rm GeV}~, \quad
|\eta(\tau)| < 2.1~, \quad
\quad |\eta(b)| < 2.4~,\\
\end{split}
\end{equation}
where $p_{T>}(\tau)$ ($p_{T<}(\tau)$) is the transverse momentum for the harder (softer) $\tau$-jet candidate, and for the semileptonic $b\bar{b}\tau^+ \tau^-$ final state ,
\begin{equation}
\begin{split}
&
p_{T} (\tau) > 45\ {\rm GeV}~, \hspace{0.2cm}
p_T (b) > 30\ {\rm GeV}~, \hspace{0.2cm}
|\eta (\tau_h)| < 2.1~, \hspace{0.2cm}
|\eta (\tau_l)| < 2.5~, \hspace{0.2cm}
\quad |\eta(b)| < 2.4~,\\
\end{split}
\end{equation}
where $\tau_h$ ($\tau_l$) denotes the hadronic (leptonic) tau and $p_T(\tau)$ cut is imposed on both the hadronic and leptonic taus.

After two tau candidates are identified, we reconstruct the resonant $h\rightarrow \tau^+\tau^-$ system. A difficulty arises due to neutrinos in the decay chain, $h\rightarrow \tau_h\tau_h\, (\tau_h\tau_l)\rightarrow 2 \nu_\tau + X\, (2 \nu_\tau + \nu_l + X$) which causes irreducible loss of the information. A well-established method used by the experimental collaborations to reconstruct the $h\rightarrow \tau^+\tau^-$ resonance utilizes the likelihood method (see~\cite{CMS-PAS-FTR-15-002} for CMS $\tau^+\tau^-$ reconstruction). Realizing this sophisticated approach in our study is beyond the scope of our work. Instead, we adopt a rather simple prescription utilizing the transverse mass type variable, proposed in~\cite{Barr:2011he}. The transverse mass in proposed in~\cite{Barr:2011he} is called $m_{\tau\tau}$ Higgs-bound which is defined as 
\begin{equation}
\label{eq:mHBound}
 m^{\rm reco}_{\tau\tau} (=  m^{\rm Higgs-bound}_{\tau\tau}) \equiv \min_{\{q_1, q_2: \chi \}} \sqrt{H^\mu H_\mu}~,
\end{equation}
where the $H_\mu$ is the total momentum of the $\tau^+\tau^-$-system,
\begin{equation}
  H^\mu = p^\mu_1 + q_1^\mu + p^\mu_2 + q_2^\mu~.
\end{equation}
The invariant mass of $H_\mu\, (=\sqrt{H_\mu H^\mu})$ is minimized over all possible assignments for the invisible four momenta, $q^\mu_1$ and $q^\mu_2$, subject to the constraint $\chi$ (see Eq.~(\ref{eq:missingEt:constraint})) as is defined in Eq.~(\ref{eq:mHBound}). The $p^\mu_1\, (=p_T(\tau_1)$) and $p^\mu_2\, (=p_T(\tau_2))$ are visible four momenta of two tau candidates. The $p_{1,\, 2}^{\mu}$ and $q_{1\, 2}^{\mu}$ are supposed to satisfy the following internal mass constraints,
\begin{align} 
\label{eq:ditau:constraints}
& q^2_1= 0~, \quad  
   q^2_2 = 0~, \quad 
  (p_1 + q_1)^2 = m^2_\tau~, \quad 
   (p_2 + q_2)^2 = m^2_\tau~,
\end{align}
where the transverse momenta of $q^\mu_1$ and $q^\mu_2$ are subject to the experimental constraint,
\begin{align} \label{eq:missingEt:constraint}
 \chi:\ \vec{q}_{1T} + \vec{q}_{2T} = \slashed{\vec{p}}_T~.
\end{align}

The minimization procedure in Eq.~(\ref{eq:mHBound}) returns the $m_{\tau\tau}^{\rm reco}$ value if there exists a solution for the constraints in Eq.~(\ref{eq:ditau:constraints}). As was pointed out in~\cite{Barr:2011he}, a solution exists for almost all events at the parton-level analysis with the truth neutrinos whereas the failure rate for a solution significantly increases when the method is applied to the hadron level events. We find that only $\sim 25$\% of the events succeed to find a solution at the hadron level analysis and the rate is sensitive to the missing transverse momentum measurement. Another cause of the failure to find a solution is the extra neutrinos from the $b\bar{b}$ system via the semileptonic decays of the $b$ hadrons. It was also pointed out in~\cite{Barr:2011he} that there exists at least a solution if and only if the following condition is satisfied:
\begin{equation}
\label{eq:MT2:inequality}
 M_{T2}(p_1,\, p_2,\, \slashed{\vec{p}}_T, m_{inv} = 0 ) < m_\tau~,
\end{equation}
where the stransverse mass $M_{T2}$~\cite{Lester:1999tx,Barr:2003rg,Cheng:2008hk} is defined as
\begin{eqnarray}
M_{T2} (p_1,\, p_2,\, \slashed{\vec{p}}_T, m_{inv} ) \equiv  \min_{ \slashed{p}_{1T} + \slashed{p}_{2T} =\, \slashed{\vec{p}}_T } \big [ \max \big\{ M_T ( p_1  , \slashed{p}_{1T} ), \; M_T ( p_2  , \slashed{p}_{2T} ) \big\}  \big ] ~,
\label{eq:mT2}
\end{eqnarray}
where $M_T$ is the transverse mass, $\slashed{p}_{1T,\, 2T}$ hypothesized transverse momenta, and $m_{inv}$ the mass of the invisible particle. 
We have confirmed that the failed events to find a solution indeed do not satisfy the constraint in Eq.~(\ref{eq:MT2:inequality}). While the parton-level events with the truth neutrinos nicely shows the end point behavior bounded by $m_\tau \sim 1.78$ GeV (on-shell mass of the tau lepton), the $M_{T2}$ distribution of the hadron level events severely overshoot above $m_{\tau} \sim 1.78$ GeV. 

%%%%%%%%%%%%%%%%%%%%%%%%%%%%%%%%%%%%%%%%%%%%%%%%%%%%%%%%%%%%
%%%%%%%%%%%%%
\begin{figure}[!htb!] %[tbp]
	\centering
	\includegraphics[width=0.45\linewidth]{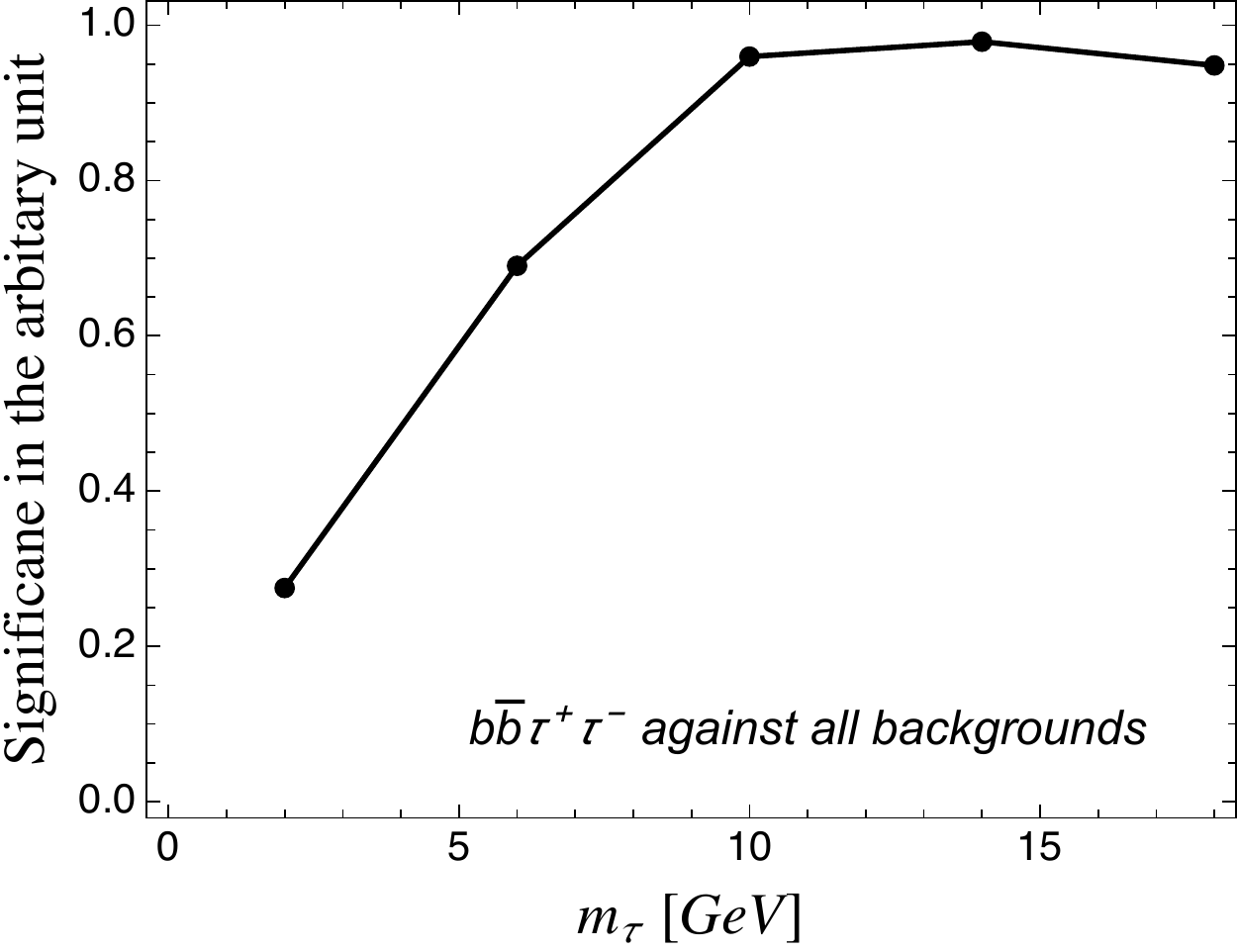} \quad
	\includegraphics[width=0.45\linewidth]{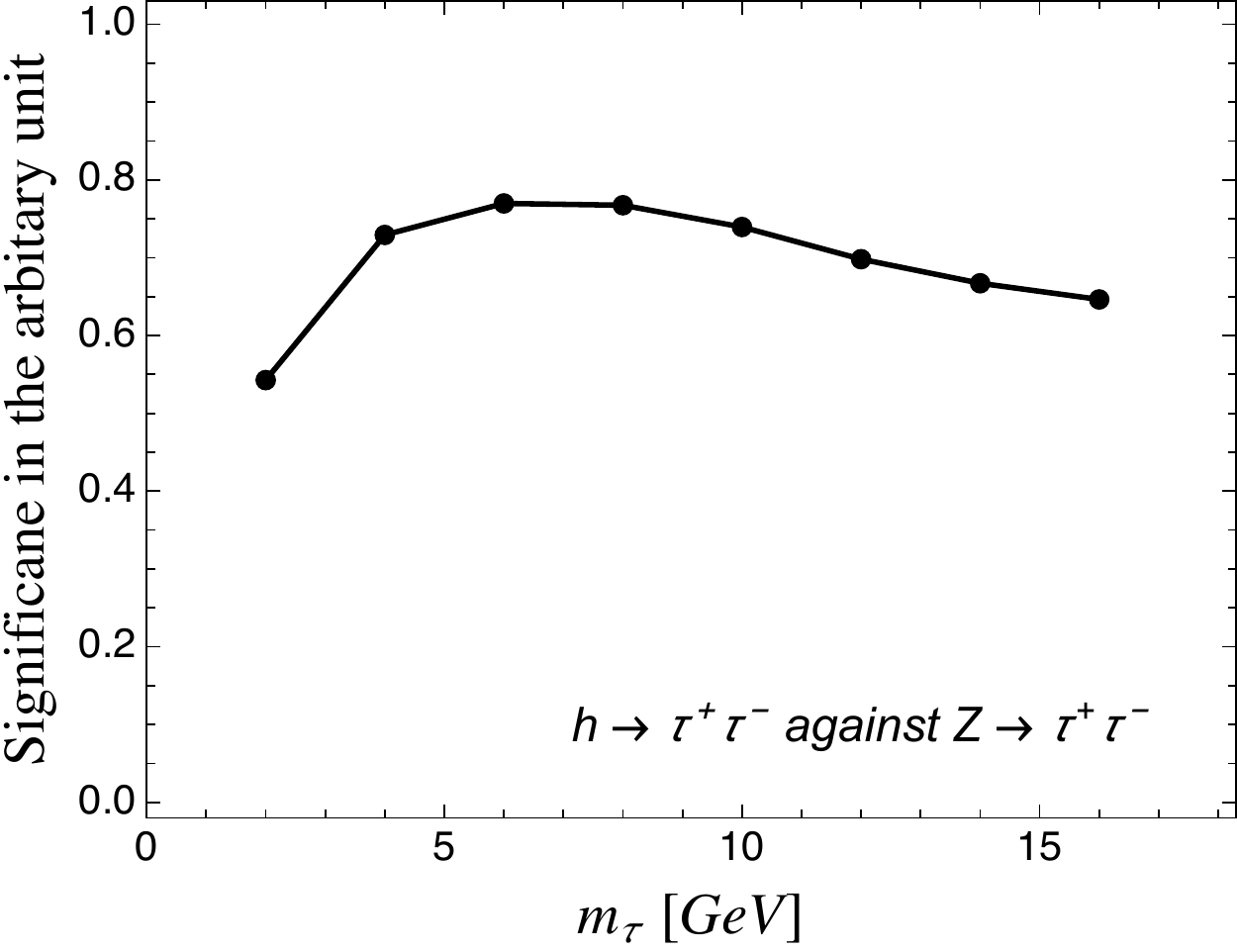}
	\caption{The significance as a function of $m_\tau$ for the $b\bar{b}\tau^+\tau^-$ process against all the backgrounds (left) and $h\rightarrow \tau^+\tau^-$ against $Z\rightarrow \tau^+\tau^-$ (right). The significance was maximized for each $m_\tau$ value by adjusting a cut on the $m_{\tau\tau}^{\rm reco}$, while the other cuts remain the same}
	\label{fig:mtau:optimize}
\end{figure}
%%%%%%%%%%%%%
%%%%%%%%%%%%%%%%%%%%%%%%%%%%%%%%%%%%%%%%%%%%%%%%%%%%%%%%%%%%
Motivated by our observation regarding the inequality in Eq.~(\ref{eq:MT2:inequality}), we take the internal mass $m_\tau$ as an independent handle that controls the success rate for the solution instead of fixing it to the on-shell tau mass. We vary the $m_\tau$ variable to retain more events that find a solution. In our analysis, we choose a $m_\tau$ value in such a way that it maximizes the significance. To this end, we estimate the significance as a function of $m_\tau$ against all the backgrounds to the $b\bar{b}\tau^+\tau^-$ decay channel of the double Higgs production.
It turns out that the significance is optimized around $m_\tau \sim 10$ GeV (see the left panel of Fig.~\ref{fig:mtau:optimize}). To measure the importance of the separation of the signal against the $Z$ + jets background, we also estimate the significance for the pure $h \rightarrow \tau^+\tau^-$ samples against the $Z\rightarrow \tau^+\tau^-$ events in a separate simulation~\footnote{The pure $h \rightarrow \tau^+\tau^-$ ($Z \rightarrow \tau^+\tau^-$) samples were generated via the $Zh$ process where the $Z$-boson (Higgs) decay was switched off, and both samples were further processed via the parton shower and hadronization. The $\tau$-jet candidates and leptons were smeared accordingly based on the description in Sec.~\ref{sec:DHiggsHL}. The significance was maximized for each $m_\tau$ value by adjusting a cut on the $m_{\tau\tau}^{\rm reco}$, while the other cuts remain the same (similarly for the left panel of Fig.~\ref{fig:mtau:optimize}).}. 
We find that the significance is maximized around $6 \sim 8$ GeV (see the right panel of Fig.~\ref{fig:mtau:optimize}) which is close to the value ($\sim 10$ GeV) obtained using the samples of the $b\bar{b}\tau^+\tau^-$ decay channel. Throughout our analysis for the $b\bar{b}\tau^+\tau^-$ decay channel, we set to $m_\tau = 10$ GeV with which we find almost all events are retained~\footnote{In the likelihood method used by CMS~\cite{Chatrchyan:2014nva}, all events are retained by reconstructing the invisible momenta from neutrinos taking into account the finite resolution of the missing transverse momentum.}.

%%%%%%%%%%%%%%%%%%%%%%%%%%%%%%%%%%%%%%%%%%
\begin{figure}[tbp]
\begin{center}
\includegraphics[width=0.45\linewidth]{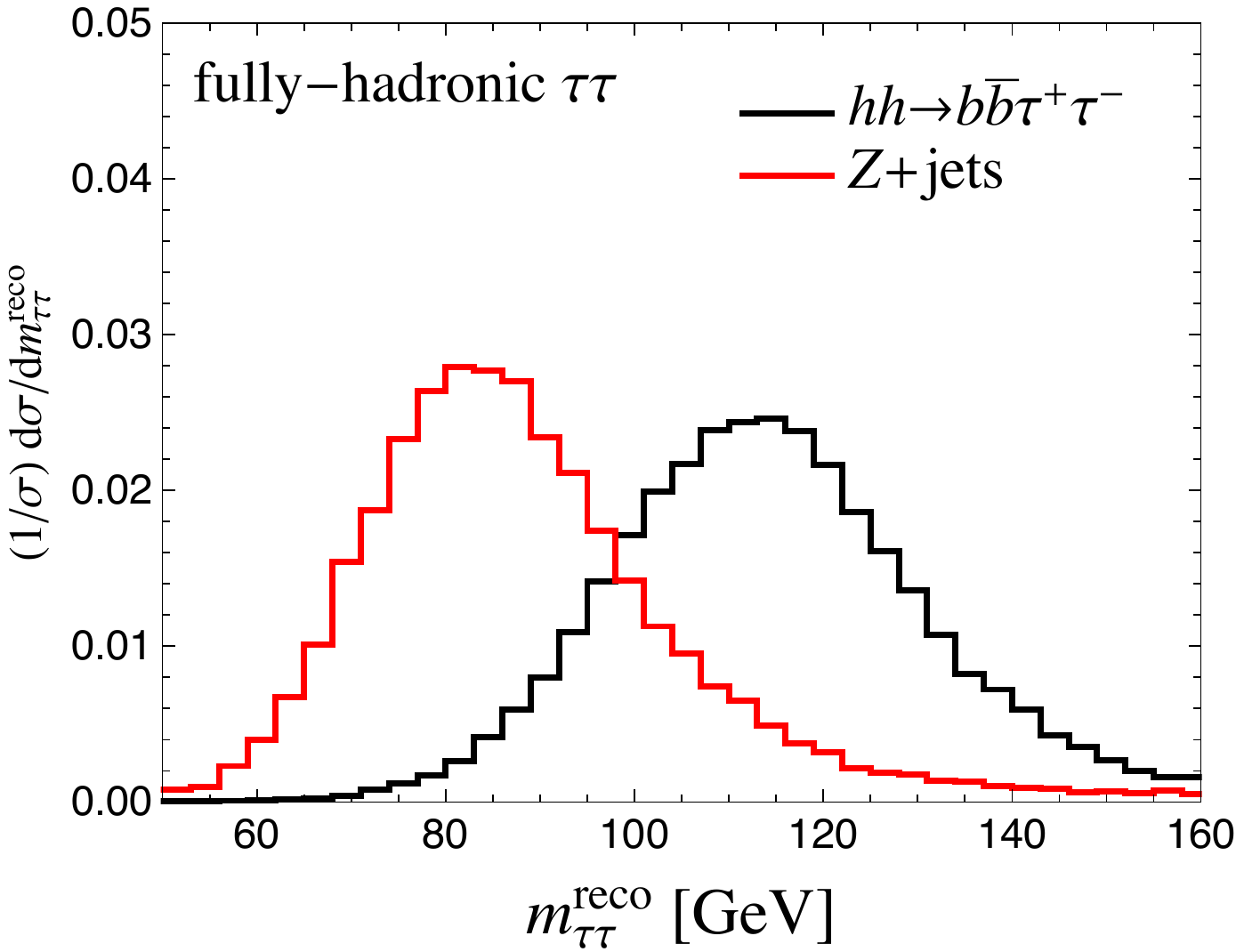}
\includegraphics[width=0.45\linewidth]{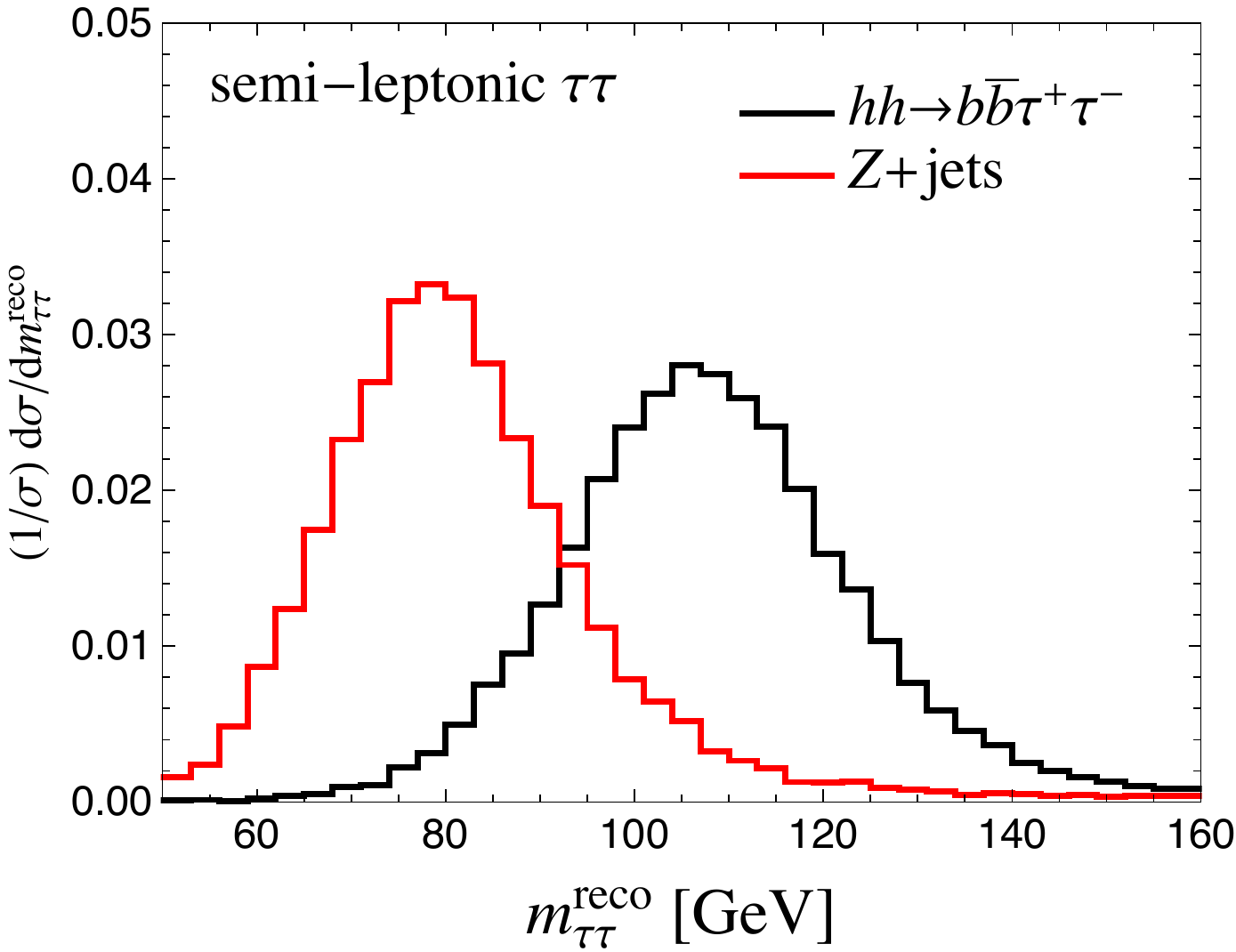}
\caption{
The distributions of $m_{\tau\tau}^{\rm reco}$ for the signal and $Z(\tau^+\tau^-)+{\rm jets}$ samples.  The left and right figures are distributions for the fully hadronic and semileptonic channels.
}
\label{fig:mtautau:HadronicSemileptonic}
\end{center}
\end{figure}
%%%%%%%%%%%%%%%%%%%%%%%%%%%%%%%%%%%%%%%%%%%

The reconstructed invariant mass $m^{\rm reco}_{\tau\tau}$ are shown in Fig.~\ref{fig:mtautau:HadronicSemileptonic} for both fully hadronic and semileptonic modes of the $\tau^+\tau^-$ system. The shape of the invariant mass $m^{\rm reco}_{\tau\tau}$ for both decay modes appear to be roughly symmetric at a lower central value than $125$ GeV (as a property of the transverse mass). Unlike the case of the likelihood method used by the experimental collaborations where the mass resolution for the fully hadronic case is better than the semileptonic case~\footnote{For instance, the {\tt SVFIT} algorithm in CMS achieves the relative $m_{\tau\tau}$ resolution of about 10\%, 15\%, and 20\% in the $\tau_h\tau_h$, $\tau_l\tau_h$, and $\tau_l\tau_{l'}$ decay channels, respectively~\cite{Chatrchyan:2014nva}.}, we instead see in Fig.~\ref{fig:mtautau:HadronicSemileptonic} that the semileptonic case has slightly better mass resolution than the fully hadronic case. This observation is an artifact caused by our $\tau$-tagging algorithm. Since our $\tau$-tagging algorithm searches for a truth $\tau$ lepton within the isolation cone around a jet, instead of requiring no hadronic activities within the annulus formed by the smaller and bigger cones around the hardest track as done in the traditional $\tau$-tagging, the resulting $\tau$-jet candidate is necessarily more contaminated than the case that passed more stringent criterion. However, we suspect that it does not change our result.

The reconstruction of the invariant mass of the $b\bar{b}$ system, on the other hand, takes the similar approach to the case of $b\bar{b} \gamma\gamma$. We restrict the events to those in the following mass windows:
\begin{align}
   &105 < m^{\rm reco}_{\tau\tau} < 130\ {\rm GeV} \text{ (fully hadronic)}~,\\
   &100 < m^{\rm reco}_{\tau\tau} < 125\ {\rm GeV} \text{ (semileptonic)}~,
\end{align}
for the $\tau^+\tau^-$ system and
\begin{align}
 95 < m^{\rm reco}_{bb} < 135\ {\rm GeV}~,
\end{align}
for the $b\bar{b}$ system.
%%%%%%%%%%%%%%%%%%%%%%%%%%%%%%%%%%%%%%%%%
\begin{figure}[!ht]
\begin{center}
\includegraphics[width=0.45\linewidth]{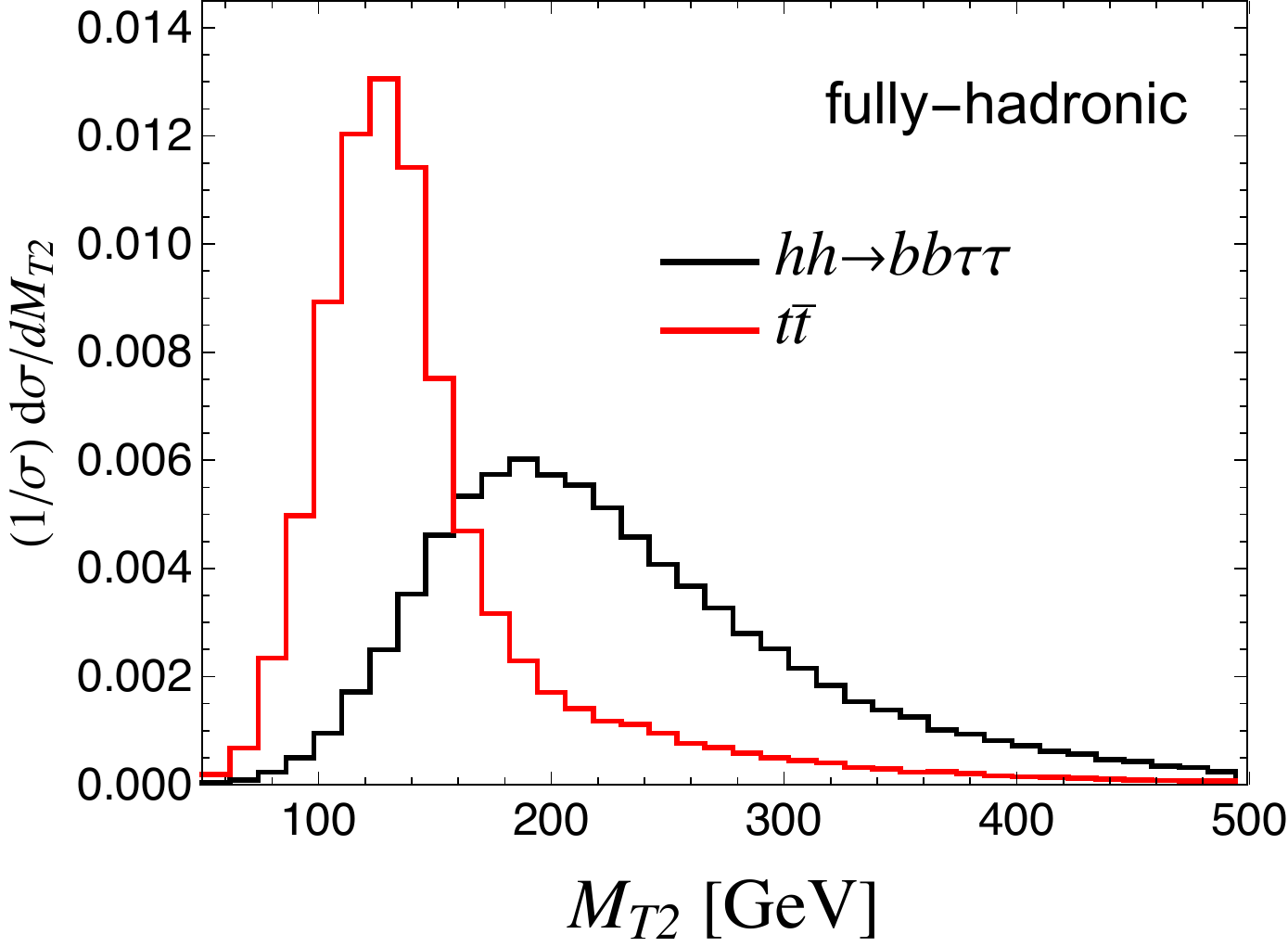}
\includegraphics[width=0.45\linewidth]{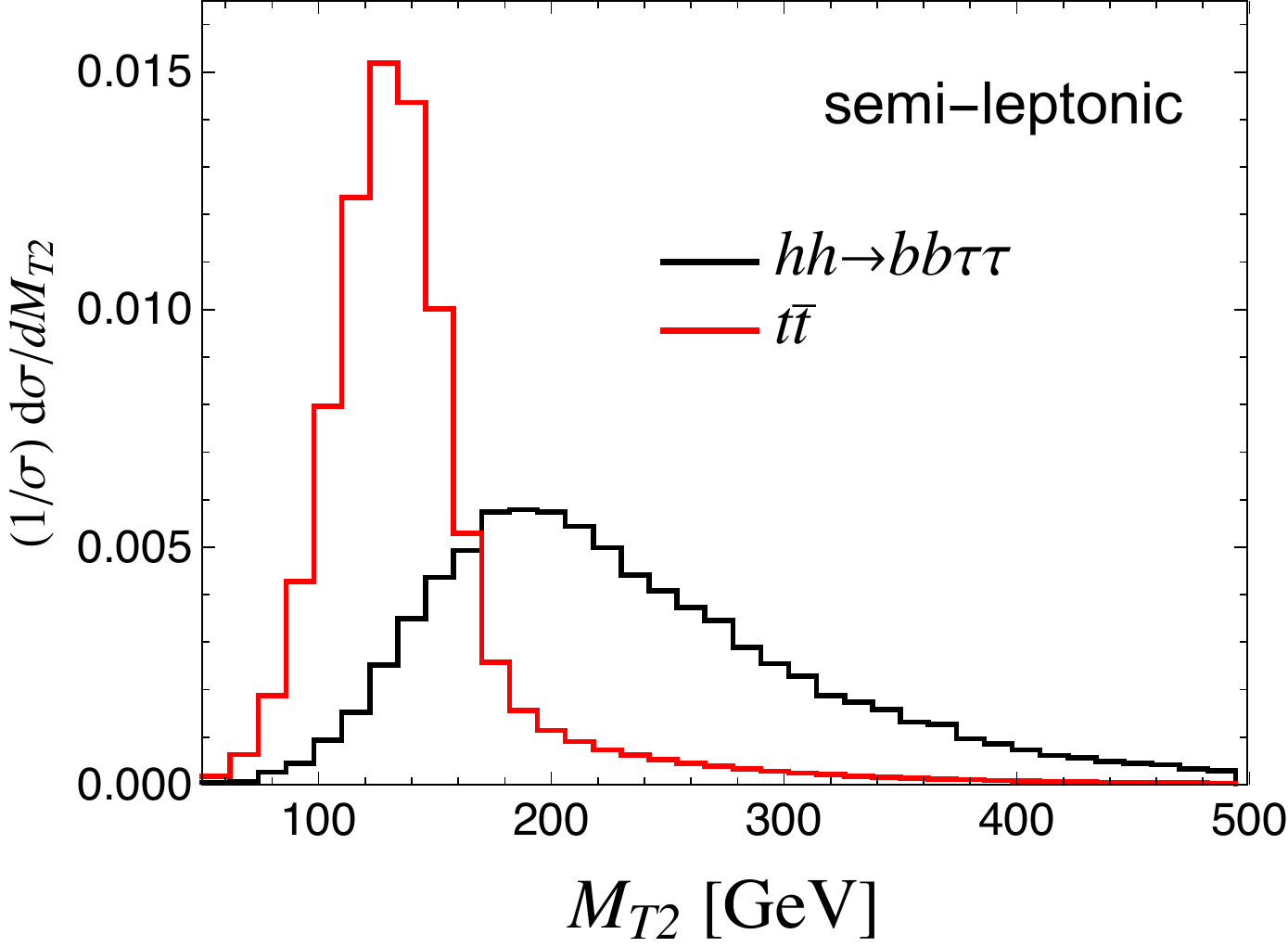}
\caption{
Distributions of $M_{T2}$ for the fully hadronic and the semileptonic channels.
}
\label{fig:mT2}
\end{center}
\end{figure}
%%%%%%%%%%%%%%%%%%%%%%%%%%%%%%%%%%%%%%%%%

While the $t \bar{t}$ is a dominant background, we introduce another $M_{T2}$ variable whose distribution is expected to be bounded from the above around top mass~\cite{Cheng:2008hk} (see~\cite{Barr:2013tda} for the related discussion). The $M_{T2}$ is computed using two $b$-jet candidates and two visible $\tau$ candidates, with the minimization over two possible combinations,
\begin{equation}
M_{T2} \equiv \min  \Big [  M_{T2}( p (b_1, \tau_{1}),\, p (b_2, \tau_{2}),\, \slashed{\vec{p}}_T , m_{inv} ), M_{T2}( p (b_1, \tau_{2}),\, p (b_2, \tau_{1}),\, \slashed{\vec{p}}_T , m_{inv} )  \Big]~,
\end{equation}
where $p (b, \tau)$ denotes the total four vector of a $b$ jet and visible $\tau$ candidate, namely $p (b, \tau) = p_T(b)+p_T(\tau)$ , and $\slashed{\vec{p}}_T$ is a total missing transverse momentum with $m_{inv} = 0$. 

The distributions of $M_{T2}$ for the fully hadronic and the semileptonic modes are shown in Fig.~\ref{fig:mT2}. The $M_{T2}$ distribution of the $t \bar{t}$ background tends to be bounded from the above near the top mass, although we observe a long high invariant mass tail beyond the end point. This, however, does not apply to the signal topology in which the $M_{T2}$ distribution can be widely spread out, and therefore, we impose the following cut on the $M_{T2}$ variable:
\begin{equation}
\label{eq:MT2:ttbar}
 M_{T2} > 180\ {\rm GeV}~.
\end{equation}

The results obtained by our cut-based analyses at the HL-LHC, assuming the data of 3 ab$^{-1}$, are presented in Table~\ref{tab:bbtautauSummary}, which looks roughly consistent with the CMS analysis~\cite{CMS-PAS-FTR-15-002}~\footnote{The validation of  our analysis by reproducing the CMS result is not straightforward as our treatment of $\tau^+\tau^-$ system differs from the CMS one. Nevertheless, we obtain the similar result.}.  
The significance of the SM for the fully hadronic (semileptonic) decay channel is estimated to be 0.9 (0.6) which is very close to 0.89 (0.55) of the CMS analysis~\cite{CMS-PAS-FTR-15-002}. 
Even though CMS has employed a MVA method for the semileptonic decay channel, we find that the MVA method improves the significance only $\sim 40\%$ for the semileptonic channel compared to the cut-based analysis (see Table~\ref{tab:bbtautauSummary}).

%%%%%%%%%%%%%%%%%%%%%%%%%%%%
\begin{table}[tbp]
\centering
\begin{tabular}{ccccc}  
\hline
 \quad Expected yields (3 ab$^{-1}$)  & \multicolumn{2}{c}{fully hadronic $\tau_h\tau_h$}  & \multicolumn{2}{c}{Semileptonic $\tau_{\mu}\tau_h$}\\
 \quad   & \quad Cut-based Analysis \quad & \quad MVA \quad & \quad Cut-based Analysis \quad & \quad MVA \quad \\
\hline \hline
%\multicolumn{5}{c}{Signal: Standard Model}\\
 $h(b\bar{b})h(\tau^+\tau^-)$  & 5.71 & 10. & 5.7  & 7.9 \\
\hline
%\multicolumn{5}{c}{Backgrounds}\\
 $t\bar{t}$    & 2.31 & 4.46  & 44.8 & 28.8 \\
 $t\bar{t}h$  & 7.63 &  7.37 & 13.1 & 12.9 \\
 $t\bar{t}V$  & 3.14 & 2.74  & 5.12 & 7.87 \\
 $tW$           & 5.37 & 7.52  & 28.3 & 12.6 \\
$Z(\tau^+\tau^-)+{\rm jets}$ & 18.4 & 25.0  & 10.1 & 32.7 \\
 $hZ$ & 1.72 & 2.22  & 1.16 & 3.8 \\
 $VV            $& 0.38 & 0.98  & 3.41 & 2.43
 \\ \hline
 Total backgrounds & 40. &  50.3 & 106 & 101 \\
\hline \hline
\end{tabular}
\caption{
The number of signal and backgrounds for the cut-based analysis in $b\bar{b}\tau^+\tau^-$ channel.  
The lepton in the semileptonic channel includes only muon.
}
\label{tab:bbtautauSummary}
\end{table}
%%%%%%%%%%%%%%%%%%%%%%%%%%%%%

%%%%%%%%%%%%%%%%%%%%%%%%%%%%%%%%%%%%%%%%
%%%%%%%%%%%%%%%%%%%%%%%%%%%%%%%%%%%%%%%%
\subsubsection{Multivariate Analysis}
\label{MVA_bbtt}

We employ the BDT algorithm for the multivariate analysis in the $b\bar{b}\tau^+\tau^-$ channel.  
First, we prepare training samples for each background.  The events are required to pass following cuts:
\begin{align}
\begin{split}
&
\text{number of $\tau$ candidate}\geq 2,~~~
\text{number of $b$ jet}\geq 2,~~~\\
&
60 < m_{bb}^{\rm reco}  <160~\text{GeV}, \quad 
60 < m_{\tau\tau}^{\rm reco}  < 160~\text{GeV}, \\
&
p_T (\tau) > 30~\text{GeV},~~~
p_T (b) > 30~\text{GeV},~~~
|\eta (\tau)| < 2.5,~~~
|\eta (b)| < 2.5, 
\label{train_condition_tautau}
\end{split}
\end{align}
where $\tau$ denotes either a hadronic or leptonic tau.
The variables used in our BDT analysis, in addition to those in Eq.~(\ref{train_condition_tautau}), are  $\Delta R(b,b)$, $\Delta R (\tau,\tau)$, $\Delta R (b_i, \tau_j)$ ($i,j=1,2$), $p_T (bb)$, $p_T (\tau\tau)$,  $m^{\rm reco}_{hh}$, $p_T(hh)$, $y (hh)$, $N_{jet}$,  $E_T^{\rm miss}$, and $m_{\rm eff}$, where an effective mass is defined by $m_{\rm eff}=\sqrt{(p(\tau_{1})+p(\tau_{2})+\slashed{\vec{p}}_T)^2}$ ($p(\tau_{1})$ and $p(\tau_{2})$ as visible four momenta of two $\tau$ candidates, as was described before).  The detail of our BDT analysis can be found in Appendix~\ref{app:MVA}.

\begin{figure}[!t]
\begin{center}
\includegraphics[width=0.45\linewidth]{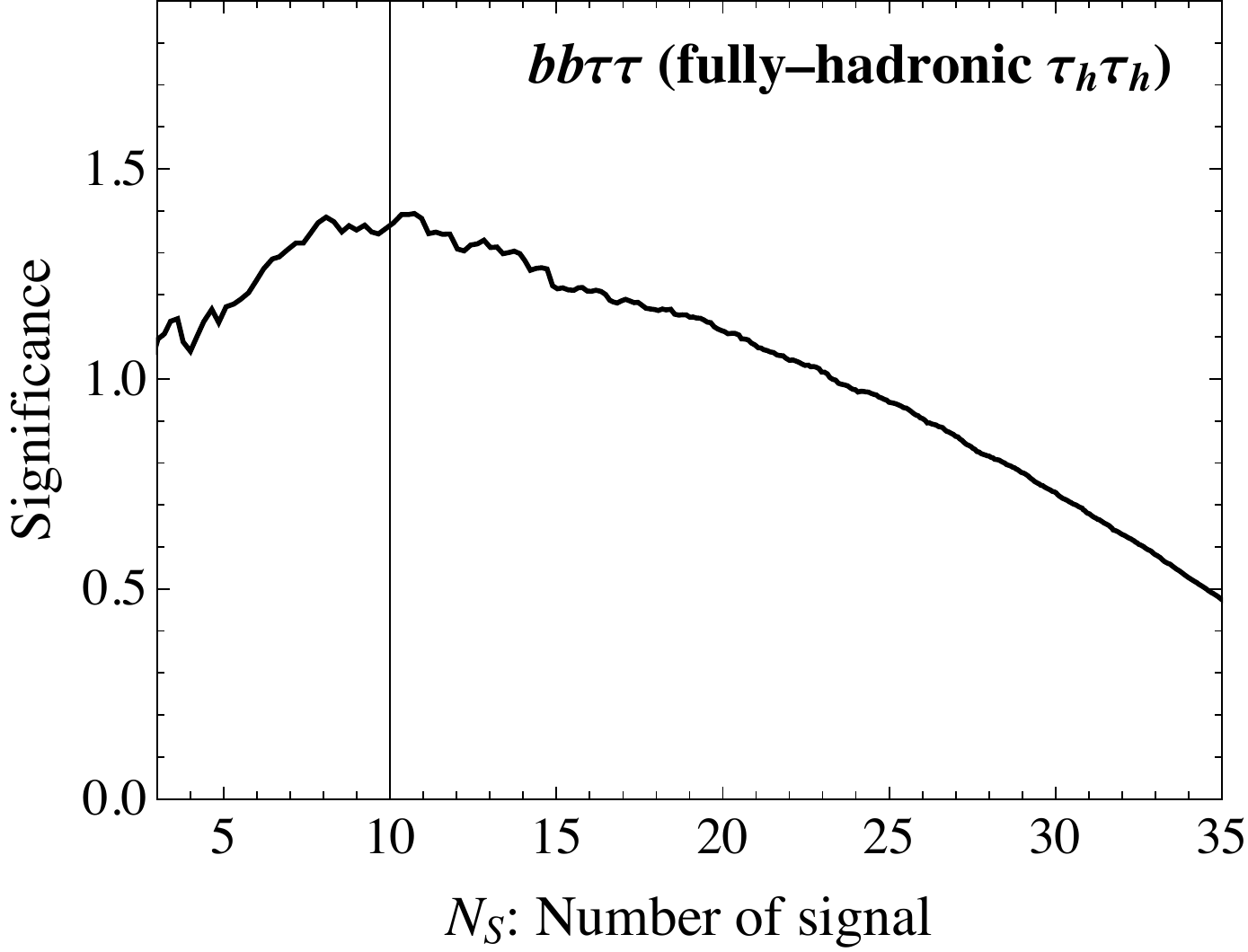}
\includegraphics[width=0.45\linewidth]{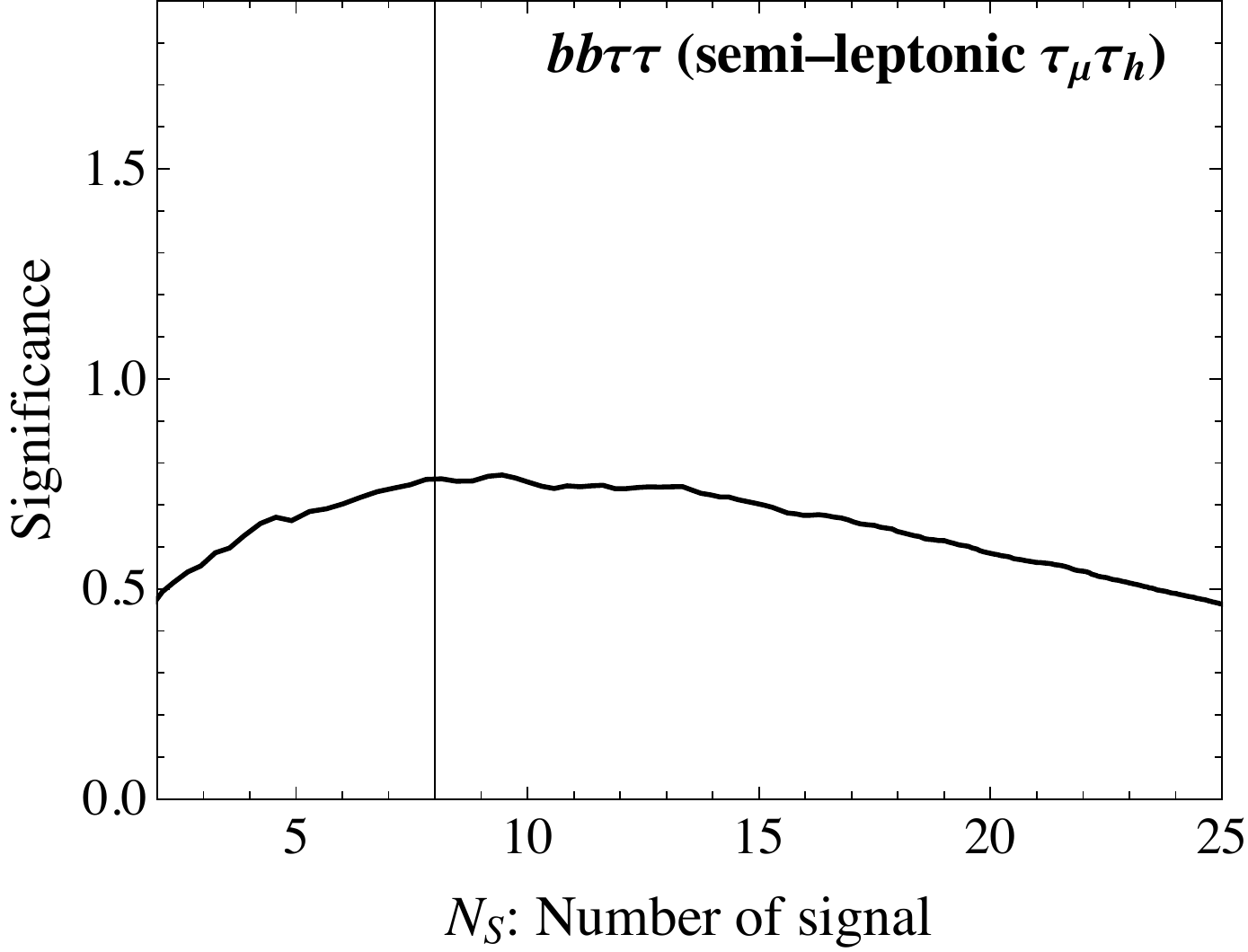}
\caption{
The significance as a function of $N_S$ (the number of signal events) for the fully hadronic $b\bar{b}\tau^+\tau^-$ mode (left) and the semileptonic $b\bar{b}\tau^+\tau^-$ mode (right).}
\label{BDT_signf_bbtt}
\end{center}
\end{figure}

The significance as a function of the signal rate is illustrated in Fig.~\ref{BDT_signf_bbtt} which shows that the improvement of the significance is more pronounced for the fully hadronic channel whereas the effect is mild for the semileptonic channel. 
The BDT cut minima are chosen so that the signal rates are 10 and 7.9 for the fully hadronic and the semileptonic channels where the significances are almost maximal.  We use these points as reference points in the BDT analysis for the $b\bar{b}\tau^+\tau^-$ channel.  Using the reference BDT cut minimum, the significances can reach 1.5 and 0.8 for the fully hadronic and the semileptonic channels.

%%%%%%%%%%%%%%%%%%%%%%%%%%%%%%%%%%%%%%%%
%%%%%%%%%%%%%%%%%%%%%%%%%%%%%%%%%%%%%%%%
%%%%%%%%%%%%%%%%%%%%%%%%%%%%%%%%%%%%%%%%
%%%%%%%%%%%%%%%%%%%%%%%%%%%%%%%%%%%%%%%%
%%%%%%%%%%%%%%%%%%%%%%%%%%%%%%%%%%%%%%%%
%%%%%%%%%%%%%%%%%%%%%%%%%%%%%%%%%%%%%%%%
%%%%%%%%%%%%%%%%%%%%%%%%%%%%%%%%%%%%%%%%
\section{Combined analysis}
\label{sec:results}

In this section, we will combine our results, obtained through the analyses in Section~\ref{sec:DHiggsHL}, for two decay channels of the double Higgs production, namely $b\bar{b}\gamma\gamma$ and $b\bar{b}\tau^+\tau^-$. For the latter, we include the semileptonic and fully hadronic decay  modes of the $\tau^+\tau^-$ system. We will perform the combined analysis of both the cut-based (with ATLAS cuts as a default choice unless specified) and multivariate type to extract the sensitivity on the coefficients of the effective Lagrangian. To this end, we will exploit the exclusive (and inclusive) analysis for the $b\bar{b}\gamma\gamma$ decay channel and two inclusive analyses for the $b\bar{b}\tau^+\tau^-$ decay channel. The types of plots presented are intended to be similar to those in~\cite{Azatov:2015oxa} to make the comparison clear. We will demonstrate several ways for further improvements, focusing on the sensitivity on the Higgs self-coupling, for simplicity. Throughout our work, we will use the Bayesian statistical method to derive the sensitivity.

%%%%%%%%%%%%%%%%%%%%%%%%%%%%%%%%%%%%%%%%
%%%%%%%%%%%%%%%%%%%%%%%%%%%%%%%%%%%%%%%%
\subsection{Sensitivity on anomalous Higgs couplings in an EFT Lagrangian}
\label{sec:SenAnomalCouplings}

 %%%%%%%%%%%%% Updated with new 5M BSM samples
\begin{figure}[!htb!] %[tbp]
	\centering
	\includegraphics[width=0.43\linewidth]{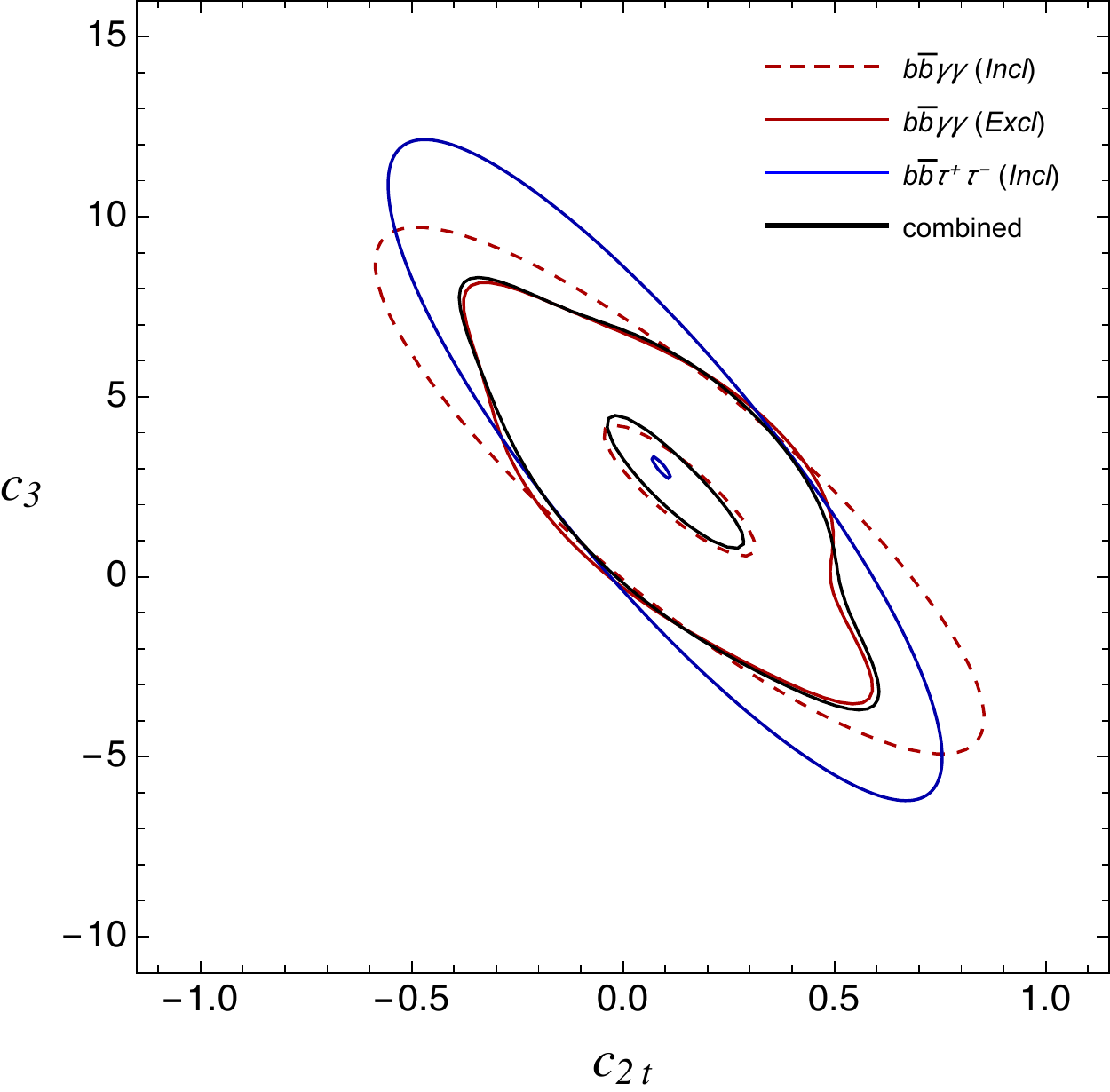}\quad
	\includegraphics[width=0.453\linewidth]{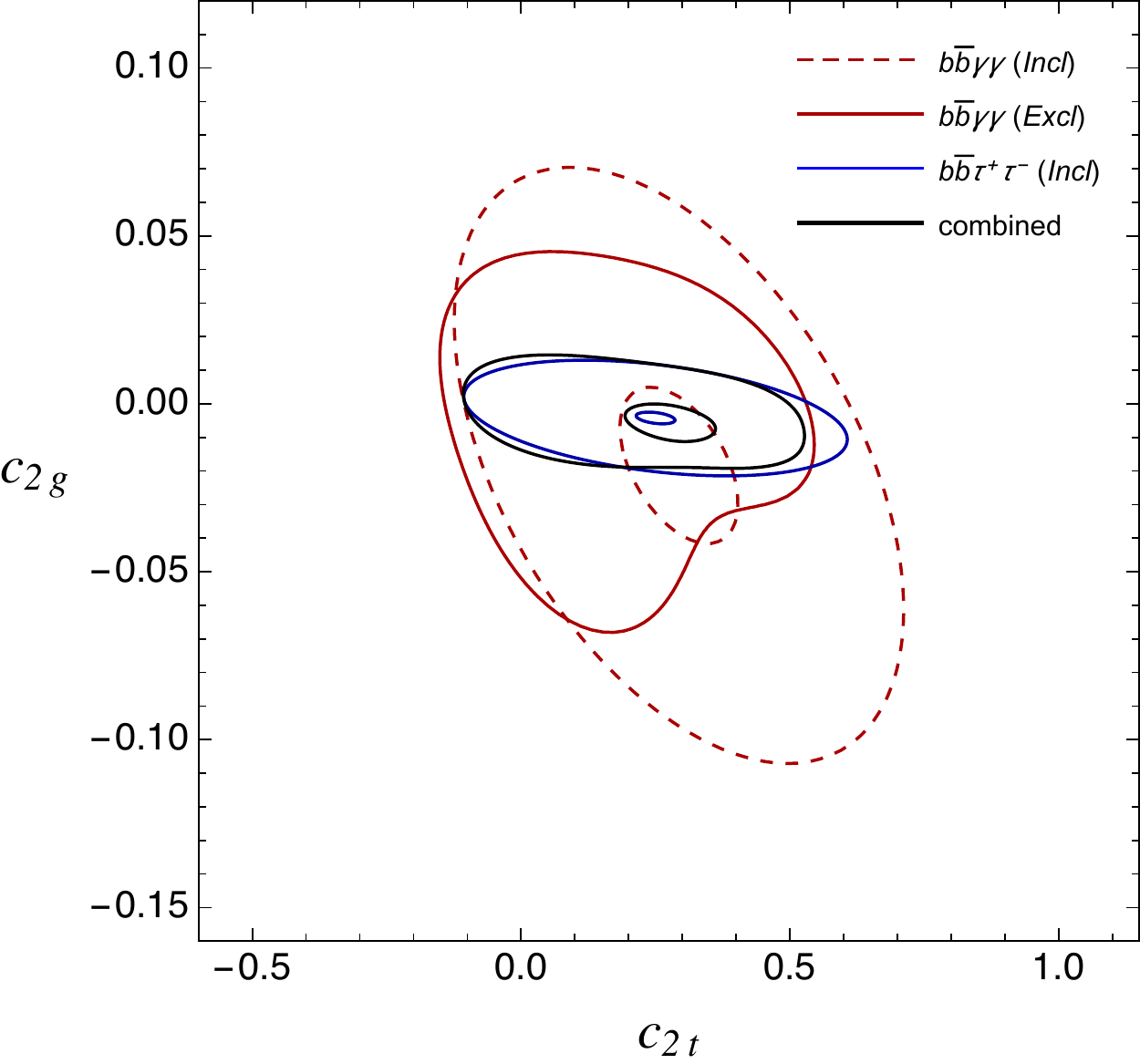}
	\caption{Left: 68\% probability contours of the likelihoods in $(c_{2t},\, c_3)$ plane using the cut-based analysis: the inclusive analysis of the double Higgs production in the $b\bar{b}\gamma\gamma$ (dashed red), the exclusive analysis of the $b\bar{b}\gamma\gamma$ (solid red), inclusive analysis of the $b\bar{b}\tau^+\tau^-$ (solid blue), and combined analysis (solid black). The combined analysis combines the exclusive analysis of the $b\bar{b}\gamma\gamma$ channel and the inclusive one of the $b\bar{b}\tau^+\tau^-$ channel. Right: similarly, 68\% probability contours of the likelihood in $(c_{2t},\, c_{2g})$ plane with the same color/line codes as the left plot. The $c_g=c_{2g} =0,\, c_t =1$ in the left plot ($c_g =0,\, c_3=c_t =1$ in the right plot) was chosen.}
	\label{fig:c2tVSxx:ATLASCMS:noERR}
\end{figure}
%%%%%%%%%%%%%

Firstly, we consider the sensitivity on the EFT coefficients in the nonlinear basis where the EFT coefficients are not related to each other. In Fig.~\ref{fig:c2tVSxx:ATLASCMS:noERR}, we illustrate 68\% probability contours of the likelihoods in $(c_{2t},\, c_3)$ and $(c_{2t},\, c_{2g})$ planes for the $b\bar{b}\gamma\gamma$ and $b\bar{b}\tau^+\tau^-$ channels using the result obtained by our cut-based analysis.  The other EFT coefficients not displayed in Fig.~\ref{fig:c2tVSxx:ATLASCMS:noERR}  were set to the SM values. The impact of the marginalization on the contours will be discussed later. In making contours of the combined analysis, we take the exclusive analysis of the $b\bar{b}\gamma\gamma$ channel as a default. We show the contours of the inclusive analysis of the $b\bar{b}\gamma\gamma$ only for the purpose of illustration.

The two plots in Fig.~\ref{fig:c2tVSxx:ATLASCMS:noERR} reveal interesting aspects of the combining various channels. The $b\bar{b}\tau^+\tau^-$ shows a similar strong anticorrelation in $(c_{2t},\, c_3)$ plane with the case of $b\bar{b}\gamma\gamma$, but the sensitivity is mainly determined by the $b\bar{b}\gamma\gamma$ due to its much stronger sensitivity. This indicates that combining two channels is the least beneficial for the determination of the Higgs self-coupling. On the contrary, the $b\bar{b}\tau^+\tau^-$ channel outperforms the $b\bar{b}\gamma\gamma$ in the $(c_{2t},\, c_{2g})$ plane, and the sensitivity is mainly set by the $b\bar{b}\tau^+\tau^-$ except the positive deviation of $c_{2t}$. Although the orientation of the contour of the $b\bar{b}\tau^+\tau^-$ appears orthogonal to that of the $b\bar{b}\gamma\gamma$, the almost region of the $b\bar{b}\tau^+\tau^-$ lies inside the contour of the $b\bar{b}\gamma\gamma$, and its benefit is mainly for the determination of $c_{2g}$. 

%%%%%%%%%%%%%  Updated with new 5M BSM samples
\begin{figure}[!htb!] %[tbp]
	\centering
	\includegraphics[width=0.430\linewidth]{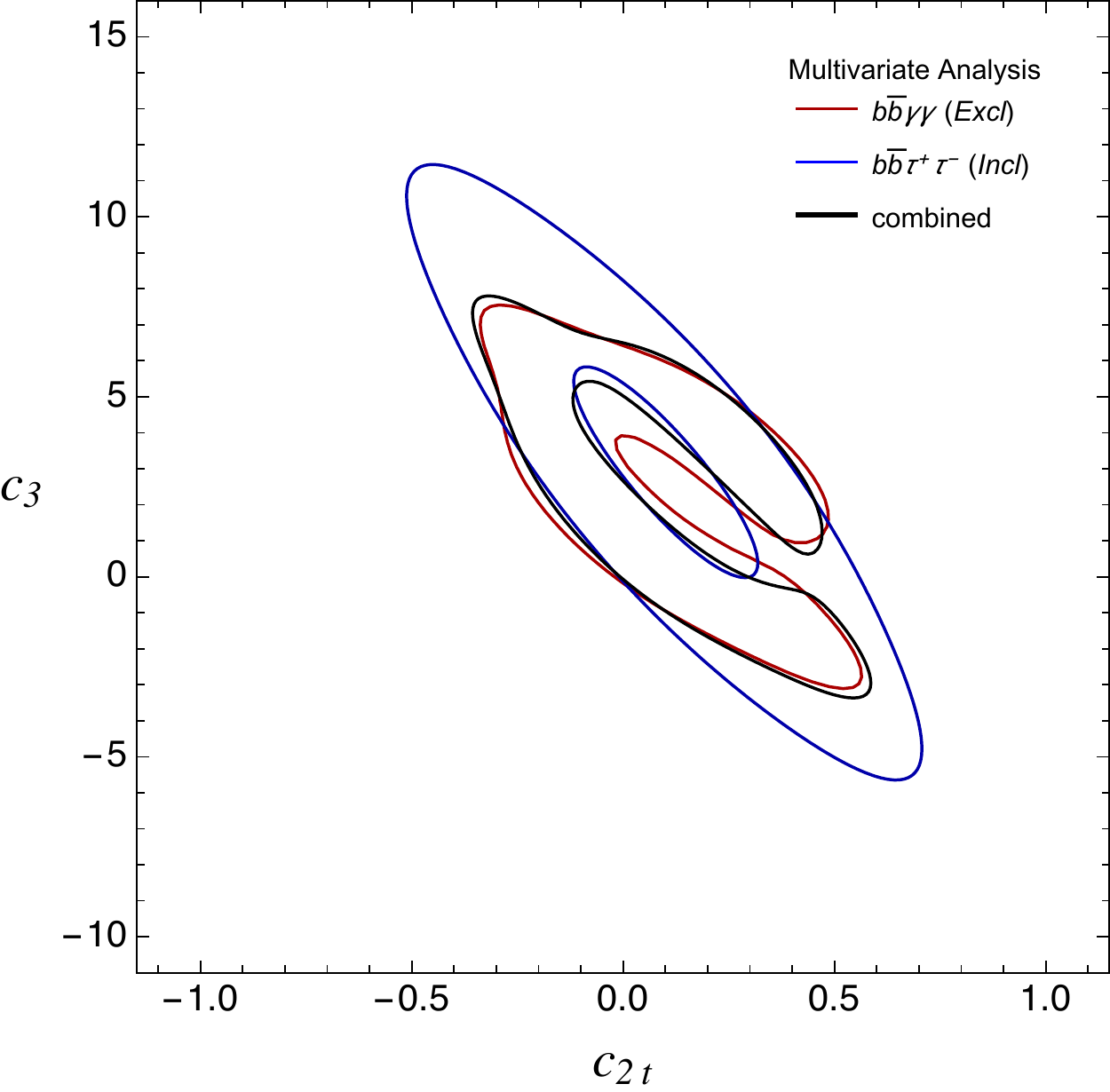}\quad
         \includegraphics[width=0.454\linewidth]{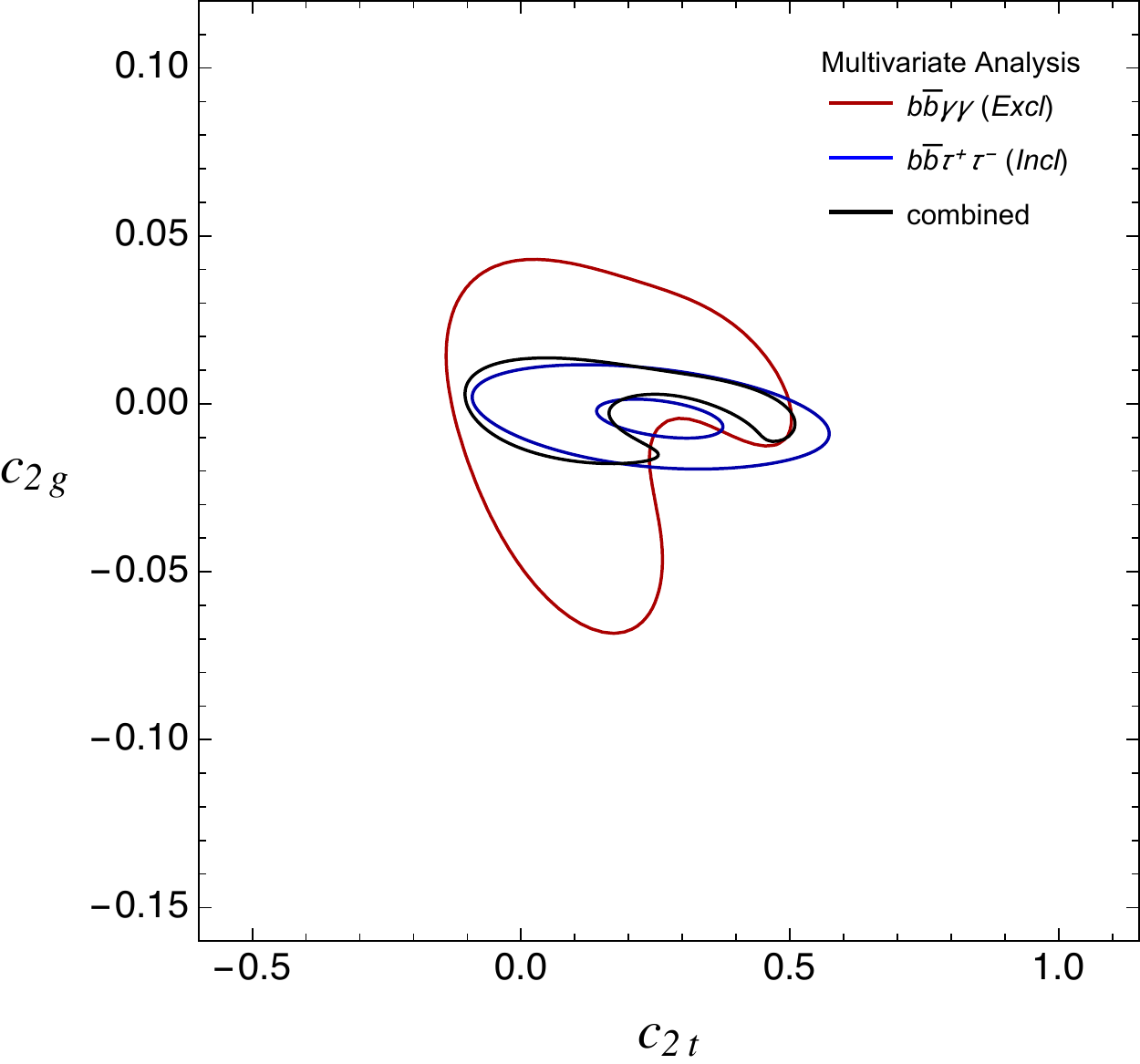}
	\caption{Left: 68\% probability contours of the likelihoods in $(c_{2t},\, c_3)$ plane using the multivariate analysis: the exclusive analysis of the double Higgs production in the $b\bar{b}\gamma\gamma$ (red), inclusive analysis of the $b\bar{b}\tau^+\tau^-$ (blue), and the combined analysis (black). Right: similarly, 68\% probability contours of the likelihood in $(c_{2t},\, c_{2g})$ plane with the same color/line codes as the left plot. The $c_g=c_{2g} =0,\, c_t =1$ in the left plot ($c_g =0,\, c_3=c_t =1$ in the right plot) was chosen.}
	\label{fig:c2tVSxx:BDT2BDT2:noERR}
\end{figure}
%%%%%%%%%%%%%

The 68\% probability contours of the likelihoods in the same $(c_{2t},\, c_3)$ and $(c_{2t},\, c_{2g})$ planes using the sophisticated multivariate analysis are presented in Fig.~\ref{fig:c2tVSxx:BDT2BDT2:noERR}. As was expected from the Tables~\ref{tab:bbaaSummary} and~\ref{tab:bbtautauSummary}, the multivariate analysis improves the significance up to the factor of 2 for each decay channel, and consequently, the contours noticeably shrink in both planes of Fig.~\ref{fig:c2tVSxx:BDT2BDT2:noERR} in such a way that the overlap between the two decay channels is reduced. The combined analysis breaks a significant amount of degeneracy.

%%%%%%%%%%%%%%%%%%  Updated with new 5M BSM samples
\begin{figure}[!htb!] %[tbp]
	\centering
        \includegraphics[width=0.43\linewidth]{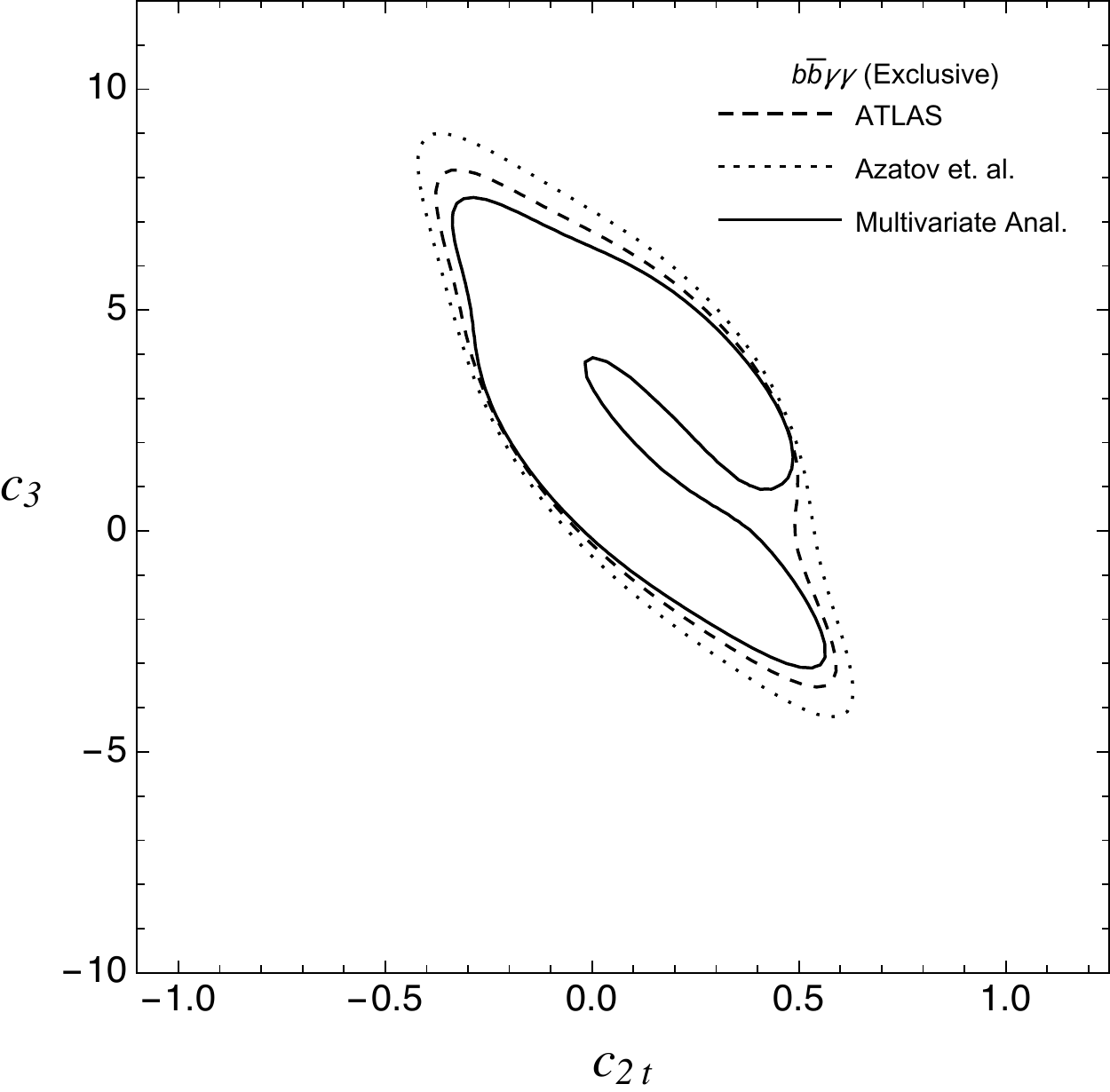}\quad
	\includegraphics[width=0.454\linewidth]{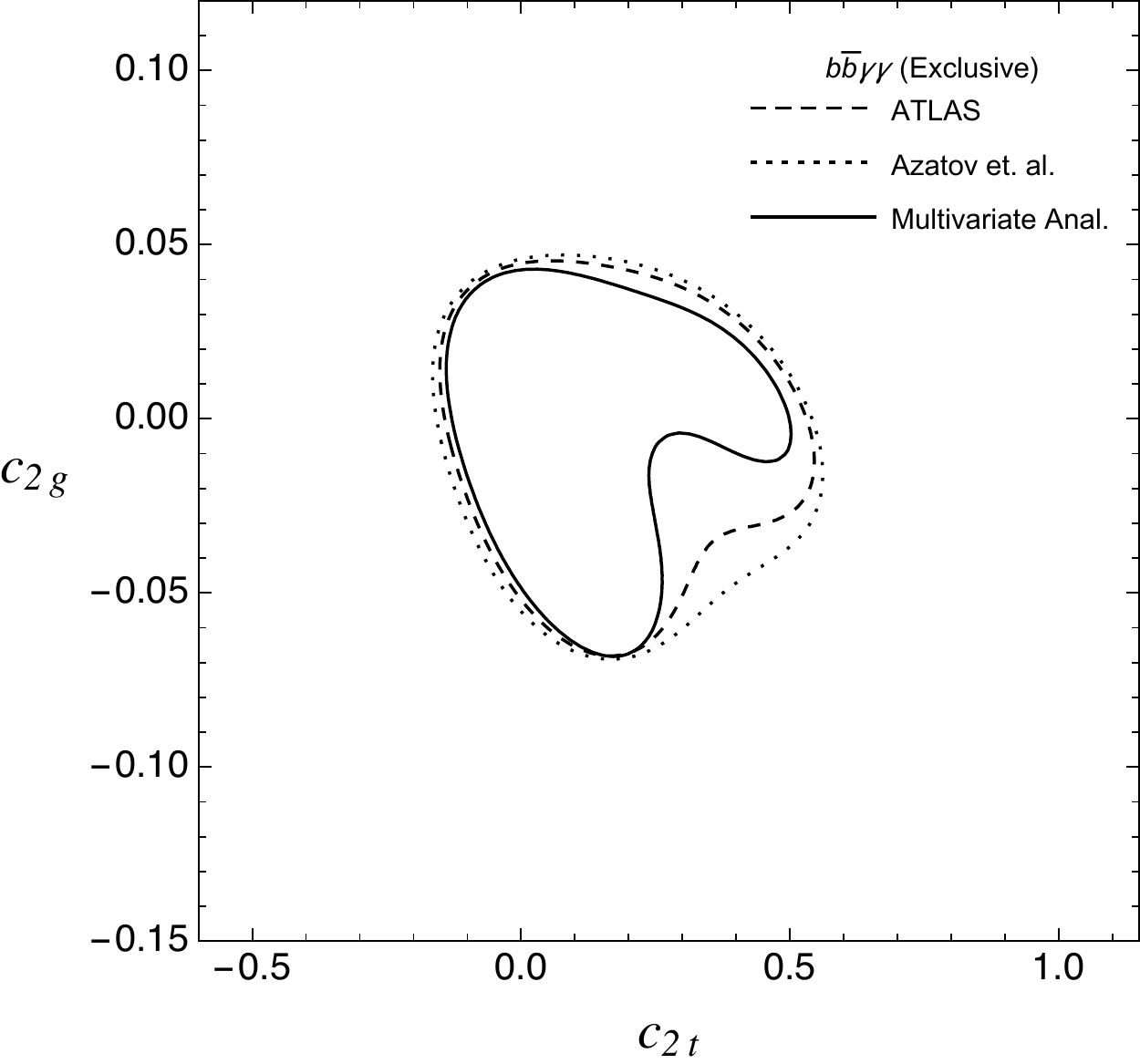}
	\caption{Left: 68\% probability contours of the likelihoods of the double Higgs production in the $b\bar{b}\gamma\gamma$ decay channel in the $(c_{2t},\, c_3)$ plane (left) and $(c_{2t},\, c_{2g})$ plane (right): with ATLAS cuts (dashed), with cuts in~\cite{Azatov:2015oxa} except the modification mentioned before (dotted), with multivariate analysis (solid). The $c_g=c_{2g} =0,\, c_t =1$ in the left plot ($c_g =0,\, c_3=c_t =1$ in the right plot) was chosen.}
	\label{fig:c2tVSxx:aabb:cutbased:noERR}
\end{figure}
%%%%%%%%%%%%%%%%%%

As is evident in Table~\ref{tab:bbaaSummary}, the significance of the SM by our newly done cut-based analysis with cuts in~\cite{Azatov:2015oxa} is worse than the result in~\cite{Azatov:2015oxa} after including a larger set of backgrounds (with a larger mistag rate for the $c$-flavor jets), applying jet smearing, and $p_T$-dependent photon reconstruction efficiency etc. We show 68\% probability contours of the likelihoods of the $b\bar{b}\gamma\gamma$ (exclusive analysis) in Fig.~\ref{fig:c2tVSxx:aabb:cutbased:noERR} using our cut-based analysis with the aforementioned modification. For the purpose of comparison, we also show 68\% probability contours of the likelihoods for the cut-based analysis with ATLAS cuts and for the multivariate analysis. For simplicity, we have not performed the marginalization over all other parameters in Fig.~\ref{fig:c2tVSxx:aabb:cutbased:noERR} where we expect only a minor effect due to it. We see in Fig.~\ref{fig:c2tVSxx:aabb:cutbased:noERR} that the sensitivity extracted from our newly done cut-based analysis is worse compared to~\cite{Azatov:2015oxa}, but the lost sensitivity is recovered by the multivariate analysis. As a result, the constrained region by the multivariate analysis look similar to the region in~\cite{Azatov:2015oxa}.

%%%%%%%%%%%%%
\begin{figure}[!htb!] %[tbp]
	\centering
	\includegraphics[width=0.44\linewidth]{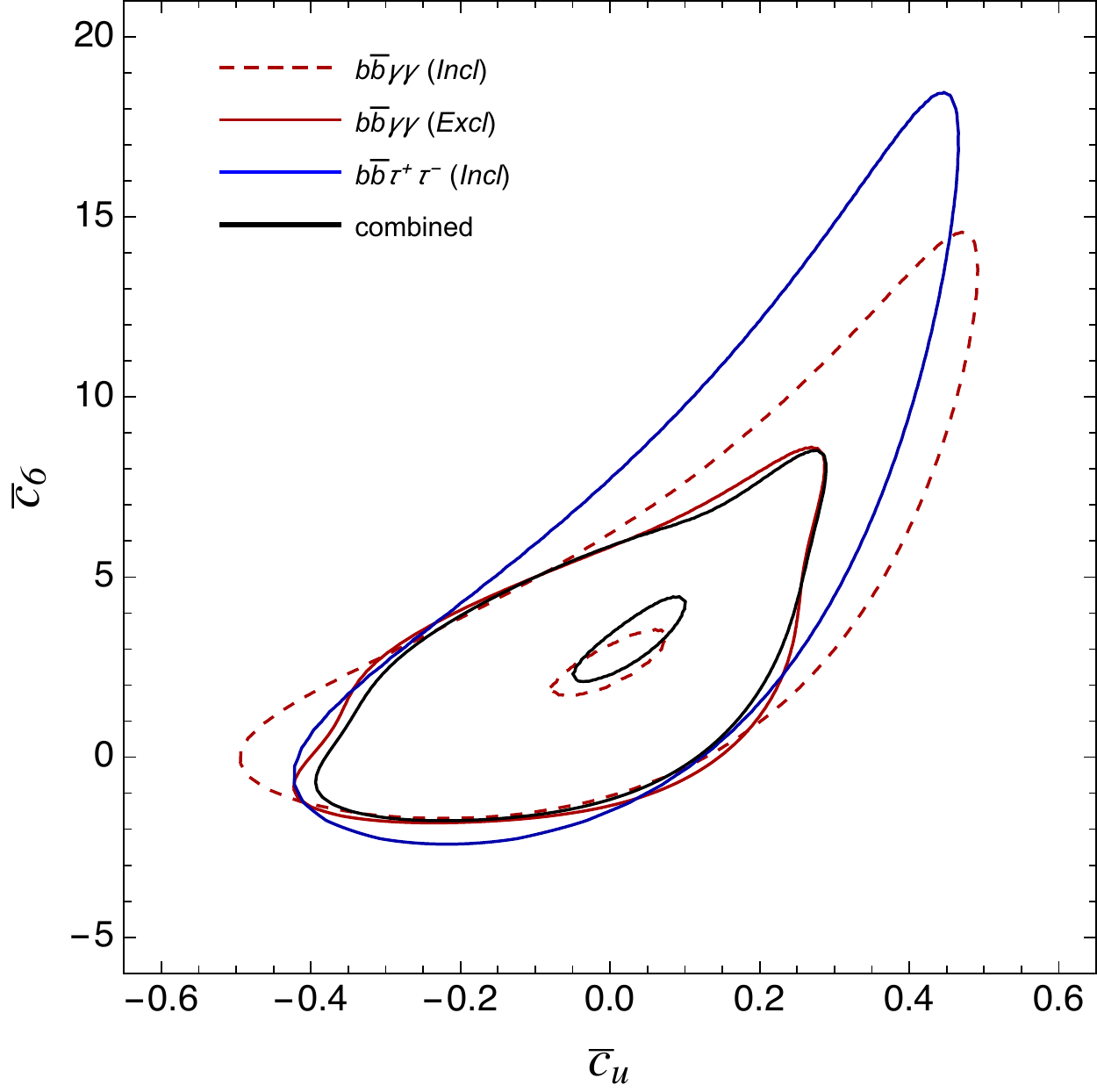}\quad
	\includegraphics[width=0.437\linewidth]{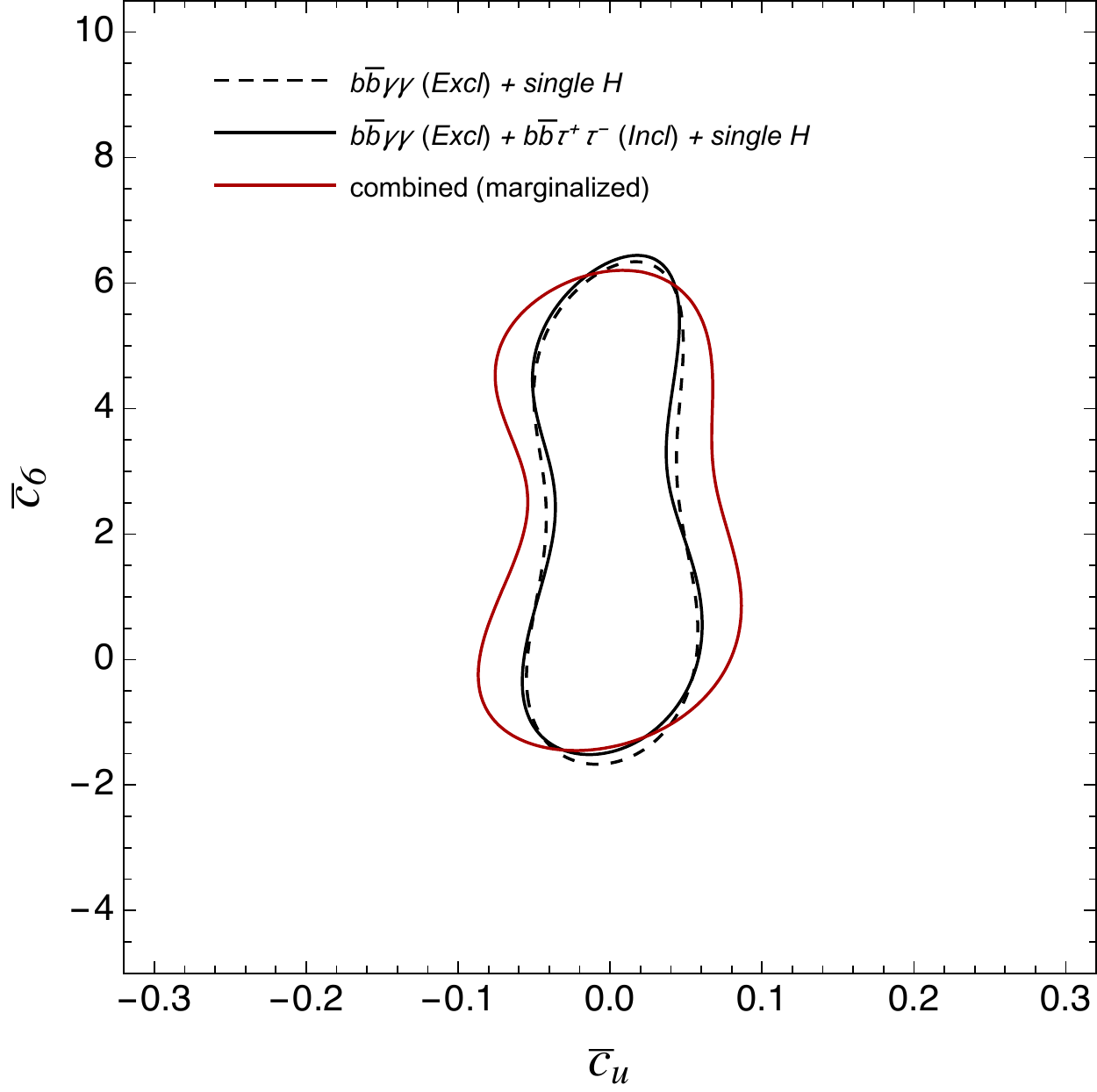}
	\caption{Left: 68\% probability contours of the likelihoods in $(\bar{c}_u,\, \bar{c}_6)$ plane using the cut-based analysis of the double Higgs production process. The combined analysis corresponds to the exclusive $b\bar{b}\gamma\gamma$ (solid red) plus the inclusive $b\bar{b}\tau^+\tau^-$ (dashed red). Right: 68\% probability contours of the likelihood of the exclusive $b\bar{b}\gamma\gamma$ plus single Higgs processes (dashed black).  The $b\bar{b}\tau^+\tau^-$ is further combined (solid black). $\bar{c}_H=\bar{c}_g=\bar{c}_d =0$ was set in all contours except the solid red line in the right plot where the likelihood was marginalized over $\bar{c}_H$, $\bar{c}_g$, and $\bar{c}_d$.}
	\label{fig:cuVSc6:HH:ATLASCMS:noERR}
\end{figure}
%%%%%%%%%%%%%

We repeat similar exercise for the coefficients of the effective Lagrangian in the linear basis. The set of the coefficients considered in this study includes $\bar{c}_H$, $\bar{c}_u$, $\bar{c}_6$, $\bar{c}_{g}$, and $\bar{c}_{d}$ while the remaining parameters such as $\bar{c}_{\gamma}$ is set to the SM values. Among those coefficients, $\bar{c}_H$, $\bar{c}_u$, $\bar{c}_g$, $\bar{c}_d$ are also constrained by the single Higgs data. We take the ATLAS projection of the single Higgs processes at the HL-LHC using 3 ab$^{-1}$~\cite{ATL-PHYS-PUB-2013-014}. So far, we have not performed the marginalization over other parameters. Since the marginalization over muti-variables is time consuming, we will examine the effect of the marginalization on the sensitivity only for limited cases in the linear basis. 

In Fig.~\ref{fig:cuVSc6:HH:ATLASCMS:noERR}, we show 68\% probability contours of the likelihoods of the double Higgs production process (with and without folding in single Higgs processes) in $(\bar{c}_u,\, \bar{c}_6)$ plane using our cut-based analysis with ATLAS cuts. While the combined analysis in the left panel of Fig.~\ref{fig:cuVSc6:HH:ATLASCMS:noERR} combines only the exclusive analysis of the $b\bar{b}\gamma\gamma$ and the inclusive one of $b\bar{b}\tau^+\tau^-$, the single Higgs processes is folded in the combined analysis in the right panel of Fig.~\ref{fig:cuVSc6:HH:ATLASCMS:noERR}. The Fig.~\ref{fig:cuVSc6:HH:ATLASCMS:noERR} indicates that having additional $b\bar{b}\tau^+\tau^-$ channel is least beneficial in $(\bar{c}_u,\, \bar{c}_6)$ plane as we have already observed a similar property in the nonlinear basis.
Comparing the two plots in Fig.~\ref{fig:cuVSc6:HH:ATLASCMS:noERR} indicates that, the single Higgs processes, mainly $t\bar{t}h$, are effective in constraining $\bar{c}_u$, namely the deviation of the up-type Yukawa coupling. We performed the marginalization for the combined analysis in the right panel of Fig.~\ref{fig:cuVSc6:HH:ATLASCMS:noERR} over $\bar{c}_g$~\footnote{We also have performed the marginalization over $\bar{c}_g$, $\bar{c}_H$, $\bar{c}_d$. We find that the effect of the marginalization over $\bar{c}_g$ dominates over the other two parameters which indicates that marginalizing over $\bar{c}_H$ and $\bar{c}_d$ causes only a negligible effect. In what follows, $\bar{c}_H$ and $\bar{c}_d$ will be set to the SM values, namely $\bar{c}_H=\bar{c}_d=0$ to save the computation time, and the marginalization will be performed only over $\bar{c}_g$ (or $\bar{c}_g$ and $\bar{c}_u$ when deriving the 1D likelihood of $\bar{c}_6$).} with the priors from the single Higgs data (shown as solid-red line). The effect of the marginalization is broadening the sensitivity on $\bar{c}_u$.

%%%%%%%%%%%%%
\begin{figure}[!htb!] %[tbp]
	\centering
	\includegraphics[width=0.44\linewidth]{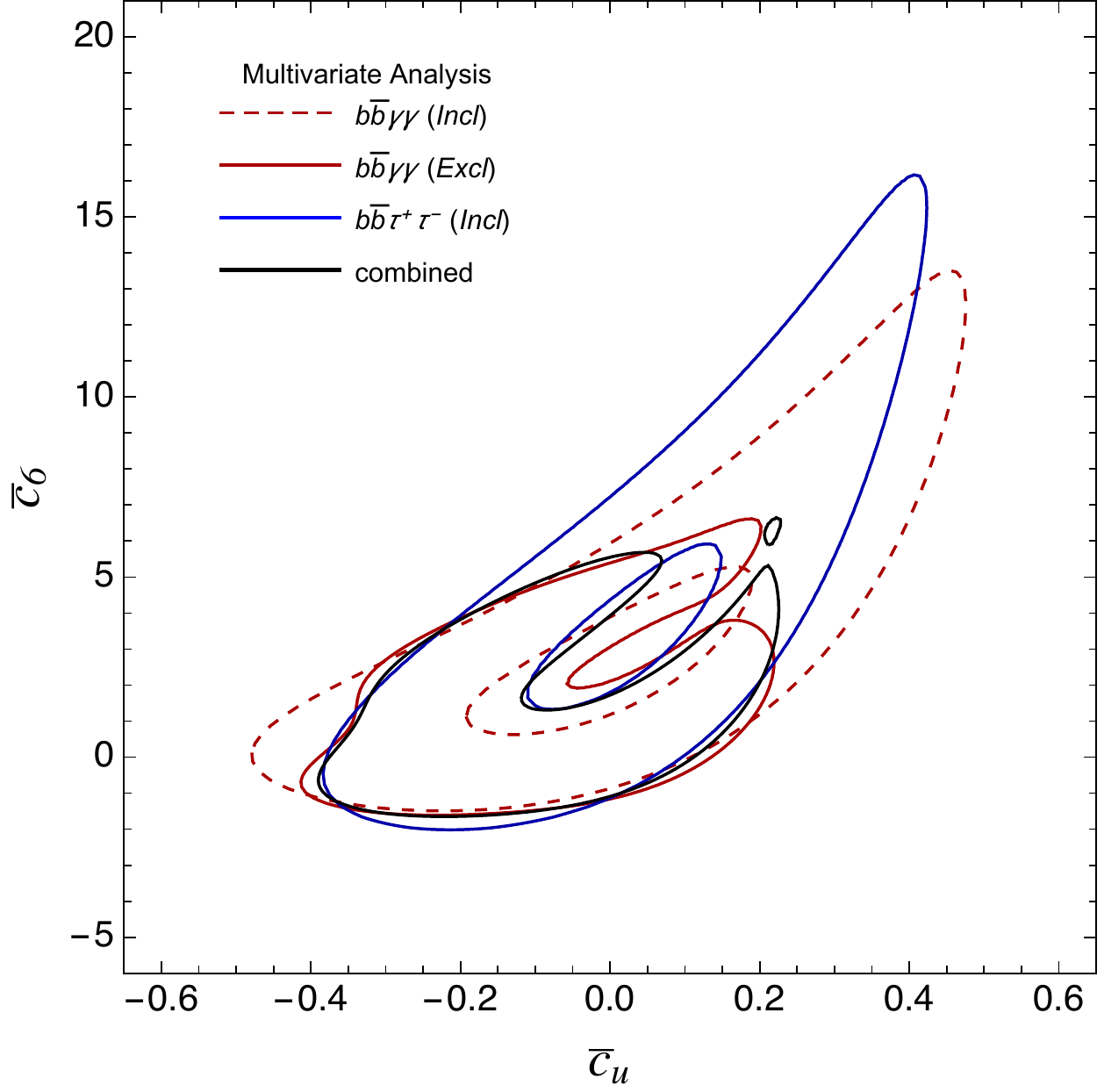}\quad
	\includegraphics[width=0.44\linewidth]{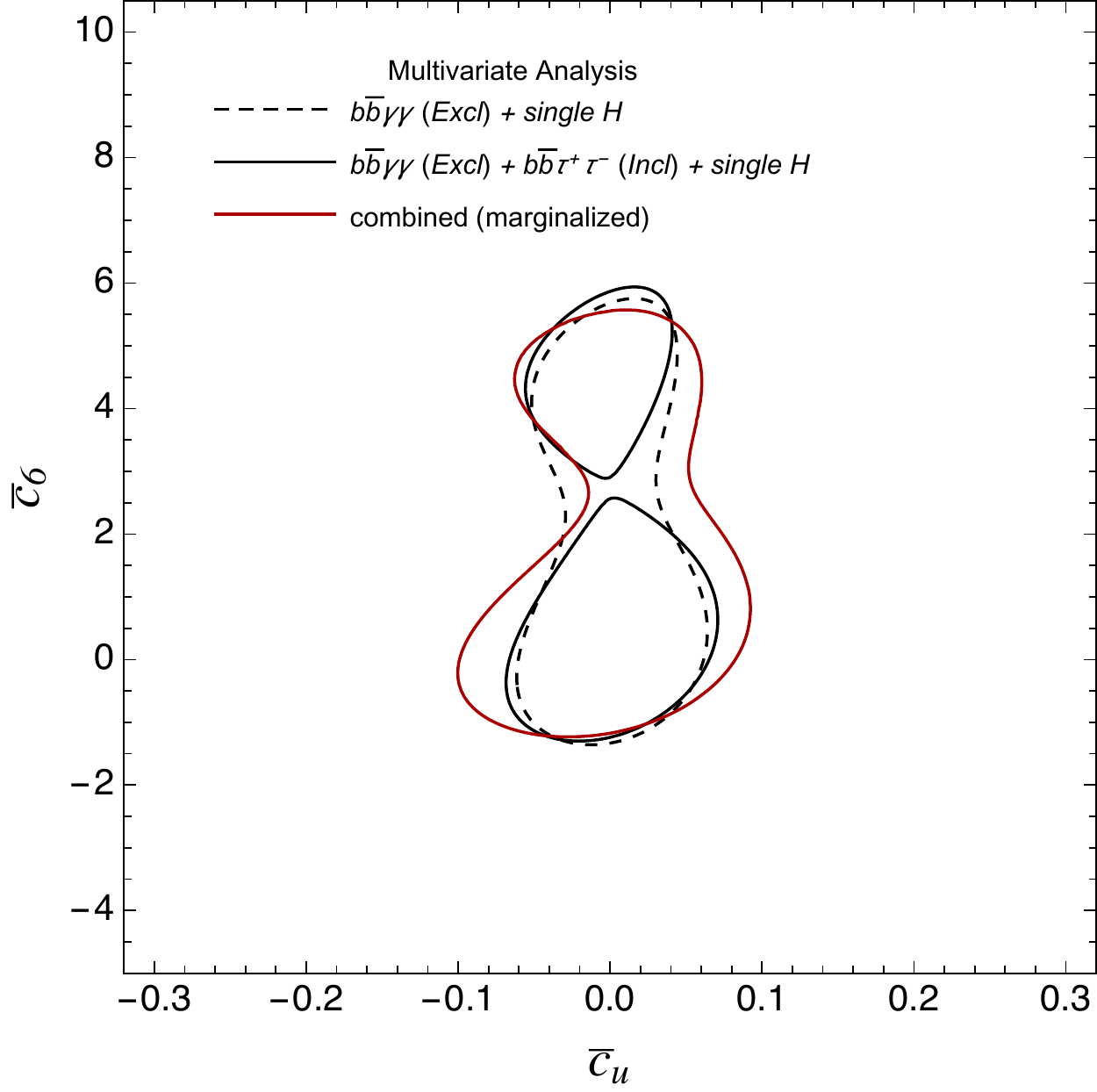}
	\caption{Left: 68\% probability contours of the likelihoods in $(\bar{c}_u,\, \bar{c}_6)$ plane using the multivariate analysis of the double Higgs production process. The combined analysis corresponds to the exclusive $b\bar{b}\gamma\gamma$ (solid red) plus the inclusive $b\bar{b}\tau^+\tau^-$ (dashed red). Right: 68\% probability contours of the likelihood of the exclusive $b\bar{b}\gamma\gamma$ plus single Higgs processes (dashed black).  The $b\bar{b}\tau^+\tau^-$ is further combined (solid black). $\bar{c}_H=\bar{c}_g=\bar{c}_d =0$ was set in all contours except solid red line in the right plot where the likelihood of the combined analysis (exclusive $b\bar{b}\gamma\gamma$, inclusive $b\bar{b}\tau^+\tau^-$, and single Higgs) was marginalized over $\bar{c}_H$, $\bar{c}_g$, and $\bar{c}_d$.}
	\label{fig:cuVSc6:HH:BDTBDT:noERR}
\end{figure}
%%%%%%%%%%%%%
We perform a similar exercise using the result by the multivariate analysis. The situation is illustrated in Fig.~\ref{fig:cuVSc6:HH:BDTBDT:noERR} where we observe a couple of changes. The allowed region noticeably shrinks with the result obtained by the multivariate analysis. Especially, as is evident in the right panel of Fig.~\ref{fig:cuVSc6:HH:BDTBDT:noERR}, the 68\% probability contour of the combined analysis having all other parameters set to SM values is split into two islands in the $(\bar{c}_u,\, \bar{c}_6)$ plane. However, when the marginalization over $\bar{c}_g$ is performed, two previously separated islands merge with the considerable change of the shape. After examining the shapes of the likelihoods, we find that the height of the second peak away from the SM point in $(\bar{c}_u,\, \bar{c}_6)$ plane is significantly reduced after the marginalization, and at the same time, the height of the first peak around the SM point as well as the middle region between two peaks is enhanced. A similar broadening effect of $\bar{c}_u$ after the marginalization is observed in this case.

%%%%%%%%%%%%%
\begin{figure}[!htb!] %[tbp]
	\centering
	\includegraphics[width=0.44\linewidth]{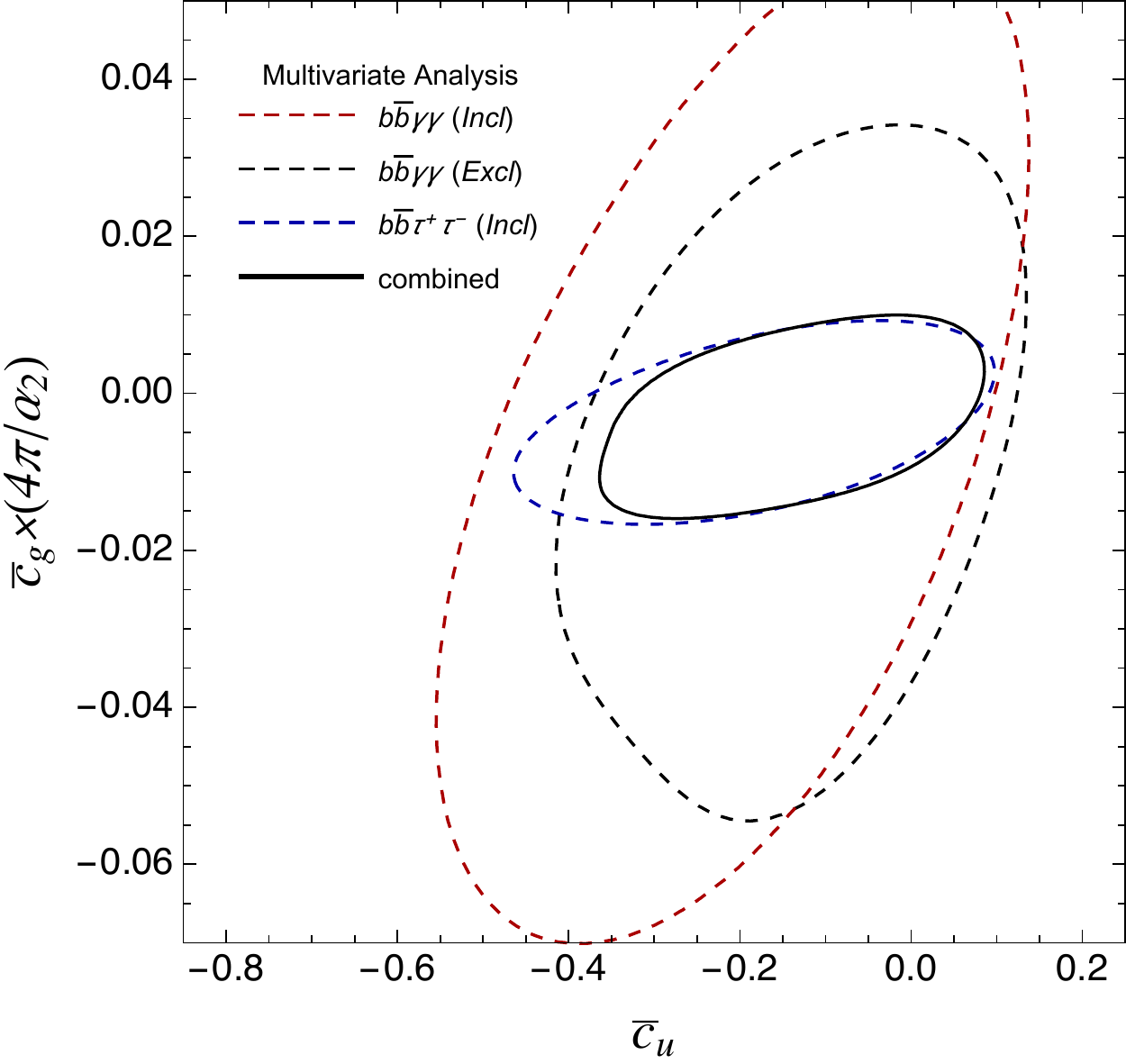}\quad
	\includegraphics[width=0.44\linewidth]{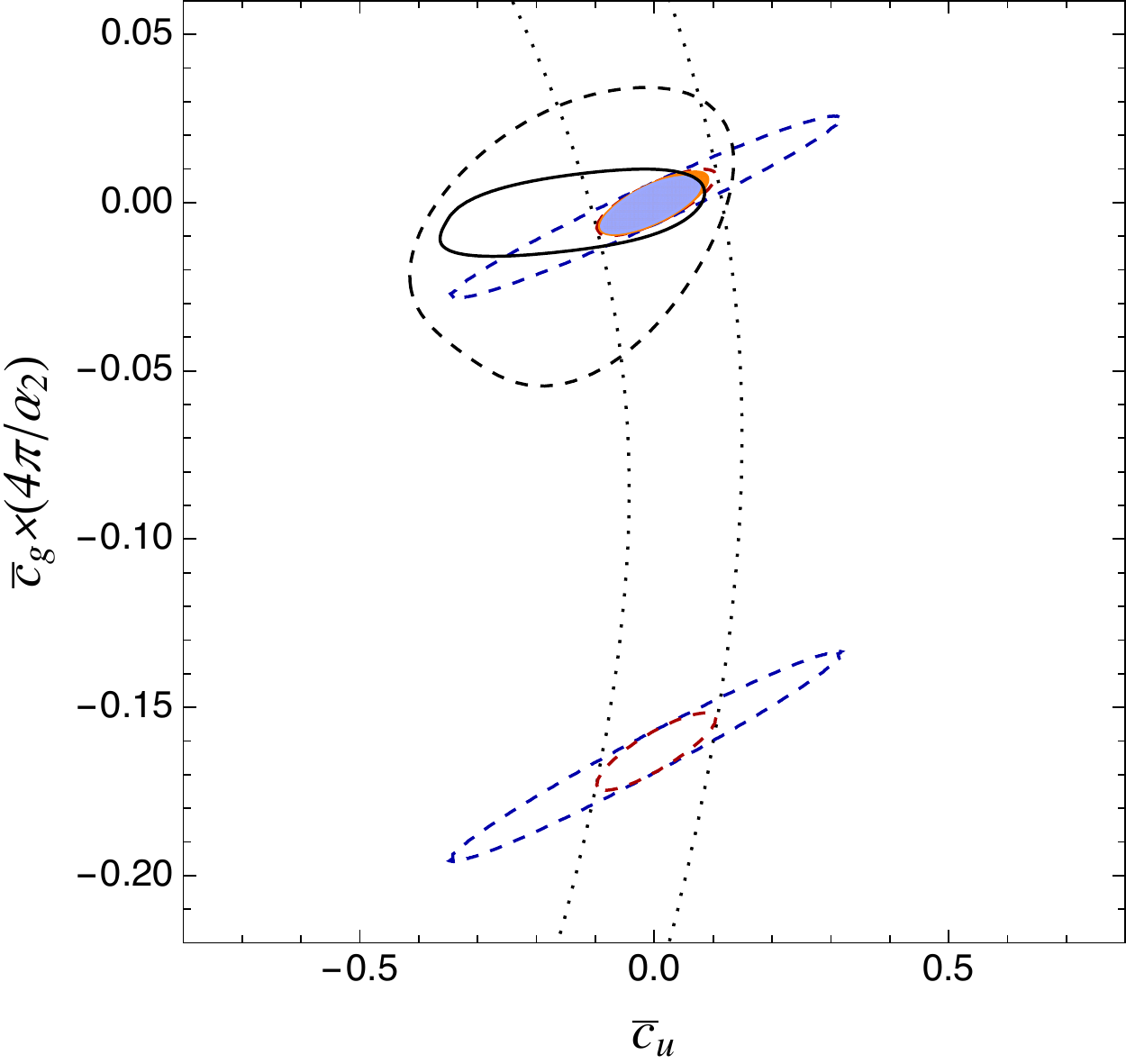}
	\caption{Left: 68\% probability contours in the $(\bar{c}_u,\, \bar{c}_g)$ plane using the result by the multivariate analysis of the double Higgs production process. The combined analysis (solid black) corresponds to the exclusive analysis of the $b\bar{b}\gamma\gamma$ (dashed red) plus the inclusive analysis of the $b\bar{b}\tau^+\tau^-$ (dashed blue). Right: 68\% probability contours of the likelihood of the various combinations. The black dashed contour is obtained by the double Higgs production in the $b\bar{b}\gamma\gamma$ (further combining with the inclusive $b\bar{b}\tau^+\tau^-$ gives a solid black contour). The single Higgs processes are split into three categories: all single Higgs processes except $t\bar{t}h$ (dashed blue), $t\bar{t}h$ alone (dotted black), and all single Higgs processes (dashed red). The double Higgs production in the $b\bar{b}\gamma\gamma$ plus single Higgs fit corresponds to the orange region. Combining all double Higgs production channels plus all single Higgs processes appears in the light blue region. $\bar{c}_H=\bar{c}_6=\bar{c}_d =0$ was chosen.}
	\label{fig:cucg:SILH}
\end{figure}
%%%%%%%%%%%%%

%%%%%%%%%%%%%
\begin{figure}[!htb!] %[tbp]
	\centering
	\includegraphics[width=0.44\linewidth]{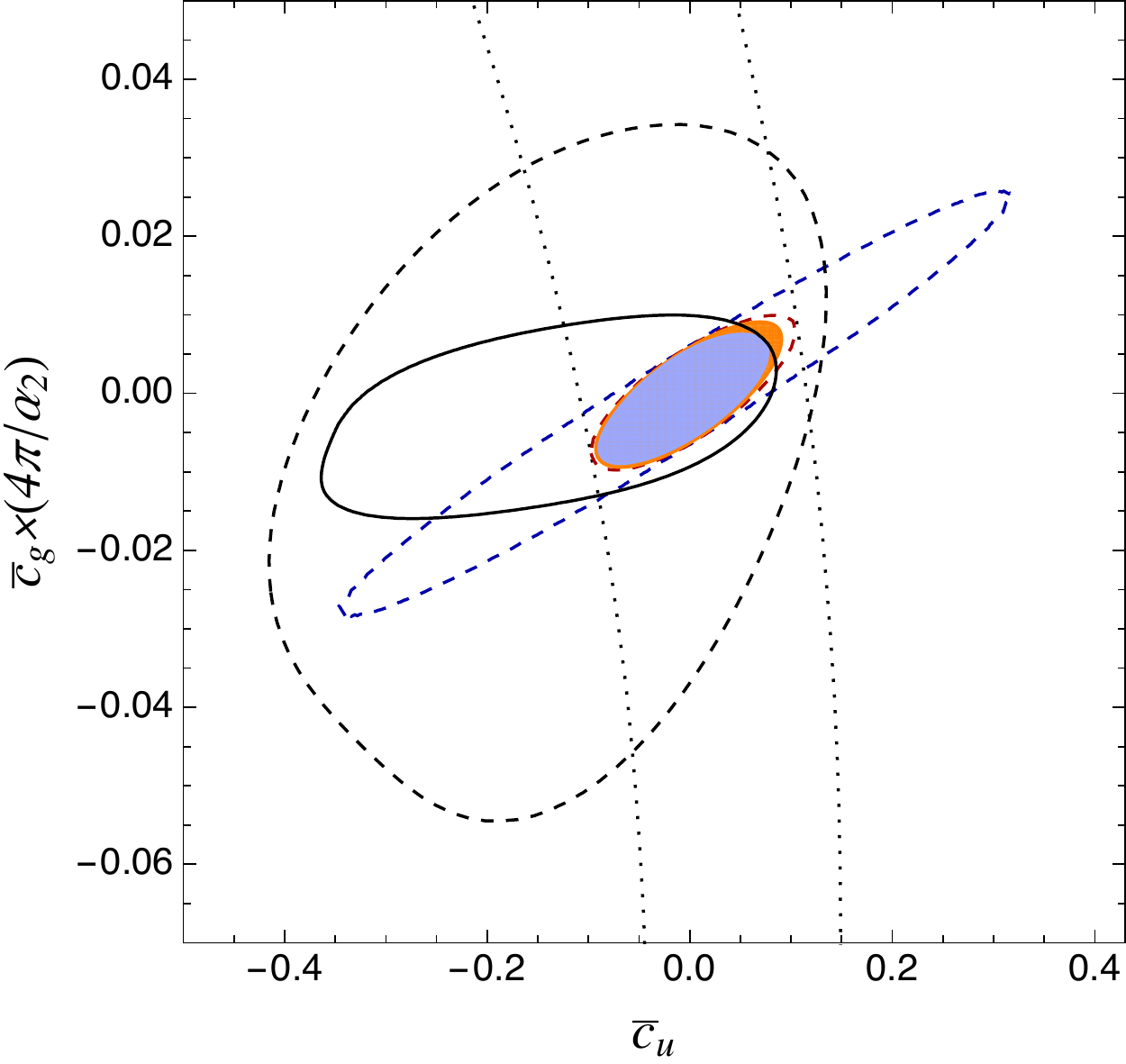}
	\caption{The surrounding region of the first island of the single Higgs process in Fig.~\ref{fig:cucg:SILH} is zoomed in. The line/color coding is the same as Fig.~\ref{fig:cucg:SILH} on the right panel.}
	\label{fig:cucg:SILH:zoom}
\end{figure}
%%%%%%%%%%%%%

We move onto the constraint in the $(\bar{c}_u,\, \bar{c}_g)$ plane. Here, we illustrate our result only for the multivariate analysis.
The sensitivity on $\bar{c}_u$ and $\bar{c}_g$ extracted by the double Higgs production process alone is illustrated in the left panel of Fig.~\ref{fig:cucg:SILH} which shows the strong benefit from the $b\bar{b}\tau^+\tau^-$ channel in constraining $\bar{c}_g$.  Regarding $\bar{c}_u$, the $b\bar{b}\gamma\gamma$ decay channel looks better for constraining a negative deviation while the positive deviation is better constrained by the $b\bar{b}\tau^+\tau^-$ process. Consequently, we see that the improvement on the positive deviation of $\bar{c}_u$ by adding the $b\bar{b}\tau^+\tau^-$ decay channel makes the double Higgs production itself comparable to the single Higgs process in constraining the positive deviation of the Yukawa coupling, as is seen in Fig.~\ref{fig:cucg:SILH:zoom}, where the sensitivities extracted by the double Higgs production alone (dashed black line by the exclusive $b\bar{b}\gamma\gamma$ and the solid black line by combining the $b\bar{b}\gamma\gamma$ and $b\bar{b}\tau^+\tau^-$ channels), $t\bar{t}h$ process (dotted black line), and the combined analyses (shaded orange region by the exclusive $b\bar{b}\gamma\gamma$ plus single Higgs processes and the light blue region by exclusive $b\bar{b}\gamma\gamma$, inclusive $b\bar{b}\tau^+\tau^-$ plus single Higgs processes) and so on are illustrated~\footnote{While we have not marginalized over other EFT coefficients in Fig.~\ref{fig:cucg:SILH:zoom} (and Fig.~\ref{fig:cucg:SILH}), see~\cite{Azatov:2015oxa} to see the effects from the marginalization.}. The shaded regions (both orange and light blue colors) in Fig.~\ref{fig:cucg:SILH:zoom} indicate the strong correlation between $\bar{c}_u$ and $\bar{c}_g$ coefficients, which means that the marginalization over $\bar{c}_g$ can significantly affect the precision on $\bar{c}_u$ and vice versa (for example, see right panels of Figs.~\ref{fig:cuVSc6:HH:ATLASCMS:noERR} and~\ref{fig:cuVSc6:HH:BDTBDT:noERR}). Another benefit of the double Higgs production is seen in the right panel of Fig.~\ref{fig:cucg:SILH}, where one of two islands in the single Higgs fit away from the SM point is disfavored by the double Higgs production process.

%%%%%%%%%%%%%
\begin{figure}[!htb!] %[tbp]
	\centering
	\includegraphics[width=0.44\linewidth]{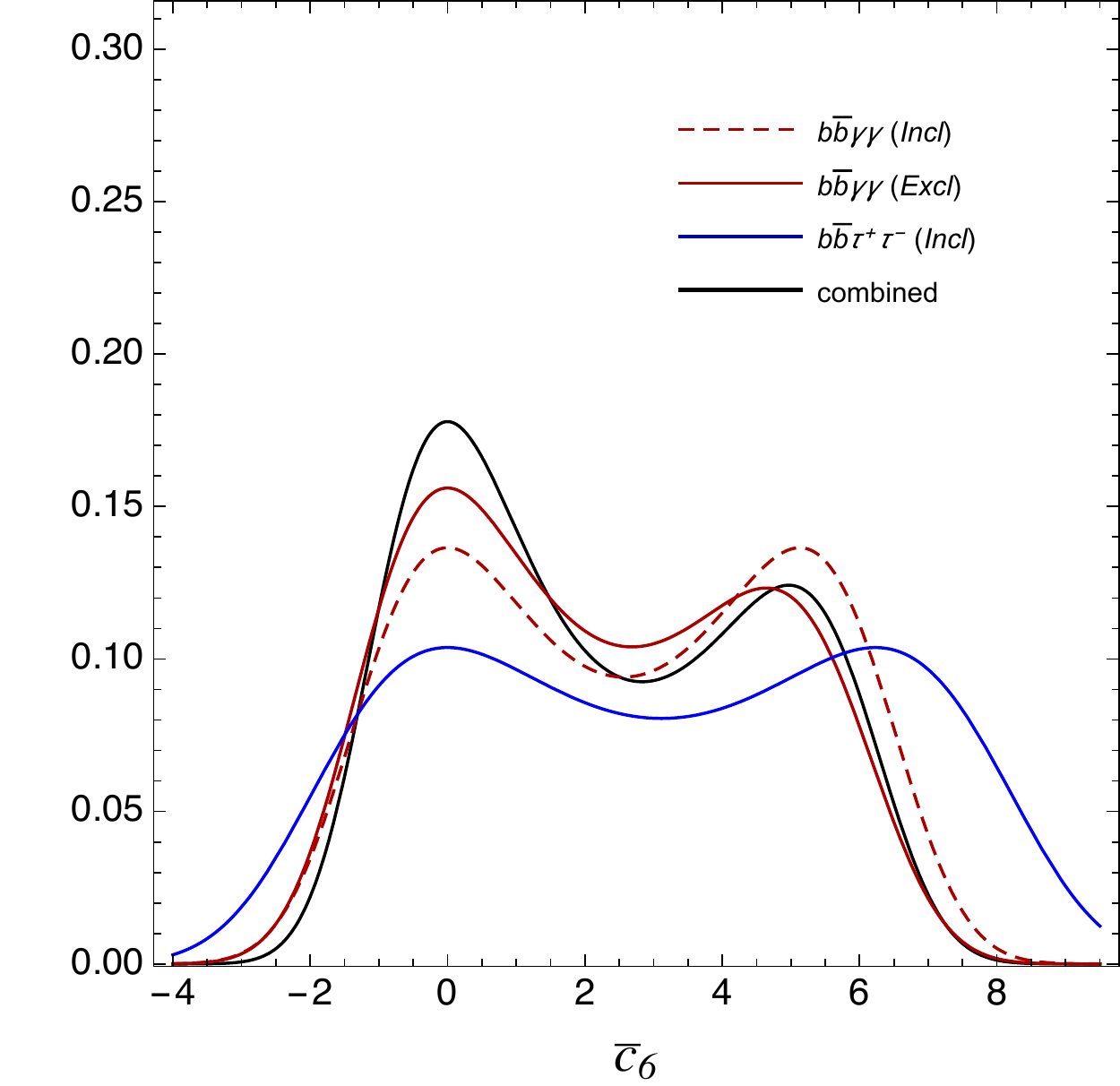}\quad
	\includegraphics[width=0.44\linewidth]{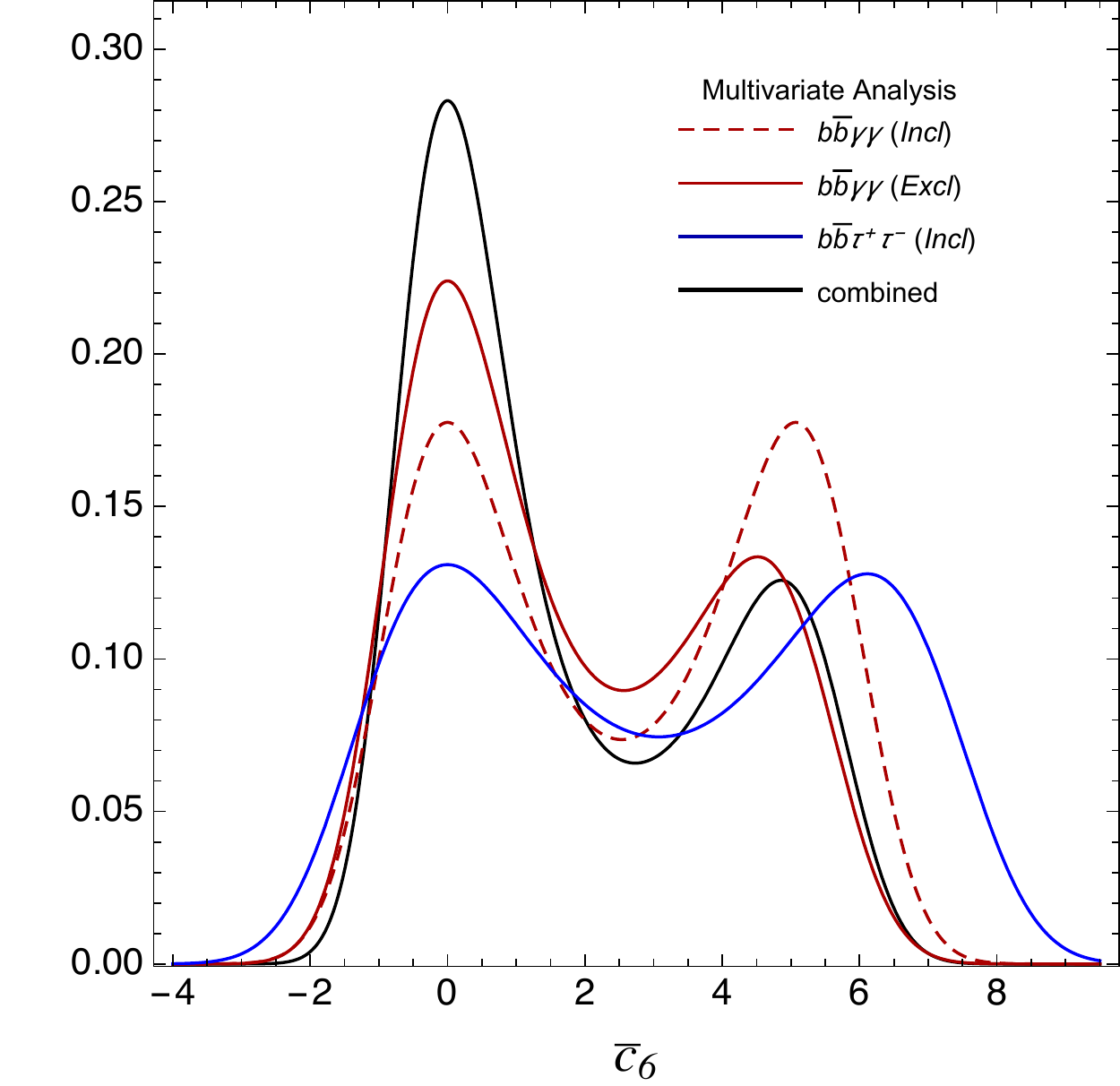}
	\caption{Likelihood distributions as a function of $\bar{c}_6$ using the cut-based analysis (left) and the multivariate analysis (right). The combined analysis corresponds to the exclusive analysis of the $b\bar{b} \gamma\gamma $ plus the inclusive analysis of the $b\bar{b} \tau^+\tau^-$. All the other parameters were set to the SM values (no marginalization).}
	\label{fig:c6:prob:noERR}
\end{figure}
%%%%%%%%%%%%%

%%%%%%%%%%%%%
\begin{figure}[!htb!] %[tbp]
	\centering
	\includegraphics[width=0.44\linewidth]{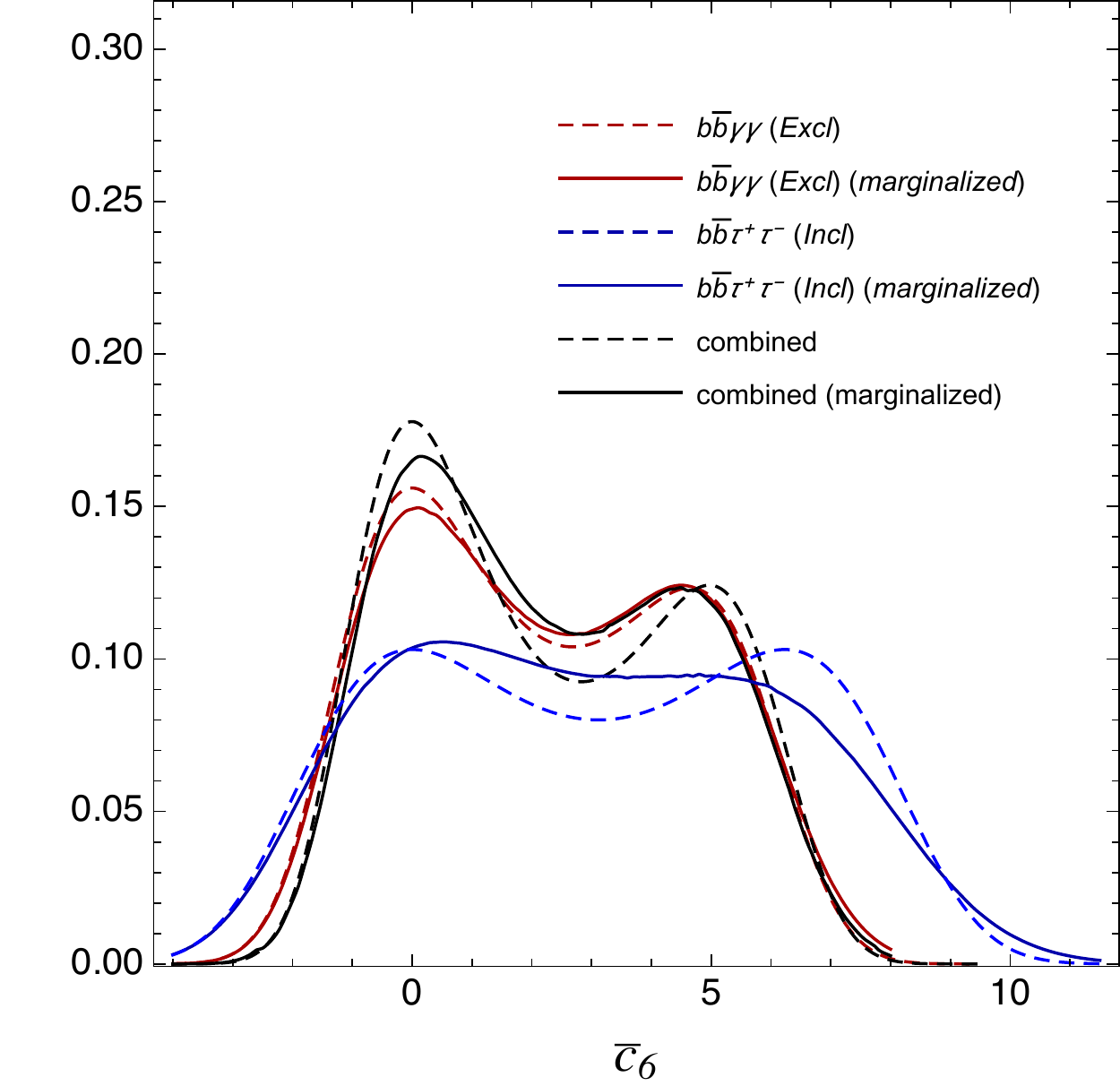}\quad
	\includegraphics[width=0.44\linewidth]{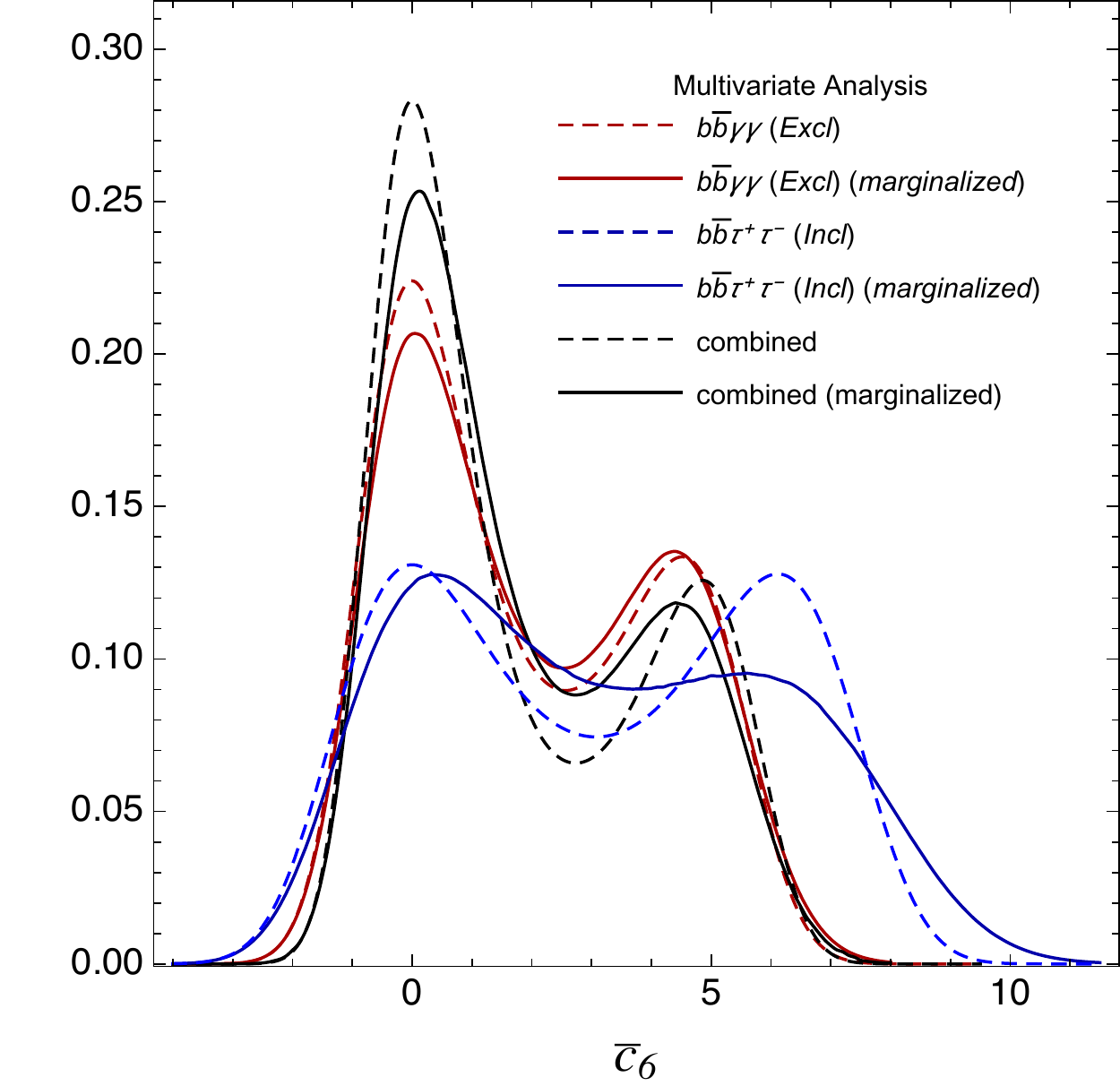}
	\caption{Likelihood distributions as a function of $\bar{c}_6$ using the cut-based analysis (left) and the multivariate analysis (right). The combined analysis combines the exclusive $b\bar{b} \gamma\gamma $, the inclusive $b\bar{b} \tau^+\tau^-$, and single Higgs processes. The marginalization was performed over $\bar{c}_g$ and $\bar{c}_u$ with the priors from the single Higgs data.}
	\label{fig:c6:prob:ERR}
\end{figure}
%%%%%%%%%%%%%

Finally, we derive 68\% and 95\% probability intervals on the Higgs cubic coupling, namely $\bar{c}_6$, from the marginalized likelihood (over $\bar{c}_u$ and $\bar{c}_g$) as a function of $\bar{c}_6$. The result is given in Table~\ref{tab:combined_table}~\footnote{\label{fn:prob}Our 68\% probability interval extracted from the exclusive $b\bar{b}\gamma\gamma$ (or combined) analysis employing the multivariate analysis is close to what has been reported in~\cite{Azatov:2015oxa}, namely $\bar{c}_6 = [-1.0,\, 1.8] \cup [3.5,\, 5.1]$. However, as was shown in Table~\ref{tab:bbaaSummary}, the newly done cut-based analysis with modifications mentioned below (but with cuts in~\cite{Azatov:2015oxa}) loses sensitivity, and the lost sensitivity is apparently recovered by the multivariate analysis.  On the other hand, our result appears more pessimistic compared to what was reported in~\cite{DiVita:2017eyz}. Note that one sigma sensitivity in~\cite{DiVita:2017eyz} was derived by a different statistical treatment, namely reading off the interval corresponding to $1 = \Delta\chi^2 = \sum_{i} {(\mu_i - 1)^2 \over \sigma_i^2}$, where $\mu_i$ is a signal strength as a function of $\bar{c}_6$ for $i$th process. When a likelihood distribution is highly non-Gaussian, as is the case for the Higgs self-coupling at the HL-LHC, the Bayesian method used in our analysis does not necessarily give a similar result to that from $1= \Delta\chi^2$. We find that the absence of the second interval in~\cite{DiVita:2017eyz} is also due to the different statistical treatment. For instance, applying $1=\Delta\chi^2$ method to the same likelihood of the combined analysis using the multivariate technique in Table~\ref{tab:combined_table} gives rise to $\bar{c}_6^{1=\Delta\chi^2} = [-0.7,\, 1.3]$ which corresponds to 43\% probability interval.}. When $\bar{c}_H =0$ is assumed, the sensitivity on $\bar{c}_6$ can be directly translated into that of $c_3$ in the nonlinear basis via the relation $c_3 = 1 + \bar{c}_6$ (see Eq.~(\ref{eq:dictionary})). As is shown in Fig.~\ref{fig:c6:prob:noERR}, the effect of the exclusive analysis for the  $b\bar{b}\gamma\gamma$ is to break degeneracy between two degenerate maxima of the likelihood. Comparison between two plots in Fig.~\ref{fig:c6:prob:noERR} demonstrates the impact of the multivariate analysis on the precision of $\bar{c}_6$. As is evident in Fig.~\ref{fig:c6:prob:noERR}, the middle region between two maxima is reduced in the multivariate analysis, and the peak around the SM point becomes more pronounced. The plots in Fig.~\ref{fig:c6:prob:ERR} illustrate the impact of the marginalization on the precision of $\bar{c}_6$. The solid (dashed) blue lines in the right panel of Fig.~\ref{fig:c6:prob:ERR} is the likelihood of the inclusive $b\bar{b} \tau^+\tau^-$ alone with the marginalization (without marginalization), and they indicate that the marginalization also breaks the degeneracy between two maxima in favor of the peak around the SM point. Overall, the net effect of the marginalization in the exclusive $b\bar{b} \gamma\gamma$ and combined analyses is a small degradation of the peak around SM point as well as the middle region for the combined analysis. The effect is less pronounced for the cut-based analysis (see the left panel of Fig.~\ref{fig:c6:prob:ERR}).
A similar observation was discussed in~\cite{Azatov:2015oxa} where a strong correlation between the precision of the Yukawa coupling and the precision of $\bar{c}_6$ at 100 TeV $pp$ collider is also discussed (see~\cite{DiVita:2017eyz} as well for related discussion).

%%%%%%%%%%%%%%%%%
\begin{table}[t]
\begin{center}
\begin{tabular}{ccc|cc}
\hline
\multicolumn{3}{c|}{HL-LHC (3 ab$^{-1}$)} & \multicolumn{2}{c}{Allowed region on $\bar{c}_6$} \\[1.5pt]
   &   &    & \quad $68\%$ probability \quad & \quad $95\%$ probability \quad \\[1.5pt]
\hline \hline
\multicolumn{5}{c}{Cut-based analysis} \\[1.5pt]
\hline 
\hspace{0.5cm} & $b\bar{b}\gamma\gamma$ (exclusive) &   &\hspace{0.5cm}  $[-0.98, 2.2] \cup [3.1, 5.3]$ \hspace{0.5cm} & \hspace{0.5cm} $[-1.8, 6.6]$ \hspace{0.5cm}  \\[1.5pt]
 & $b\bar{b}\tau^+\tau^-$ &   &  $[-0.87, 6.1 ]$  &  $[-2.5, 8.8 ]$  \\[1.5pt]
 & Combined &              &   $[-0.91, 2.3] \cup [3.4, 5.3]$  &  $[-1.6, 6.5]$  \\[1.5pt]
\hline
\multicolumn{5}{c}{Multivariate analysis}\\[1.5pt]
\hline
 & $b\bar{b}\gamma\gamma$ (exclusive) & &  $[-0.99, 1.8] \cup [3.4, 5.1]$  &  $[-1.4,5.9 ]$  \\[1.5pt] 
 & $b\bar{b}\tau^+\tau^-$&   & $[-0.89, 3.3] \cup [4.1, 6.4]$  &  $[-1.8, 8.5]$  \\ 
 & Combined &           &  $[-0.96, 1.9] \cup [3.8, 5.0]$  &  $[ -1.3,5.8 ]$  \\[1.5pt]
\hline
\end{tabular}
\caption{
The allowed region on $\bar{c}_6$ for the exclusive $b\bar{b} \gamma\gamma $, inclusive $b\bar{b} \tau^+\tau^-$, and combined channels. The 68\% and 95\% probability intervals are extracted from the marginalized likelihoods (over $\bar{c}_g$ and $\bar{c}_u$) for the cut-based and the multivariate analyses.  The interval on $\bar{c}_6$ can be translated to that of $c_3$ via the relation $c_3 = 1 + \bar{c}_6$, assuming $\bar{c}_H =0$.
}
\label{tab:combined_table}
\end{center}
\end{table}
%%%%%%%%%%%%%

%%%%%%%%%%%%%%%%%%%%%%%%%%%%%%%%%%%%%%%%
%%%%%%%%%%%%%%%%%%%%%%%%%%%%%%%%%%%%%%%%
\subsection{On the effect of future phenomenological studies}
\label{sec:varyingFakes}
\begin{figure}[!t]
\begin{center}
\includegraphics[width=0.32\linewidth]{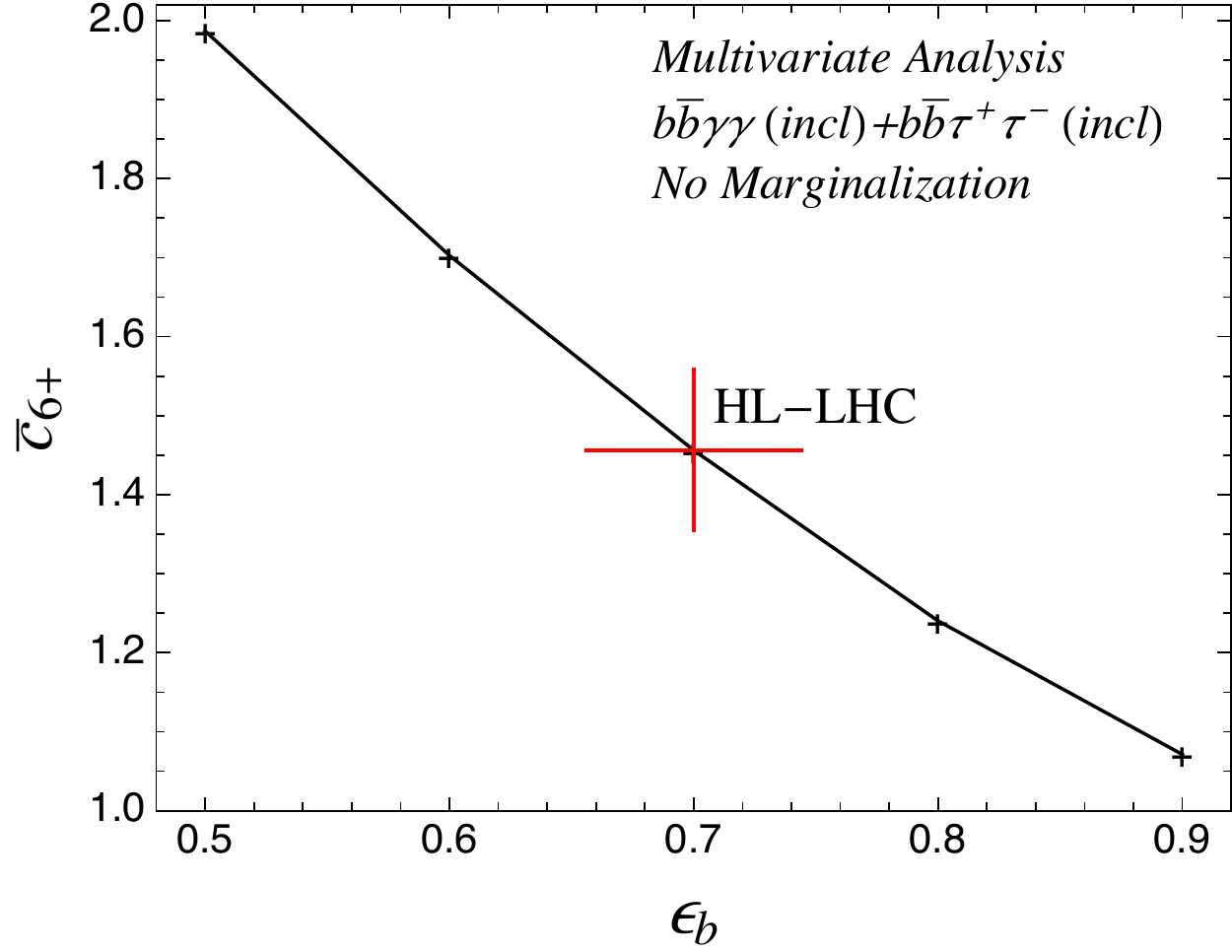}\quad
\includegraphics[width=0.32\linewidth]{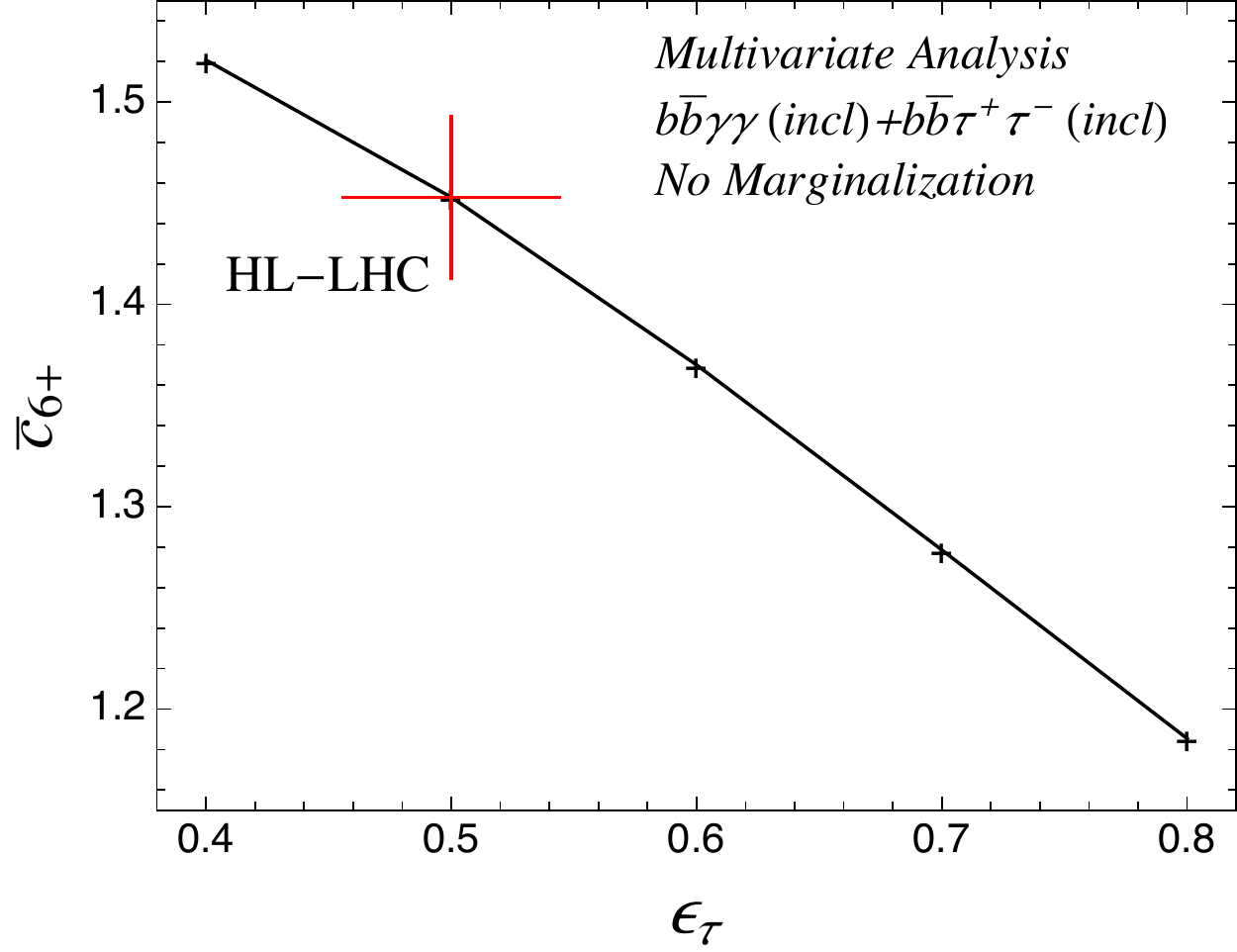}
\caption{
The upper value of the 68\% probability interval of $\bar{c}_6$ $(= c_3 - 1)$ around the SM point, $\bar{c}_{6+}$, as a function of $\tau$-tag rate (left) and $b$-tag rate (right) at the HL-LHC. The red-cross line (in an arbitrary size) corresponds to the HL-LHC.
}
\label{fig:tag}
\end{center}
\end{figure}

\begin{figure}[!t]
\begin{center}
\includegraphics[width=0.32\linewidth]{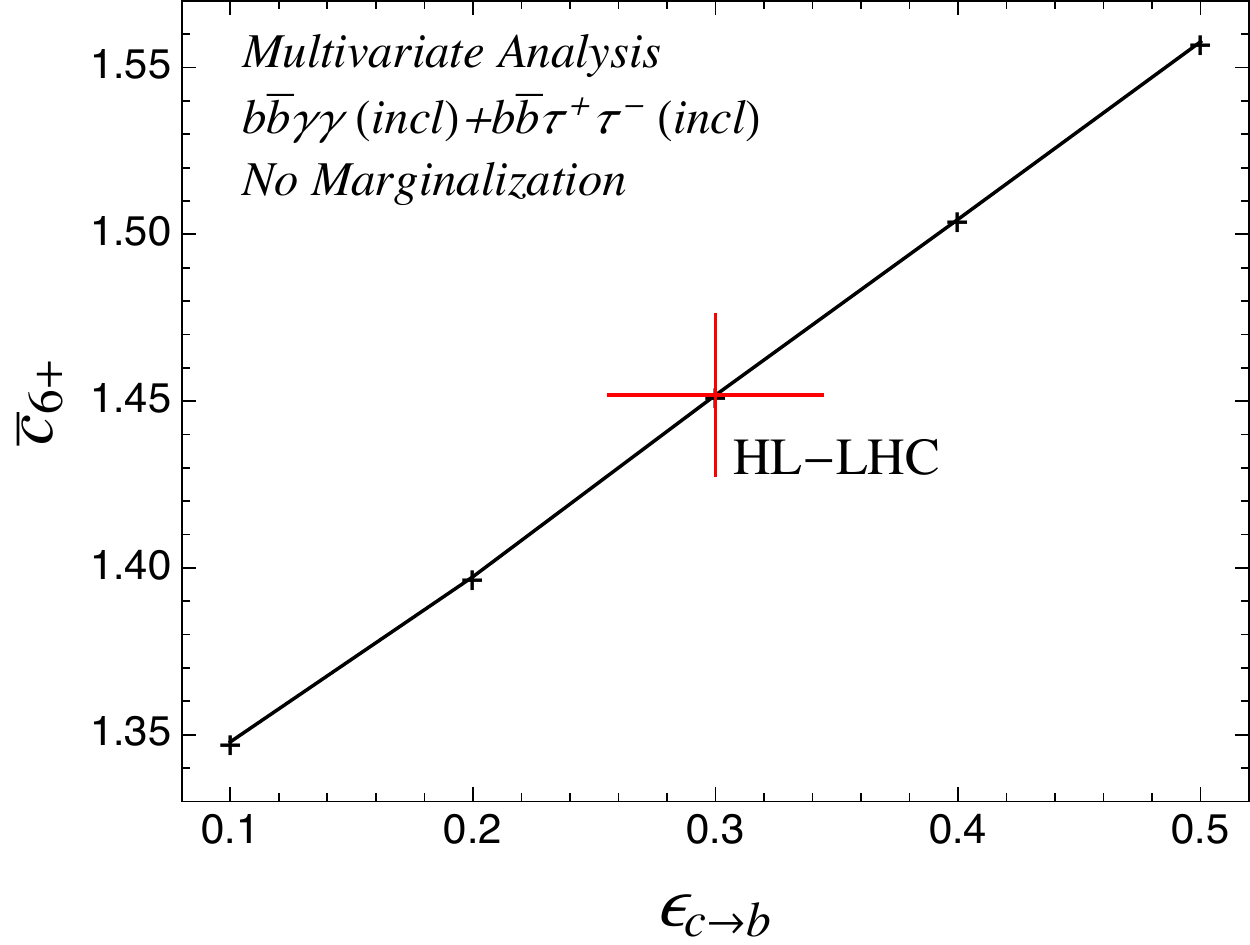}
\includegraphics[width=0.32\linewidth]{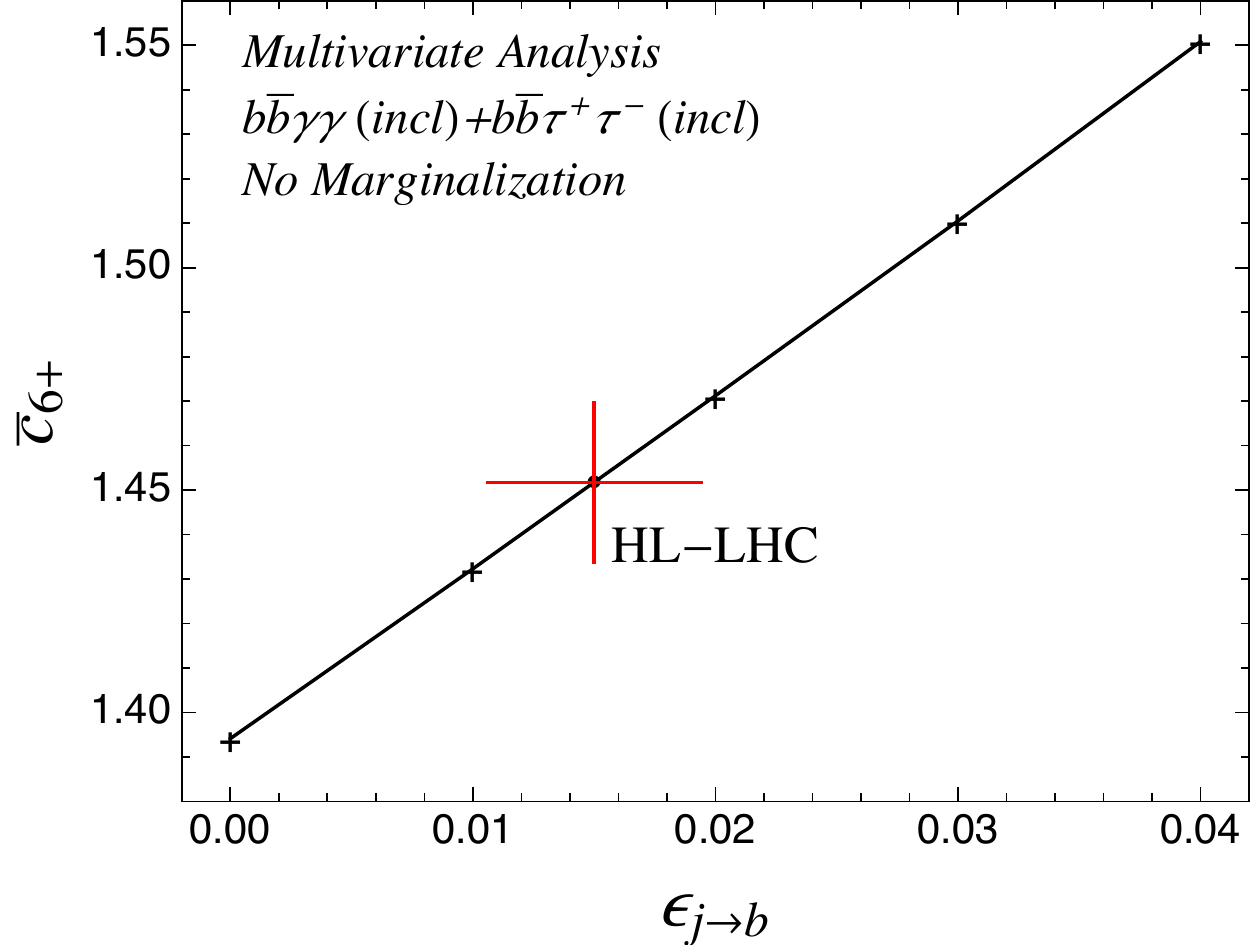}
\includegraphics[width=0.32\linewidth]{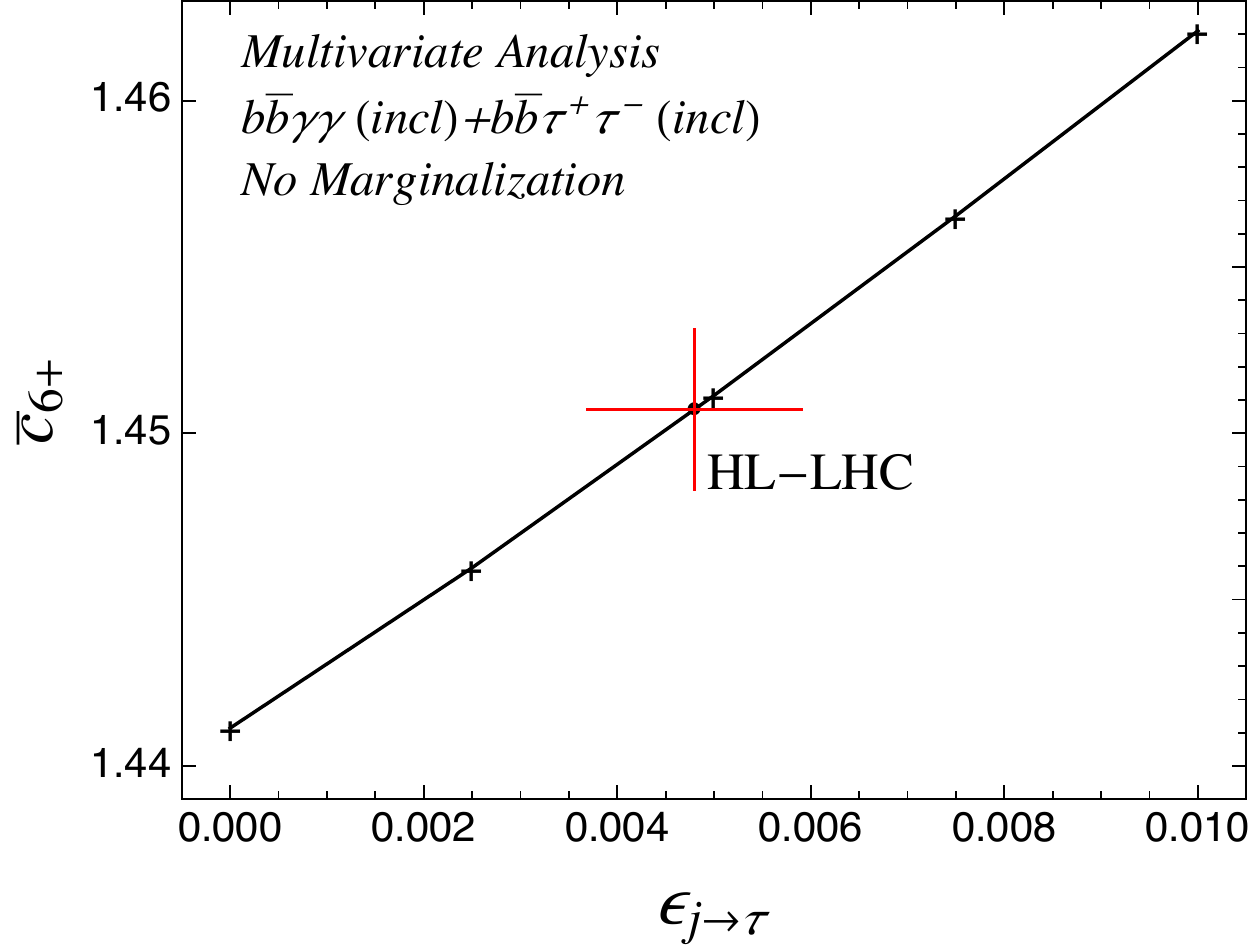}
\caption{ 
The upper value of the 68\% probability interval of $\bar{c}_6$ $(= c_3 - 1)$ around the SM point, $\bar{c}_{6+}$, as a function of fake rate at the HL-LHC. The red-cross line (in an arbitrary size) corresponds to the HL-LHC.  
}
\label{fig:fake}
\end{center}
\end{figure}

By the time the HL-LHC starts operating, there will be significant improvements in many factors that are beneficial to the performance of the double Higgs production via various independent phenomenological studies. The most important factors will be the improved $\tau$, $c$, $b$-tagging along with the reduced mistag rates, $\epsilon_{j \to \tau}$, $\epsilon_{j\to b}$, $\epsilon_{c\to b}$ and the improved photon identification, $\epsilon_\gamma$, and invariant mass resolutions. To estimate the impact of those factors on the precision of the Higgs self-coupling as our special interest (also for simplicity), we vary them one-by-one to obtain the precision of the Higgs self-coupling as a function of each improvable factor. Our result is illustrated in Figs.~\ref{fig:tag} and ~\ref{fig:fake} where we show the upper value of the 68\% probability interval of the Higgs self-coupling around the SM point, $\bar{c}_{6+}$, as a function of each improvable factor using the multivariate analysis.  To save computational time, we do not perform marginalization over other correlated couplings in Figs.~\ref{fig:tag} and~\ref{fig:fake}. We vary only $\bar{c}_6$ (or equivalently $c_3 =1 + \bar{c}_6$) while setting other EFT coefficients to the SM values. Also, for simplicity, we take the combined analysis of the inclusive analyses of the $b\bar{b}\gamma\gamma$ and $b\bar{b}\tau^+\tau^-$ decay channels in Figs.~\ref{fig:tag} and ~\ref{fig:fake}. The first interval of $\bar{c}_6$ around the SM point among two peaks of the likelihood (for instance, see Fig.~\ref{fig:c6:prob:noERR}) is selected for illustration. We do this exercise for the purpose of illustration.

%%%%%%%%%%%%%%%%%%%%%%%%%%%%%%%%%%%%%%
%%%%%%%%%%%%%%%%%%%%%%%%%%%%%%%%%%%%%%

%%%%%%%%%%%%%%%%%%%%%%%%%%%%%%%%%%%%%%
%%%%%%%%%%%%%%%%%%%%%%%%%%%%%%%%%%%%%%

The dependence of the positive deviation of $\bar{c}_6$ on the $\epsilon_{b}$ and $\epsilon_{\tau}$ is illustrated in Fig.~\ref{fig:tag}.  
The signal rate scales like $\sim \epsilon^2_{b}$ and $\sim \epsilon^2_{\tau}$ for the $b\bar{b}\gamma\gamma$  and $b\bar{b}\tau^+\tau^-$ decay channels, respectively, which means that the significance scales with one less power of tag rate, namely $\sim \epsilon_{b}$ and $\sim \epsilon_{\tau}$.  On the other hand, the dependence on the fake rates which can widely change the size of the backgrounds are shown in Fig.~\ref{fig:fake}. The improvements are less pronounced compared to those by $\epsilon_{b}$ and $\epsilon_{\tau}$. 

\begin{figure}[!t]
\begin{center}
\includegraphics[width=7.0cm, bb=0 0 400 313]{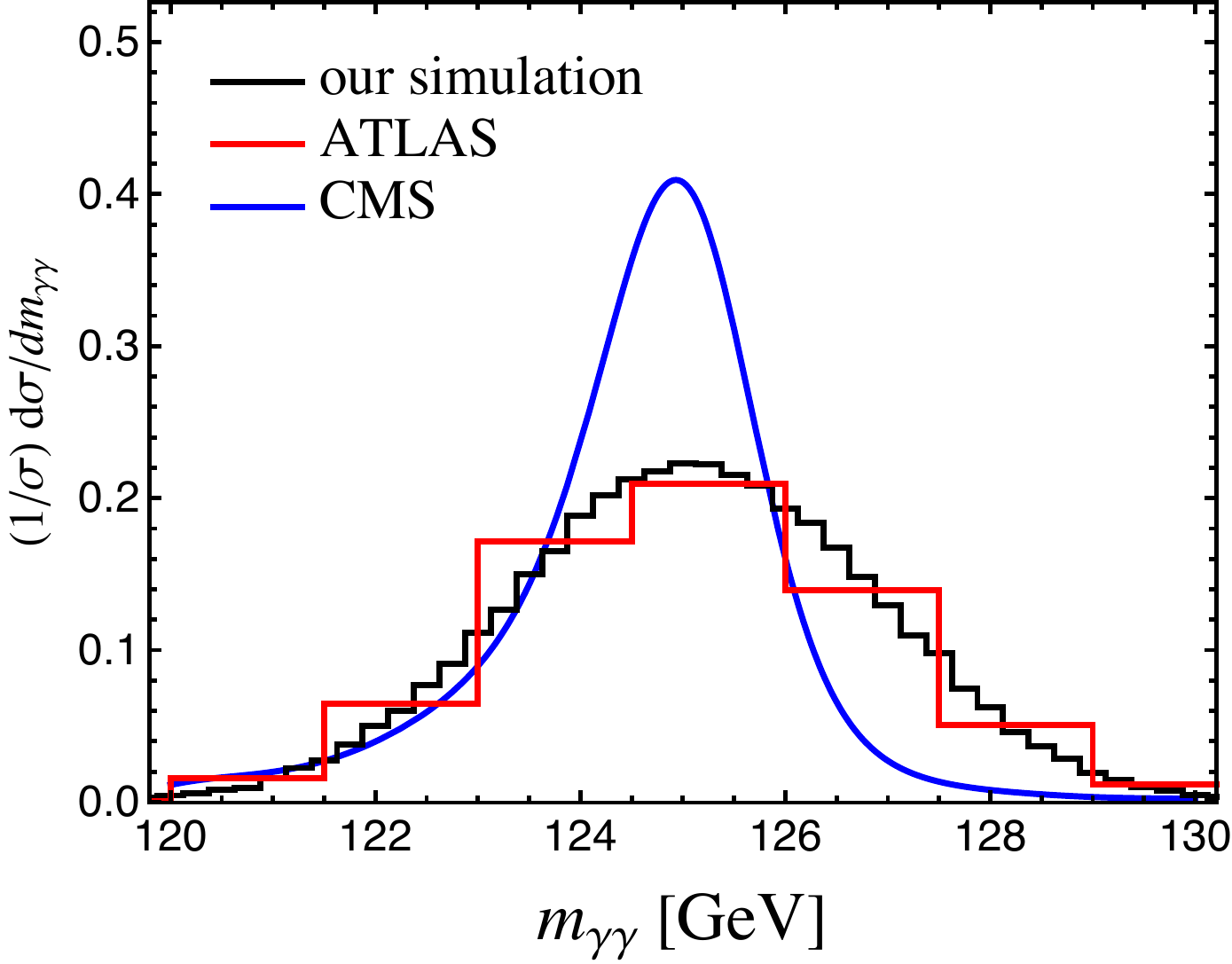}
\caption{
Distribution of $m_{\gamma\gamma}$ from ATLAS \cite{ATL-PHYS-PUB-2014-019}, CMS \cite{CMS-PAS-HIG-17-008}, and our simulation.
}
\label{maa}
\end{center}
\end{figure}

The resolution of $m_{\gamma\gamma}$ is an another important factor for the improvement of the performance. As a nonresonant background has a featureless $m_{\gamma\gamma}$ distribution unlike the sharp peak of the signal, the gaining in the signal-background discrimination can be significant. In Fig.~\ref{maa}, we show the $m_{\gamma\gamma}$ distribution from ATLAS \cite{ATL-PHYS-PUB-2014-019}, CMS \cite{CMS-PAS-HIG-17-008}, and our simulation.  The standard deviation of $m_{\gamma\gamma}$ distribution (SD($m_{\gamma\gamma}$)) in our simulation, which is roughly 1.74 GeV, is similar with the ATLAS simulation.   The standard deviation for the recent CMS simulation is about 25\% smaller (${\rm SD}(m_{\gamma\gamma})=1.3$ GeV) than ATLAS and ours.  At the OPT-HL-LHC (see Eq.~(\ref{eq:OPT-HL-LHC})), we adopt this CMS value. The net effect of the smaller width is the 25\% relative reduction of $b\bar{b}\gamma\gamma$, $c\bar{c}\gamma\gamma$, $b\bar{b} j\gamma$, $jj\gamma\gamma$ and $t\bar{t}\gamma$ backgrounds. It is because that the $m_{\gamma\gamma}$ distribution for the background processes is almost flat in the range $120 < m_{\gamma\gamma} < 130~{\rm GeV}$.

Based on the previous exercise and the recent progress on the performance of the $b$- and $\tau$-tagging algorithms~\cite{CMS-PAS-TAU-16-002,ATL-PHYS-PUB-2017-013}, we select the following benchmark scenario as an optimistic situation at the HL-LHC (we call it OPT-HL-LHC)  to estimate the precision on $\bar{c}_6$,
\begin{equation} \label{eq:OPT-HL-LHC}
\begin{split}
&{\rm \hbox{Optimistic HL-LHC (OPT-HL-LHC)}}  \\[3pt]
= &  
\left [ 
\begin{array} {l}
\hspace{0.14in} \epsilon_{b}=0.8~,\quad 
\epsilon_{\tau}=0.7~,\quad
\epsilon_{c\to b}=0.1~,\quad 
\epsilon_{j\to b}=0.01,  \quad
\epsilon_{j\to \tau}=0.001~, \\[3pt]
\hspace{0.14in} \hbox{25\% improvement of}\ m_{\gamma\gamma}\ \hbox{resolution}~,\\[3pt]
\hspace{0.14in} \hbox{20\% improvement of jet energy resolution}~,\\[3pt]
\end{array}
\right.
\end{split}
\end{equation}
where we still include only muons as leptonic taus.

%%%%%%%%%%%%%%%%%%%%%%%%%%%%%%
%%%%%%%%%%%%%%%%%%%%%%%%%%%%%%

\begin{figure}[!htb!] %[tbp]
	\centering
	\includegraphics[width=0.44\linewidth]{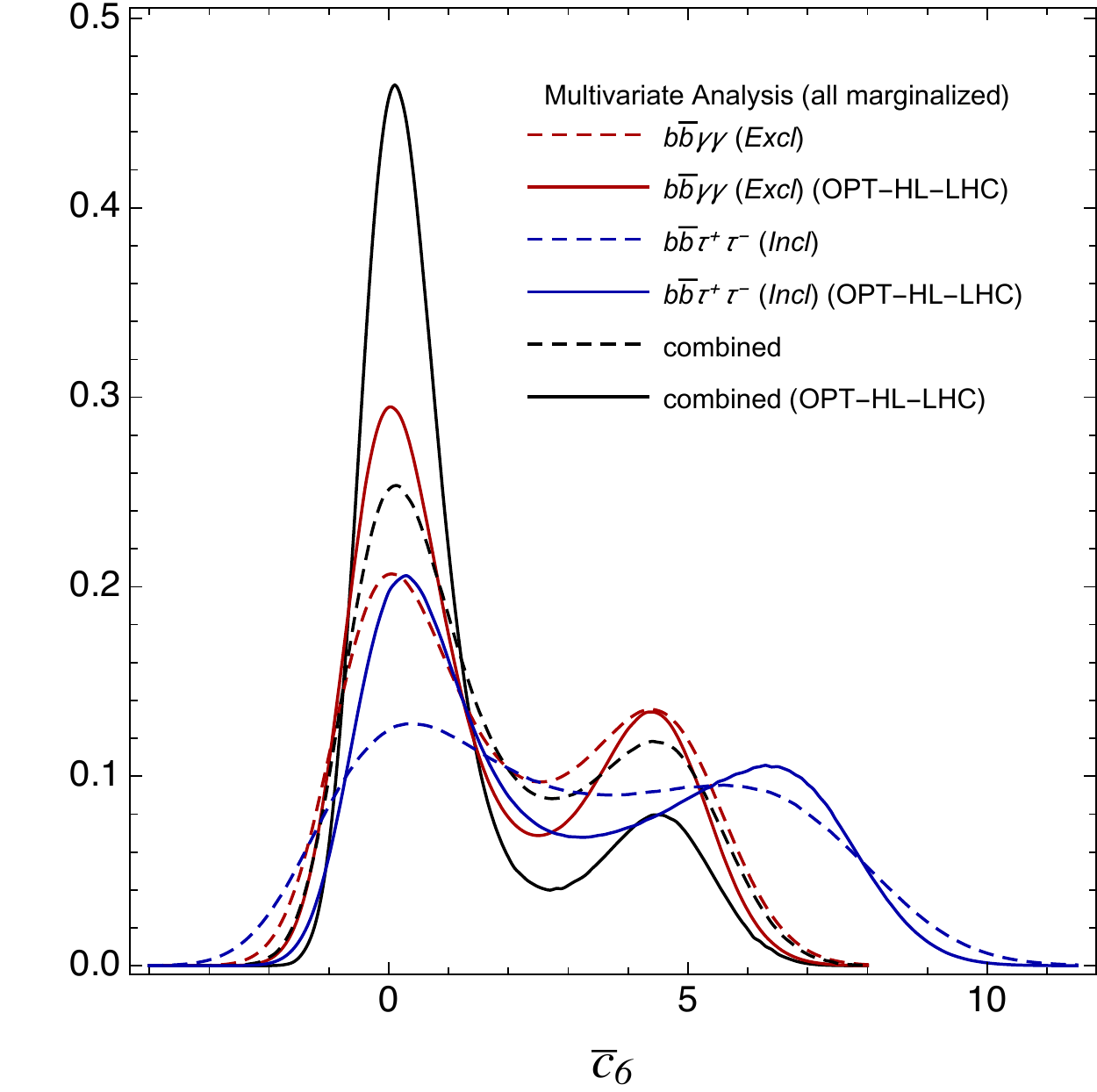}
	\caption{Likelihood distributions as a function of $\bar{c}_6$ using the multivariate analysis at the HL-LHC (dashed) and OPT-HL-LHC (solid). The combined analysis combines the exclusive $b\bar{b} \gamma\gamma $, the inclusive $b\bar{b} \tau^+\tau^-$, and single Higgs processes. The marginalization was performed over $\bar{c}_g$ and $\bar{c}_u$ with the priors from the single Higgs data.}
	\label{fig:c6:prob:Benchmark:ERR}
\end{figure}

The significance of the SM at the OPT-HL-LHC for $b\bar{b}\gamma\gamma$, $b\bar{b}\tau^+\tau^-$ (fully hadronic), and $b\bar{b}\tau^+\tau^-$ (semileptonic) decay channels are estimated to be 3.0, 3.4, and 1.5 respectively. The likelihood of $\bar{c}_6$ at the HL-LHC (for the purpose of comparison) and OPT-HL-LHC are illustrated in Fig.~\ref{fig:c6:prob:Benchmark:ERR}. Our estimate of the precision on $\bar{c}_6$ using the multivariate analysis is reported in Table~\ref{tab:benchmark} where we used the same BDT cut minima as the HL-LHC. We extract the numbers in Table~\ref{tab:benchmark} taking into account the marginalization over $\bar{c}_u$ and $\bar{c}_g$ with the priors from the single Higgs data. As is evident in Table~\ref{tab:benchmark}, the second interval of the 68\% probability intervals at the OPT-HL-LHC for the combined analysis has disappeared, and the previous 95\% probability interval got split into two intervals. If one focuses on the first interval, a meaningful $\mathcal{O}$(1) determination of the Higgs cubic coupling is possible even at the 95\% probability level.

 %%%%%%%%%%%%%
%%%%%%%%%%%%%

%%%%%%%%%%%%%%%%%
\begin{table}[!t]
\begin{center}
\begin{tabular}{ccc|cc}
\hline
\multicolumn{3}{c|}{OPT-HL-LHC (3 ab$^{-1}$)} & \multicolumn{2}{c}{Allowed region on $\bar{c}_6$} \\[1.5pt]
   &   &    & \quad $68\%$ probability \quad & \quad $95\%$ probability \quad \\[1.5pt]
\hline \hline
 \hspace{0.5cm} & $b\bar{b}\gamma\gamma$ (exclusive) & & \hspace{0.5cm}  $[-0.97,\, 1.5] \cup [3.8,\, 4.9]$  \hspace{0.5cm} &  \hspace{0.05cm} $[-1.2,\, 5.6 ]$  \hspace{0.5cm}\\[1.5pt] 
 & $b\bar{b}\tau^+\tau^-$&   & $[-0.80,\, 2.1] \cup [4.9,\, 7.3]$  & \hspace{0.5cm}  $[-1.1,\, 8.0]$ \hspace{0.5cm}  \\ 
 & Combined                   &   & $[-0.8,\, 1.3]$  & \hspace{0.5cm}  $[-1.1,\, 2.5] \cup [3.0,\, 5.5]$ \hspace{0.5cm} \\[1.5pt]
\hline
\end{tabular}
\caption{
The allowed region on $\bar{c}_6$ for the exclusive $b\bar{b} \gamma\gamma $, inclusive $b\bar{b} \tau^+\tau^-$, and combined channels at the OPT-HL-LHC. The 68\% and 95\% probability intervals are extracted from the marginalized likelihoods (over $\bar{c}_g$ and $\bar{c}_u$) for the multivariate analyses.  The interval on $\bar{c}_6$ can be translated to that of $c_3$ via the relation $c_3 = 1 + \bar{c}_6$, assuming $\bar{c}_H =0$.
}
\label{tab:benchmark}
\end{center}
\end{table}
%%%%%%%%%%%%%

In the OPT-HL-LHC, we also included the jet energy resolution as an improvable factor that can affect the performance. 
We smeared jet momenta according to the parametrization by ATLAS \cite{ATL-PHYS-PUB-2013-004}.  The fractional jet energy resolution is described by three parameters, namely noise ($N(\langle\mu\rangle)$), stochastic ($S$), and constant ($C$) terms as and it is given by
\begin{align} \label{eq:JER}
\frac{\sigma_{\rm res}}{p_T} =  f_{\rm res} \sqrt{
\frac{N(\langle\mu\rangle)^2}{p_T^2}  +\frac{S^2}{p_T}  +C^2~.
}
\end{align}
The noise parameter depends on an averaged pile-up $\langle\mu\rangle$. As was stated in Section~\ref{sec:DHiggsHL}, we take $N(140)=13.15~{\rm GeV}$, $S=0.74~{\rm GeV}^{1/2}$, and $C=0.05$ in this work.  As is seen in Eq.~(\ref{eq:JER}), we introduced a new parameter $f_{\rm res}$ in the jet energy resolution to measure the impact of the improved overall jet energy resolution on the precision of the Higgs self-coupling. The smaller $f_{\rm res}$ will lead to a better jet energy resolution.

%%%%%%%%%%%%%%%%%%%%%%%%%%%%%%%%%%%%%%%
\begin{figure}[!t]
\begin{center}
\includegraphics[width=0.44\linewidth]{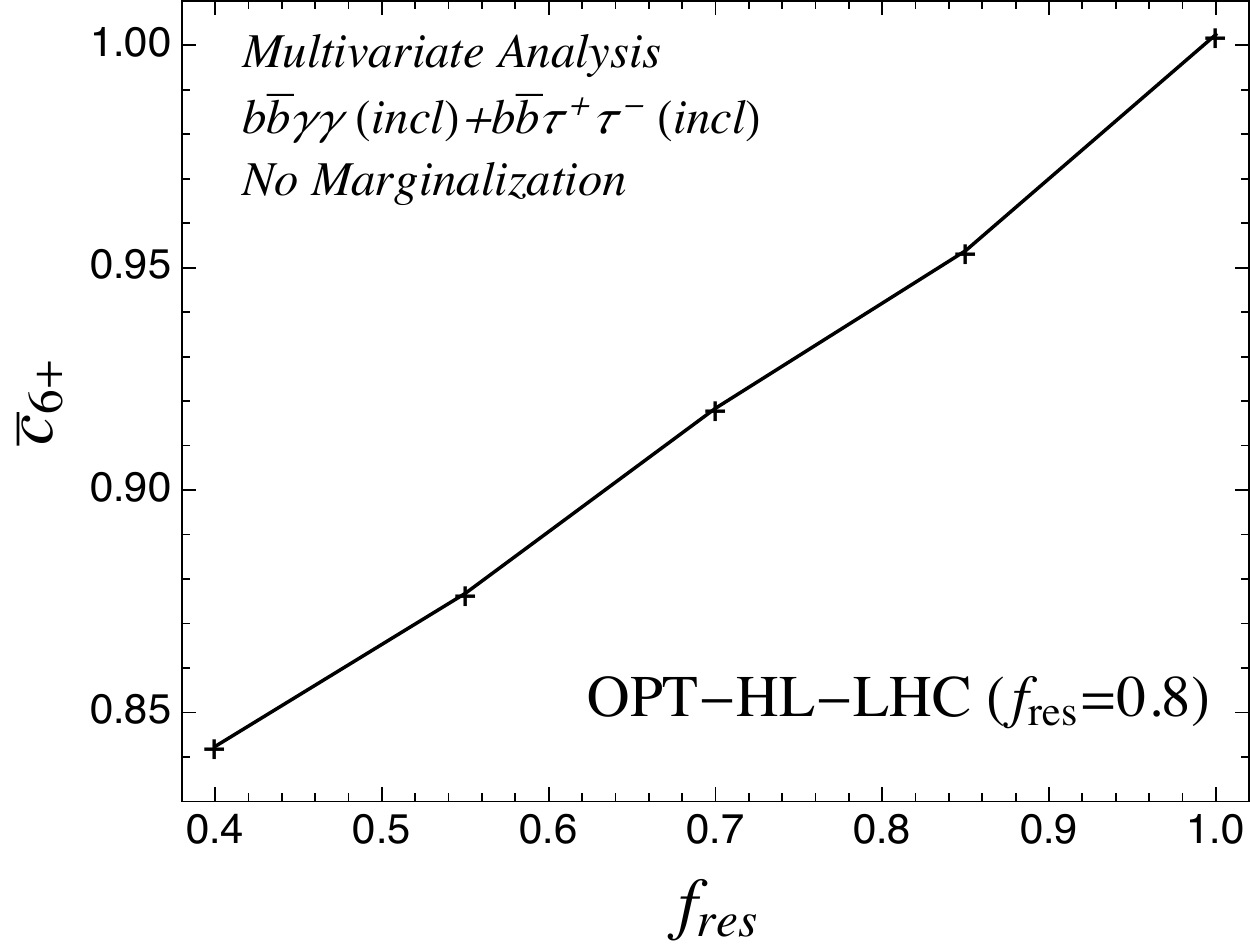}
\caption{ 
The upper value of the 68\% probability interval of $\bar{c}_6$ $(= c_3 - 1)$ around the SM point, $\bar{c}_{6+}$, as a function of $f_{\rm res}$ at the OPT-HL-LHC. The $f_{res}=1$ corresponds to the OPT-HL-LHC.}
\label{fig:FJES}
\end{center}
\end{figure}
\begin{figure}[!t]
\begin{center}
\includegraphics[width=0.44\linewidth]{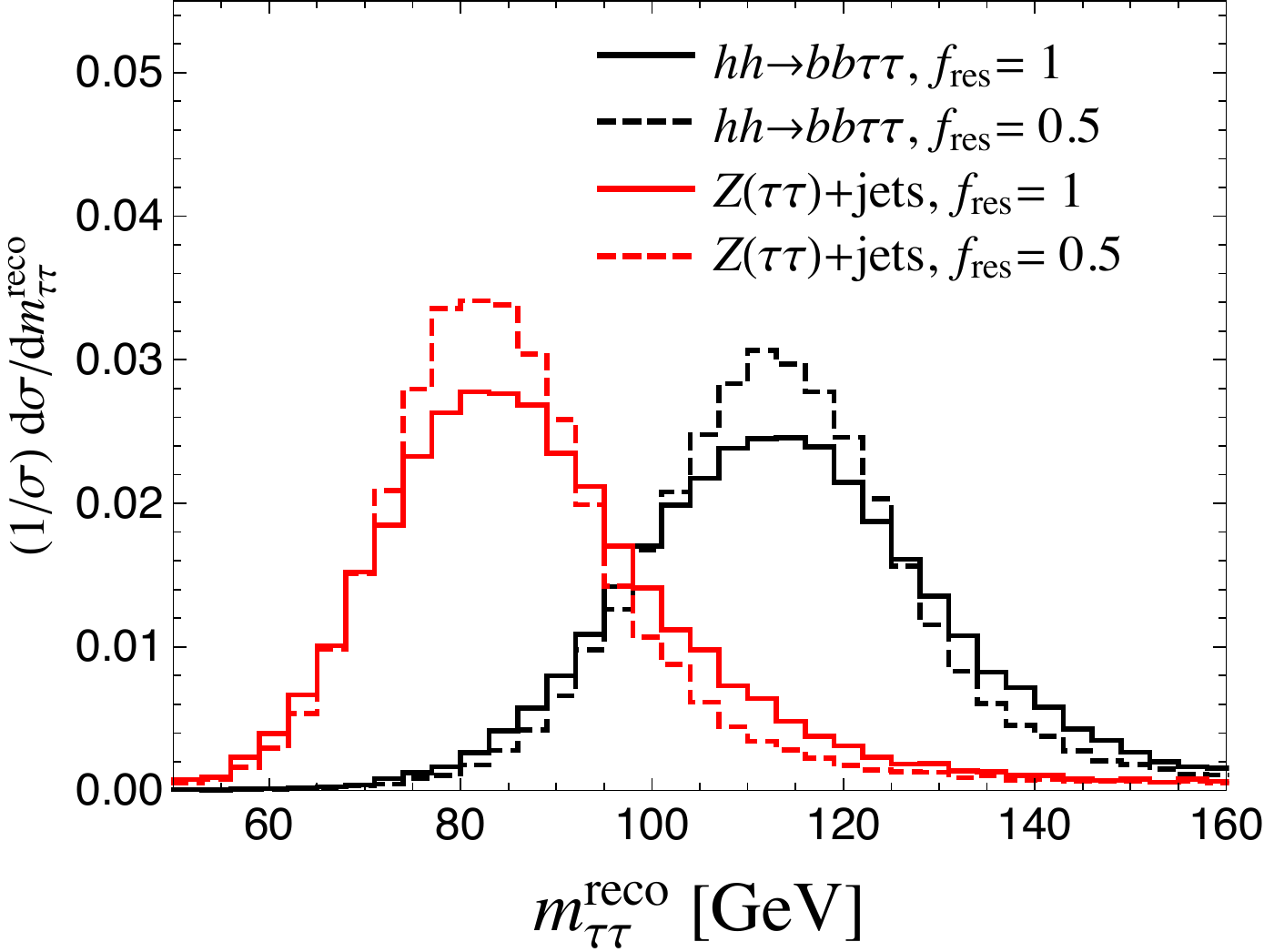}\quad
\includegraphics[width=0.45\linewidth]{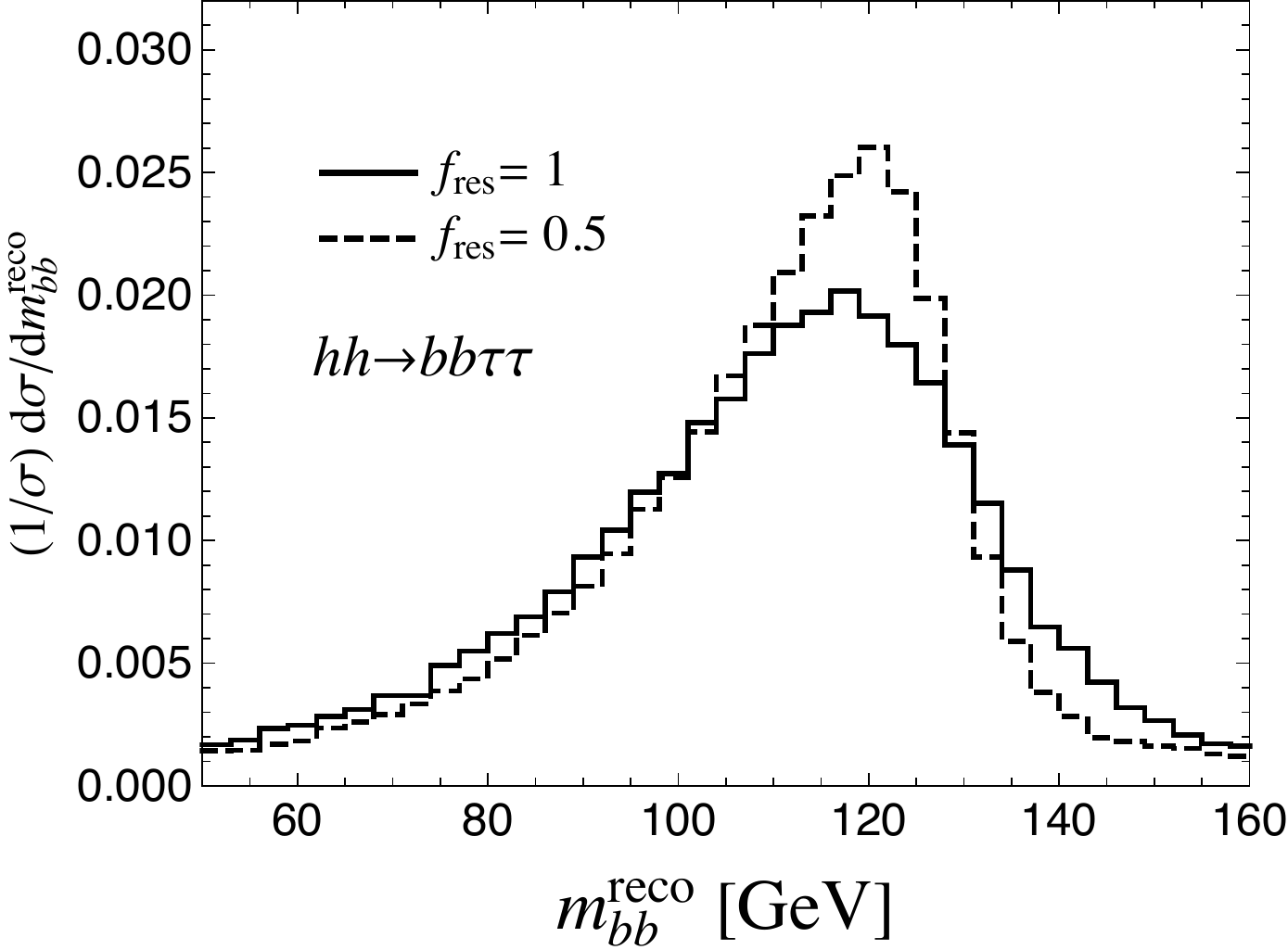}
\caption{
The $m_{\tau\tau}^{\rm reco}$ and $m_{bb}^{\rm reco}$ distributions for $f_{\rm res}=1$  (default) and $0.5$ (50\% improved).
}
\label{fig:fres_m}
\end{center}
\end{figure}

%%%%%%%%%%%%%%%%%%%%%%%%%%%%%%%%%%%%%%%
The $f_{\rm res}$ dependence on the positive deviation of the Higgs self-coupling around the SM point, $\bar{c}_{6+}$, at the OPT-HL-LHC is illustrated in Fig.~\ref{fig:FJES}~\footnote{The 68\% sensitivity of $\bar{c}_{6+}$ at the OPT-HL-LHC in Fig.~\ref{fig:FJES}, or $\bar{c}_{6+} \sim 1$, is much better than $\bar{c}_{6+} = 1.7$ in Table~\ref{tab:benchmark}. This seems to be a characteristic of the likelihood at the HL-LHC, namely highly non-Gaussian with two peaks. When only inclusive analyses of the $b\bar{b} \gamma\gamma$ and $b\bar{b} \tau^+\tau^-$ decay channels are combined, the second peak of the likelihood of $\bar{c}_6$ away from the SM point becomes much more pronounced than the case using the exclusive analysis of the $b\bar{b} \gamma\gamma$ (as in Table~\ref{tab:benchmark}). This implies that the relative probability, or the area, of the second (first) peak increases (decreases), and this makes the 68\% probability interval of the first peak narrower. Skipping the marginalization also makes the peaks narrower than the marginalized case.}. 
As for the plots in Figs.~\ref{fig:tag} and~\ref{fig:fake}, for the purpose of illustration, we use the combined analysis of the inclusive analyses of the $b\bar{b}\gamma\gamma$ and $b\bar{b}\tau^+\tau^-$ decay channels and we do not perform the marginalization over other EFT coefficients in Fig.~\ref{fig:FJES}. We find that the improved jet energy resolution is especially beneficial to the fully hadronic $b\bar{b} \tau^+\tau^-$ decay channel. For instance, it improves the discrimination of the signal, $h\rightarrow \tau^+\tau^-$, against $Z\rightarrow \tau^+\tau^-$ of $Z$ + jets in the $\tau^+\tau^-$ system as well as the improved resolution of the $b\bar{b}$ system, as is evident in Fig.~\ref{fig:fres_m}.

%%%%%%%%%%%%%%%%%%%%%%%%%%%%%%%%%%%%%%%%
%%%%%%%%%%%%%%%%%%%%%%%%%%%%%%%%%%%%%%%%
%%%%%%%%%%%%%%%%%%%%%%%%%%%%%%%%%%%%%%%%
%%%%%%%%%%%%%%%%%%%%%%%%%%%%%%%%%%%%%%%%
%%%%%%%%%%%%%%%%%%%%%%%%%%%%%%%%%%%%%%%%
%%%%%%%%%%%%%%%%%%%%%%%%%%%%%%%%%%%%%%%%
%%%%%%%%%%%%%%%%%%%%%%%%%%%%%%%%%%%%%%%%
\section{Summary}
\label{sec:Summary}

In this work, we have performed a combined analysis of the double Higgs production in the $b\bar{b} \gamma\gamma$ and $b\bar{b}\tau^+\tau^-$ decay channels at the HL-LHC in the EFT approach. We have validated our cut-based analysis of the $b\bar{b} \gamma\gamma$ decay channel by reproducing the ATLAS result. We have also provided the cut-based analysis following the cuts in~\cite{Azatov:2015oxa} (with the modification described in Section~\ref{sec:bbaa}) for the purpose of comparison. For the $b\bar{b}\tau^+\tau^-$ decay channel, we have obtained a similar CMS result. While the CMS analysis utilized the maximum likelihood fit method, called the {\tt SVFIT} algorithm, we have implemented $m_{\tau\tau}^{\rm Higgs-bound}$~\cite{Barr:2011he} in our analysis to reconstruct the invariant mass of the $\tau^+\tau^-$ system. We have shown that increasing the $m_\tau$ value (as an independent variable) with respect to the truth tau-lepton mass improves the signal efficiency resulting in the better significance. We have explored the multivariate analysis employing the BDT technique, and we found that the BDT technique improves the significance of the SM by roughly a factor of 2 in the both the $b\bar{b}\gamma\gamma$ and $b\bar{b}\tau^+\tau^-$ decay channels. 

Regarding the EFT approach, we have demonstrated that the different decay channels can constrain different regions of the parameter space. For instance, we have shown that the $b\bar{b}\tau^+\tau^-$ channel plays an important role in constraining the energy-growing higher-dimensional operators, while it is least beneficial for improving the precision of the Higgs self-coupling where the sensitivity is dominantly determined by the $b\bar{b} \gamma\gamma$ channel. We have illustrated that the double Higgs production can be more efficient to constrain the positive deviation of the top-Yukawa coupling compared to the single Higgs process.
 
The performance of the double Higgs production in the $b\bar{b} \gamma\gamma$ and $b\bar{b}\tau^+\tau^-$ decay channels relies on the efficient identification of tau leptons and heavy flavor jets against the fake rates as well as the performance of the calorimeters. These factors can be improved in the future throughout numerous independent phenomenological studies. 
%Therefore, it is informative to estimate the precision of the anomalous couplings or the significance of the SM as a function of those improvable parameters. 
In this work, we have illustrated the dependence of the precision of the Higgs self-coupling on the improvable factors, namely 
the tag and mistag rates of tau jets and heavy flavor jets, photon identification efficiency, and invariant mass resolution.

As a benchmark scenario, we have considered a situation at the HL-LHC with a set of improved parameters, what we call an optimistic HL-LHC or OPT-HL-LHC. We have shown that the second 68\% probability interval can be removed at the OPT-HL-LHC, while the 95\% probability interval is being split into two intervals whose the first one around the SM gives a meaningful $\mathcal{O}(1)$ determination of the Higgs cubic coupling. For instance, given the choice of parameters at the OPT-HL-LHC, the 68\% and 95\% probability intervals of the Higgs self-coupling, $\lambda_3/\lambda_{3}^{SM}$, can reach $[0.2,\, 2.3]$ and $[-0.1,\, 3.5] \cup [4.0,\, 6.5]$, respectively.
%Any further improvements on parameters or more tailored analysis may have the potential to remove the second  68\% probability interval. 
On the other hand, the 68\% and 95\% probability intervals of the Higgs self-couplings at the HL-LHC assuming similar set of parameters to the ATLAS and CMS analyses is found to be $[0.04,\, 2.9] \cup [4.8,\, 6.0]$ and $[-0.3,\, 6.8]$, respectively~\footnote{As was mentioned in footnote~\ref{fn:prob}, the sensitivity on the Higgs self-coupling can depend on the statistical treatment when a likelihood is far from the Gaussian, and it makes the comparison among literature ambiguous.}.\\

{\bf Note Added:} While our work is being completed, we have noticed~\cite{Adhikary:2017jtu} where the authors combined various final decay channels of the double Higgs production at the HL-LHC. While~\cite{Adhikary:2017jtu} focuses on the combination of various channels for the SM as well as the consideration of various BSM signatures, we have performed the combination of two decay channels in the EFT approach keeping all the EFT coefficients.

%%%%%%%%%%%%%%%%%%%%%%%%%%%%%%%%%%%%%%%%
%%%%%%%%%%%%%%%%%%%%%%%%%%%%%%%%%%%%%%%%
%%%%%%%%%%%%%%%%%%%%%%%%%%%%%%%%%%%%%%%%
%%%%%%%%%%%%%%%%%%%%%%%%%%%%%%%%%%%%%%%%
%%%%%%%%%%%%%%%%%%%%%%%%%%%%%%%%%%%%%%%%
%%%%%%%%%%%%%%%%%%%%%%%%%%%%%%%%%%%%%%%%
%%%%%%%%%%%%%%%%%%%%%%%%%%%%%%%%%%%%%%%%
% change__________________________________________
\section*{Acknowledgments}
%\acknowledgments
%__________________________________________
We would like to thank Roberto~Contino, Kyoungchul Kong, Olivier Mattelaer, Andreas Papaefstathiou, and Weiming~Yao for many insightful discussions and help on technical issues. We would like to thank Giuliano Panico for the clarification on their work in~\cite{DiVita:2017eyz}. MS thanks the Hong Kong Institute for Advanced Study for the hospitality during the last part of this
work. MS and YS were supported by the Samsung Science and Technology Foundation under Project Number SSTF-BA1602-04. JHK is supported in part by US-DOE (DE-SC0017988) and by the University of Kansas General
Research Fund allocation 2302091

%%%%%%%%%%%%%%%%%%%%%%%%%%%%%%%%%%%%%%%%
%%%%%%%%%%%%%%%%%%%%%%%%%%%%%%%%%%%%%%%%
%%%%%%%%%%%%%%%%%%%%%%%%%%%%%%%%%%%%%%%%
%%%%%%%%%%%%%%%%%%%%%%%%%%%%%%%%%%%%%%%%
%%%%%%%%%%%%%%%%%%%%%%%%%%%%%%%%%%%%%%%%
%%%%%%%%%%%%%%%%%%%%%%%%%%%%%%%%%%%%%%%%
%%%%%%%%%%%%%%%%%%%%%%%%%%%%%%%%%%%%%%%%
\appendix

%%%%%%%%%%%%%%%%%%%%%%%%%%%%%%%%%%%%%%%%
%%%%%%%%%%%%%%%%%%%%%%%%%%%%%%%%%%%%%%%%
%%%%%%%%%%%%%%%%%%%%%%%%%%%%%%%%%%%%%%%%
%%%%%%%%%%%%%%%%%%%%%%%%%%%%%%%%%%%%%%%%
%%%%%%%%%%%%%%%%%%%%%%%%%%%%%%%%%%%%%%%%
%%%%%%%%%%%%%%%%%%%%%%%%%%%%%%%%%%%%%%%%

\section{Simulation Details}
\label{app:sec:simdetails}

%%%%%%%%%%%%%%%%%%%%%
\subsection{Background simulation of $b\bar{b}\gamma\gamma$ decay channel}
\label{app:subsec:BKG:bbaa}
The $\gamma\gamma$ + jets samples, as the major backgrounds, are matched~\footnote{Matching $\gamma\gamma+$jets is nontrivial. To guarantee enough statistics, we simulated $\gamma\gamma+n$ jets (with $n=2,3$) at the ME, allowing extra jets from the parton shower. QCD partons in this process are not necessarily originated from QCD splitting with the strength of $\alpha_s$. When a branching proceeds via the electroweak splitting (with $\alpha_{EW}$), the matching between ME and parton shower becomes ambiguous, for instance, some $\gamma\gamma+3j$ amplitudes can not be generated via $\gamma\gamma+2j$ at the ME plus an extra jet at the PS (similarly for the case with lower jet multiplicities). We also have simulated $\gamma\gamma+n$ jets (with $n=1,2,3$) samples and checked that our result with the samples with only $n=2,3$ multiplicity is not affected. A similar issue exists in $Z$ + jets samples too. We thank Olivier Mattelaer for pointing out this issue.} 
up to one additional $j$ using the $k_T$-jet MLM matching~\cite{Alwall:2007fs} to partially take into account the NLO effects.  To enhance the statistics of the $b$-enriched samples, we simulated the events through two processes, $b\bar{b}\gamma\gamma$ and $jj\gamma\gamma$ at the matrix element (ME) level where $j$ represents partons in the four-flavor scheme. In both processes, we imposed a cut on the invariant mass of two photons, $110~{\rm GeV}<m_{\gamma\gamma}<140~{\rm GeV}$, at the generation for better statistics. For the $b\bar{b}\gamma\gamma$ background, we also imposed a cut on the invariant mass of the $b\bar{b}$ system, $60~{\rm GeV}<m_{b\bar{b}}<300~{\rm GeV}$ at the generation. In the four-flavor scheme, the contribution from the $c\bar{c}\gamma\gamma$ process is included in the $jj\gamma\gamma$ process. It has been shown in~\cite{Azatov:2015oxa} via the NLO estimate performed by \textsc{MadGraph}5\_aMC$@$NLO that the $k$-factor $\sim 2$ of the $b\bar{b}\gamma\gamma$ background is mainly originated by the real emission. 
The resonant backgrounds, $Z(b\bar{b})h(\gamma\gamma)$ and $b\bar{b}h(\gamma\gamma)$, are matched to allow an additional jet at the ME~\footnote{The cross sections of $Z(b\bar{b})h(\gamma\gamma)$ and $b\bar{b}h(\gamma\gamma)$ backgrounds are estimated to the NLO in~\cite{ATL-PHYS-PUB-2014-019}. In our simulation, the NLO effect is partially included via a matching.}.
Whereas a resonant $t\bar{t}h(\gamma\gamma)$ background is generated without the matching, the tree-level cross section is normalized by NLO k-factor~\cite{Dittmaier:2011ti} (k-factor = 1.435).  
The ATLAS analysis~\cite{ATL-PHYS-PUB-2014-019} has shown two more non-negligible backgrounds, $b\bar{b}j\gamma$ and $t\bar{t}\gamma$ backgrounds. Both samples are similarly matched up to an additional jet at the ME~\footnote{As far as we can tell, the $t\bar{t}\gamma$ sample in~\cite{ATL-PHYS-PUB-2014-019} has not been matched.}. For the $b\bar{b}j\gamma$ background, we restrict the events to the window, $60~{\rm GeV}<m_{b\bar{b}}<300~{\rm GeV}$, at the generation to improve statistics. The $b$ quarks (photons) in all backgrounds are required to satisfy $p_T(b,\, \gamma) > 20$ GeV and $|\eta(b)| < 3$ ($|\eta(\gamma)| < 2.5$) at the generation.
The simulation details are summarized in Table~\ref{tab:gen_bbaa}. 
%%%%%%%%%%%%%%%%%%
\begin{table}[t]
\begin{center}
\begin{tabular}{c|c|c|c|c}
\hline
ME  
& Matching
& {\tt xqcut}/{\tt Qcut} (GeV)
& $\sigma \cdot {\rm BR(fb)}$
& Generated Events
\\[1.5pt]
\hline
$h(\gamma\gamma)h(b\bar{b})$  &   $-   $       &   $  -  $    &  $9.7 \times 10^{-2} $   &  $1.0\times 10^6$  \\[1.5pt]
$b\bar{b}\gamma\gamma      $  &   $\surd$       &   $10/20$    &  $28            $   &  $7.9\times 10^5$  \\[1.5pt]
$jj\gamma\gamma      $  &   $\surd$      &   $10/20$    &  $2.9\times 10^3$   &  $2.0\times 10^6$  \\[1.5pt]
$b\bar{b} j\gamma           $  &   $\surd$     &   $10/20$    &  $3.3\times 10^5$   &  $1.9\times 10^6$  \\[1.5pt]
$b\bar{b}h(\gamma\gamma)   $  &   $\surd$     &   $20/30$    &  $0.14          $   &  $4.9\times 10^5$  \\[1.5pt]
$t\bar{t} \gamma           $  &   $\surd$     &   $20/30$    &  $1.5\times 10^3$   &  $2.6\times 10^6$  \\[1.5pt]
$t\bar{t}h(\gamma\gamma)   $  &   $-$       &   $  -  $    &  $0.99          $   &  $4.0\times 10^5$  \\[1.5pt]
$z(b\bar{b})h(\gamma\gamma)$  &   $\surd$     &   $20/30$    &  $0.23          $   &  $2.4\times 10^5$  \\[1.5pt]
\hline
\end{tabular}
\caption{ The summary of the backgrounds to the $b\bar{b}\gamma\gamma$ decay channel of the double Higgs production process. The {\tt xqcut} and {\tt Qcut} set the matching scale in the $k_T$-jet MLM matching. The matching is performed in the four-flavor scheme. The $\sigma \cdot \rm BR$ denotes the signal rate before applying k-factors. In the $jj\gamma\gamma$ background, we required at least two leading jets with $p_T(j) > 25$ GeV (higher than the matching scale).}
\label{tab:gen_bbaa}
\end{center}
\end{table}
%%%%%%%%%%%%%%%%%

In the validation of our simulation against the ATLAS result in Table~\ref{tab:bbaaSummary}, we still need a few more steps in the background estimation. In our simulation of the $\gamma\gamma$ + jets samples (done by two processes, $b\bar{b}\gamma\gamma$ and $jj\gamma\gamma$ at the ME) matched in the four-favor scheme, the contribution to the $bj\gamma\gamma$ and $bc\gamma\gamma$ at the hadron level are underestimated. This issue is partly related to that, in the matching procedure in the four-flavor scheme, a proton does not include $b$ quark as a initial parton and the gluon splitting into $b\bar{b}$ pair, $g\to b\bar{b}$, is prohibited in the parton shower for the consistency of the four-flavor scheme. 
While the $jj\gamma\gamma$ category in Table~\ref{tab:bbaaSummary} includes $jj\gamma\gamma$, $cj\gamma\gamma$, $bj\gamma\gamma$, and $bc\gamma\gamma$ at the hadron level, the contributions from $jj\gamma\gamma$ and $cj\gamma\gamma$ at the hadron level are not big enough to agree with the ATLAS estimate. For the clarification, we made a separate set of $\gamma\gamma$ + jets samples matched in the five-flavor scheme and estimated the sizes of the $bj\gamma\gamma$, and $bc\gamma\gamma$ at the hadron level (along with the good agreement of $jj\gamma\gamma$, $cj\gamma\gamma$ with those obtained in four-flavor scheme). It turns out that the dominant contributions to $jj\gamma\gamma$ category in Table~\ref{tab:bbaaSummary} come from the $bc\gamma\gamma$ followed by $bj\gamma\gamma$, $cj\gamma\gamma$ at the hadron level.

Since we chose the matching in the four-flavor scheme with the separate simulation of the $b\bar{b}\gamma\gamma$ process at the ME to achieve a better statistics of the heavy flavor jets, we adopt a simple (somewhat {\it ad hoc}) trick to take into account the contributions to the $bc\gamma\gamma$ and $bj\gamma\gamma$ at the hadron level. We estimate the conversion probability of $j\rightarrow b$ (not confused with the mistag rate) by comparing matched ($g\to b\bar{b}$ forbidden) and unmatched ($g\to b\bar{b}$ allowed) $jj\gamma\gamma$ samples.  We find that the conversion probability is $0.9\%$ and $1.3\%$ for the default and Perugia-2012~\cite{Skands:2010ak} parton shower tuning, respectively.  Although there is some parton shower dependence, we use ${\cal P}_{j\rightarrow b} = 1\%$ in our analysis.  
%The conversion effect increase the number of backgrounds 5\% (????).

%%%%%%%%%%%%%%%%%%
\subsection{Background simulation of $b\bar{b}\tau^+\tau^-$ decay channel}
\label{app:bbtautau}

%%%%%%%%%%%%%%%%%%%%%%%%%%%%%%%%
\begin{table}[t]
\begin{center}
\begin{tabular}{c|c|c|c|c}
\hline
ME  
& Matching (scheme)
& {\tt xqcut}/{\tt Qcut} (GeV)
& $\sigma \cdot {\rm BR(fb)}$
& Generated Events
\\[1.5pt]
\hline
$ h(\tau^+\tau^-)h(b\bar{b}) $ & $  -     $ & $   -    $ & $ 2.7              $ & $  5.0 \times 10^6 $ \\[1.5pt]
$ t\bar{t}               $ & $\surd$ & $ 20/30  $ & $ 6.5  \times 10^4 $ & $  2.4 \times 10^7 $ \\[1.5pt]
$ t\bar{t}h              $ & $  -     $ & $   -    $ & $ 4.3  \times 10^2 $ & $  4.0 \times 10^6 $ \\[1.5pt]
$ t\bar{t}V              $ & $\surd$ & $ 15/25  $ & $ 1.3  \times 10^3 $ & $  4.2 \times 10^6 $ \\[1.5pt]
$ tW                     $ & $\surd$ (5-flavor)& $ 20/30  $ & $ 6.9  \times 10^4 $ & $  2.2 \times 10^7 $ \\[1.5pt]
$ Z j j              $ & $\surd$ & $ 15/25  $ & $ 1.1  \times 10^5 $ & $  6.1 \times 10^6 $ \\[1.5pt]
$ Z b\bar{b}             $ & $ \surd$ & $ 15/25  $ & $ 3.3  \times 10^3 $ & $  2.4 \times 10^6 $ \\[1.5pt]
$ hZ                     $ & $\surd$ & $ 20/30  $ & $ 23               $ & $  9.3 \times 10^6 $ \\[1.5pt]
$ VV                     $ & $\surd$ (5-flavor) & $  5/10  $ & $ 1.6  \times 10^5 $ & $  7.0 \times 10^6 $ \\[1.5pt]
\hline
\end{tabular}
\caption{The summary of the backgrounds to the $b\bar{b}\tau^+\tau^-$ decay channel of the double Higgs production process. The {\tt xqcut} and {\tt Qcut} set the matching scale in the $k_T$-jet MLM matching. The matching is performed in the four-flavor scheme unless specified explicitly. The $\sigma \cdot \rm BR$ denotes the signal rate before applying k-factors. In the $Zjj$ background, we required at least two leading jets with $p_T(j) > 25$ GeV (higher than the matching scale)}
\label{tab:gen_bbtt}
\end{center}
\end{table}
%%%%%%%%%%%%%%%%%%%%%%%%%%%%%%%%

The relevant background to the $b\bar{b}\tau^+\tau^-$ decay channel includes $Z$ + jets, $t \bar{t}$, $t \bar{t} h$, $t \bar{t} V$ (with $V = W^{\pm}, Z$), $h Z(b\bar{b}\tau^+\tau^-)$, $tW$, and $VV$. Among them, the $Z$ + jets are generated by two processes, $Z b\bar{b}$ and $Z jj$, at the ME to obtain enough statistics of the $b$-enriched events. For the $Z b\bar{b}$ background, we imposed a cut on the invariant mass of the $b\bar{b}$ system, namely $60~{\rm GeV}<m_{b\bar{b}}<300~{\rm GeV}$ at the generation. All the samples except for $t \bar{t} h$ are matched up to an additional $j$ at the ME using the $k_T$-jet MLM matching in either a four- or five-flavor scheme to partially take into accounts for NLO effects. The four-flavor scheme was chosen for the matching procedures of $Z$ + jets, $t \bar{t}$, $t \bar{t} V$, and $hZ$ processes. 
The tree-level cross section of the $t \bar{t} h$ background was rescaled to the NLO value (with the k-factor of 1.435). On the other hand, the single top with a $W$ boson, $tW$, is matched up to one additional $j$ in the five-flavor scheme excluding diagrams that overlap with the $t \bar{t}$ process to avoid a double counting~\cite{CMS-PAS-TOP-11-022}.
 
Similarly, the diboson background $VV$ is matched allowing an additional $j$ in the five-flavor scheme excluding diagrams that overlap with the $tW$ process.
The $t \bar{t}$ is a dominant background for the semileptonic $\tau^+\tau^-$ channel. We include only the leptonic decays of both tops for better statistics, or $t\bar{t}\rightarrow (bW^+)(\bar{b}W^-)\rightarrow (bl^+\nu)(bl^-\bar{\nu})$, where $l=e,\, \mu,\, \tau$. For the $h Z$ background, both $h$ and $Z$ were forced to decay into either two $b$ quarks or two $\tau$'s to improve statistics. 
The simulation details are summarized in Table~\ref{tab:gen_bbtt}. 

%%%%%%%%%%%%%%%%%%
\section{Multivariate Analysis}
\label{app:MVA}
In this section, we provide some detail of our BDT analyses discussed in Sections~\ref{MVA_bbaa} and~\ref{MVA_bbtt}. There are a few things that help us to understand the situation better. The improvements by BDT analyses in our $b\bar{b}\gamma\gamma$ and $b\bar{b}\tau^+\tau^-$ channels are marginal (less than a factor of 2), which implies that the performances of the cut-and-count analyses are not bad. The set of cuts in our cut-and-count analyses are not based on a more sophisticated optimization, and in principle, an optimization over multiparameter space could lead to a smaller discrepancy between the BDT and optimized cut-and-count analysis. In that situation, the remaining discrepancy would be purely due to the nonredundant discrimination from the BDT analysis. We suspect that the improvement by the BDT analysis is the accumulated effect of a series of small improvements in variables due to more efficient signal isolation as will be partly explained below. 

%%%%%%%%%%%%%%%%%%%
\begin{figure}[!htp!] %[tbp]
	\centering
	\includegraphics[width=0.30\linewidth]{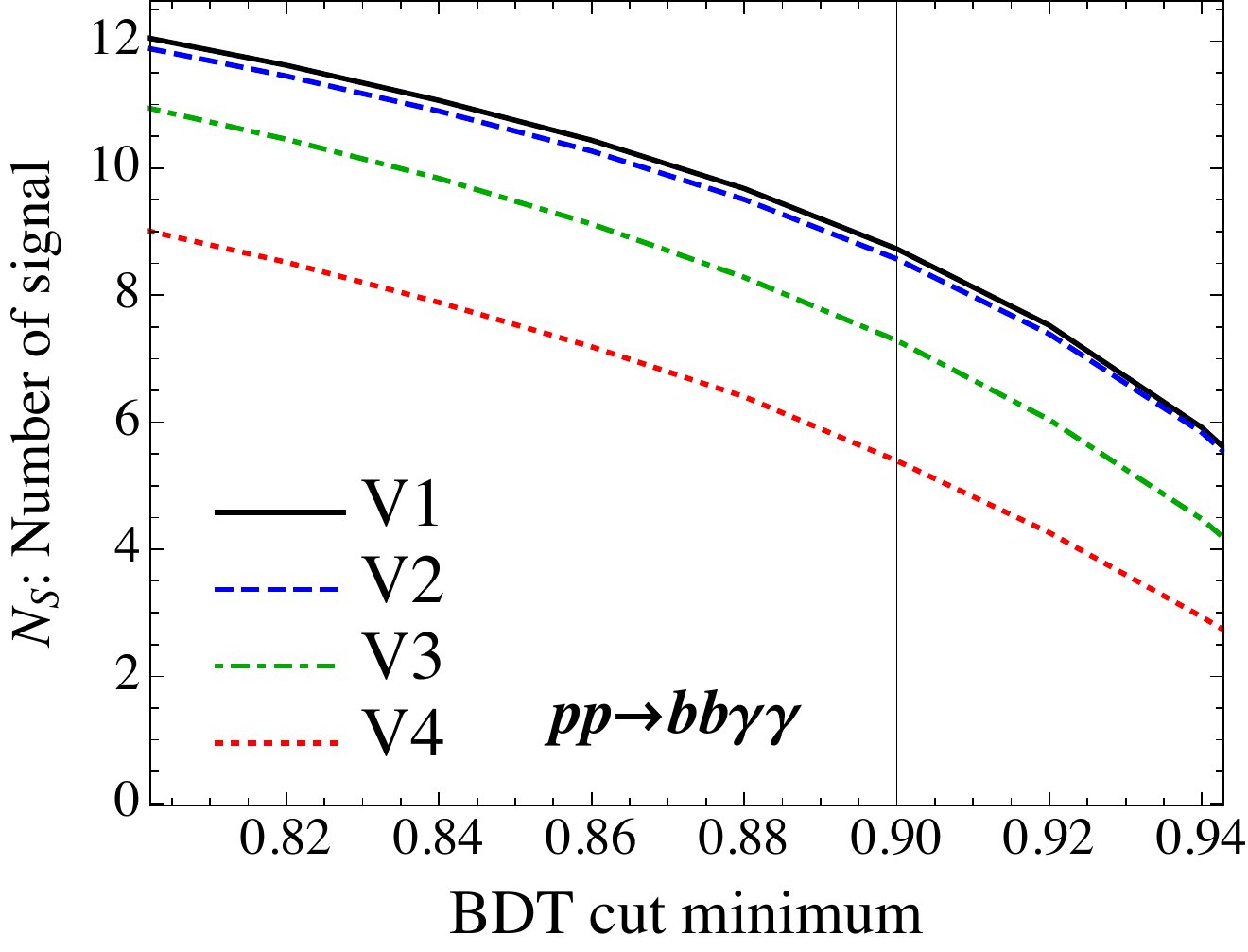}
	\includegraphics[width=0.30\linewidth]{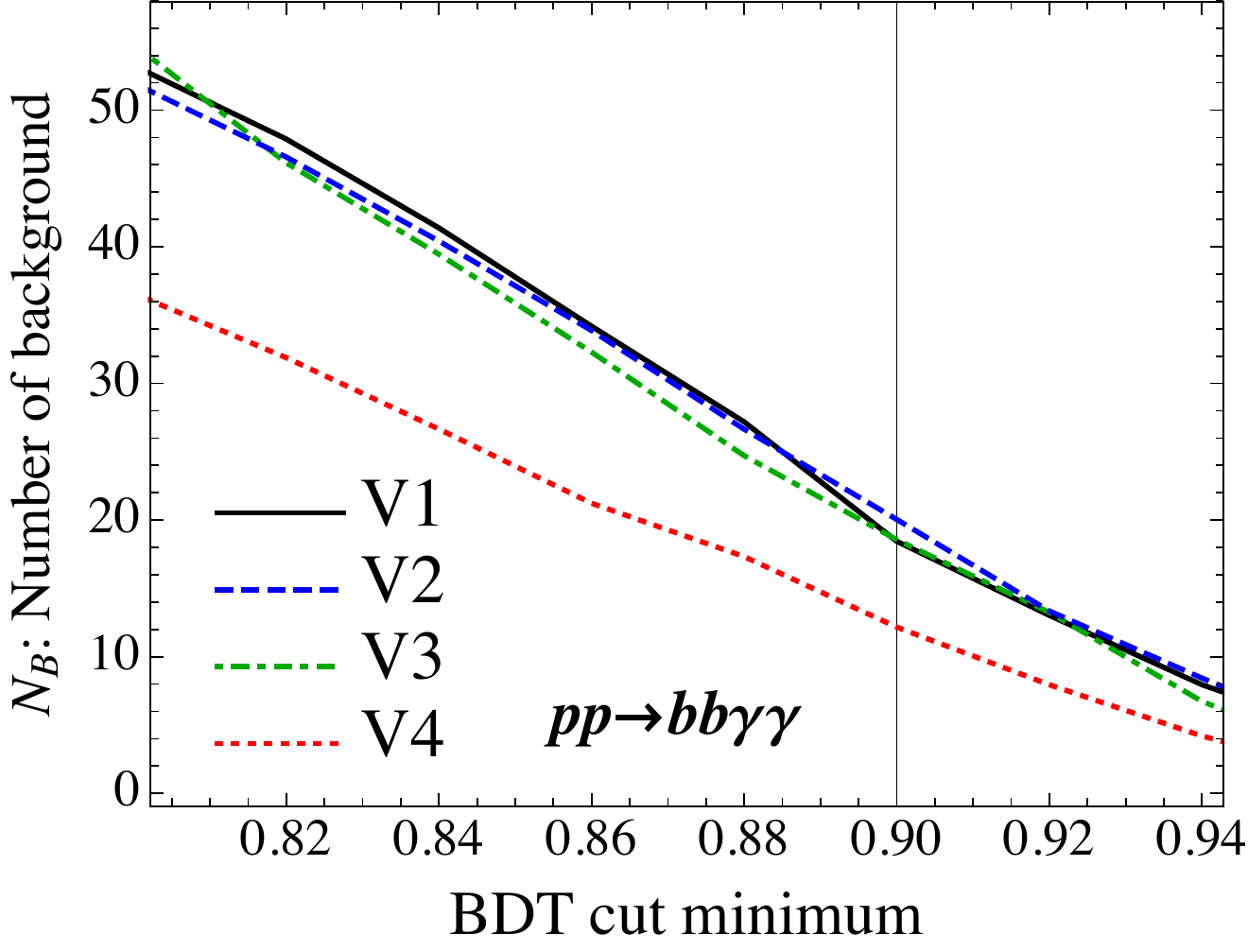}
	\includegraphics[width=0.30\linewidth]{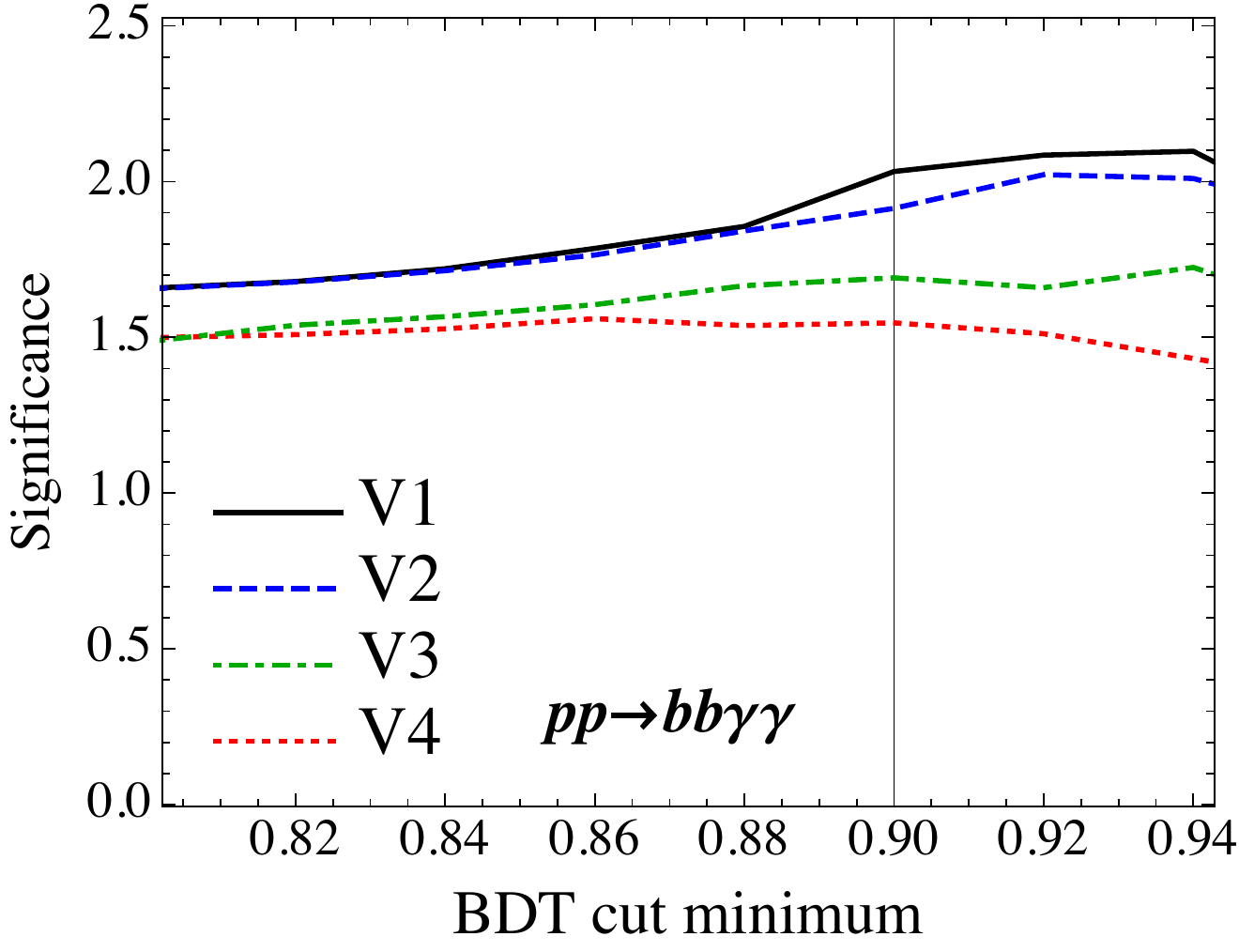}
	\caption{Similar plots to Fig.~\ref{fig:BDT_bbaa} for more BDT analyses wth the smaller set of variables. }
	\label{fig:app:bbaa}
\end{figure}
\begin{figure}[!htp!] %[tbp]
	\centering
	\includegraphics[width=0.40\linewidth]{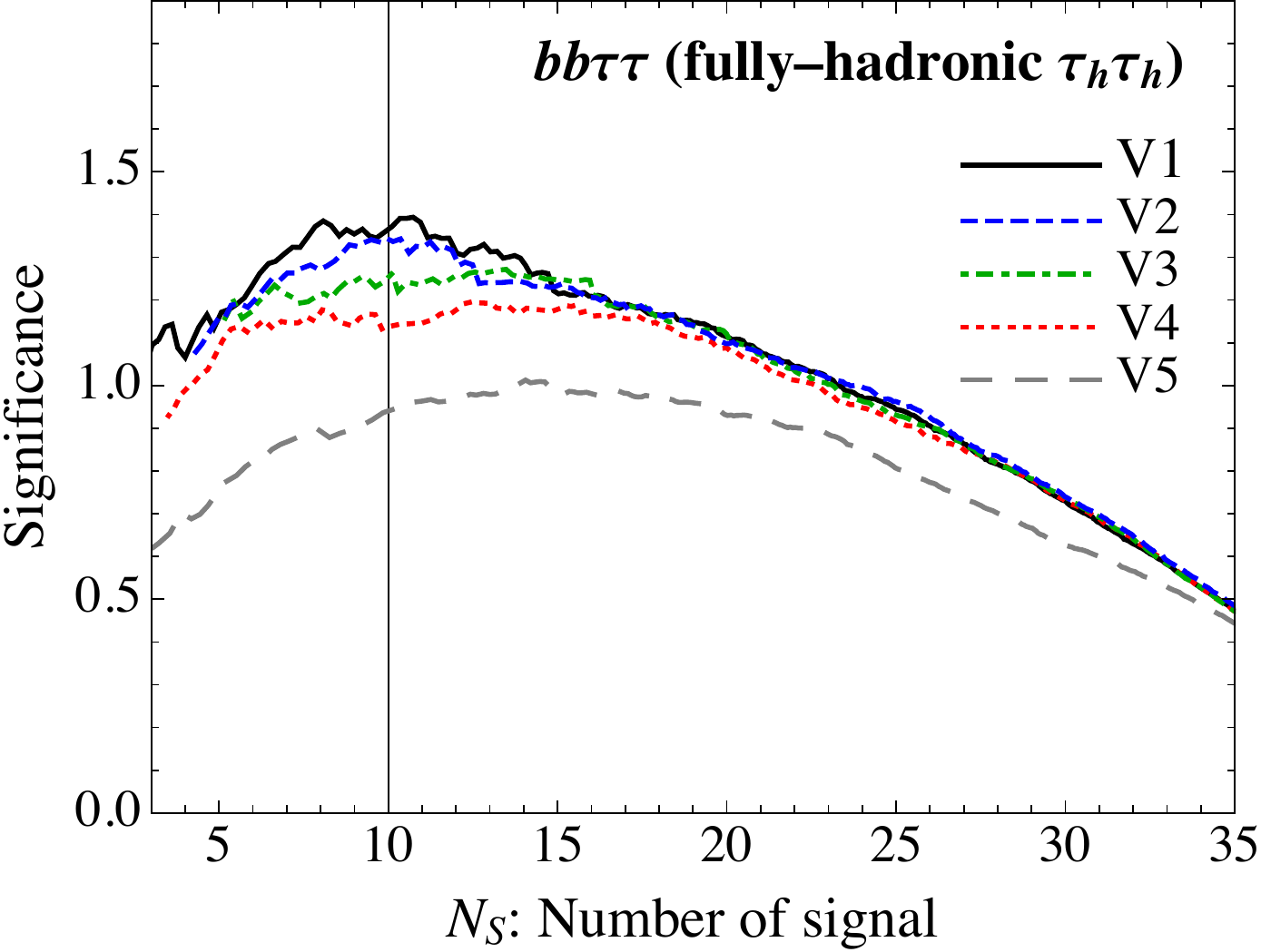}
	\includegraphics[width=0.40\linewidth]{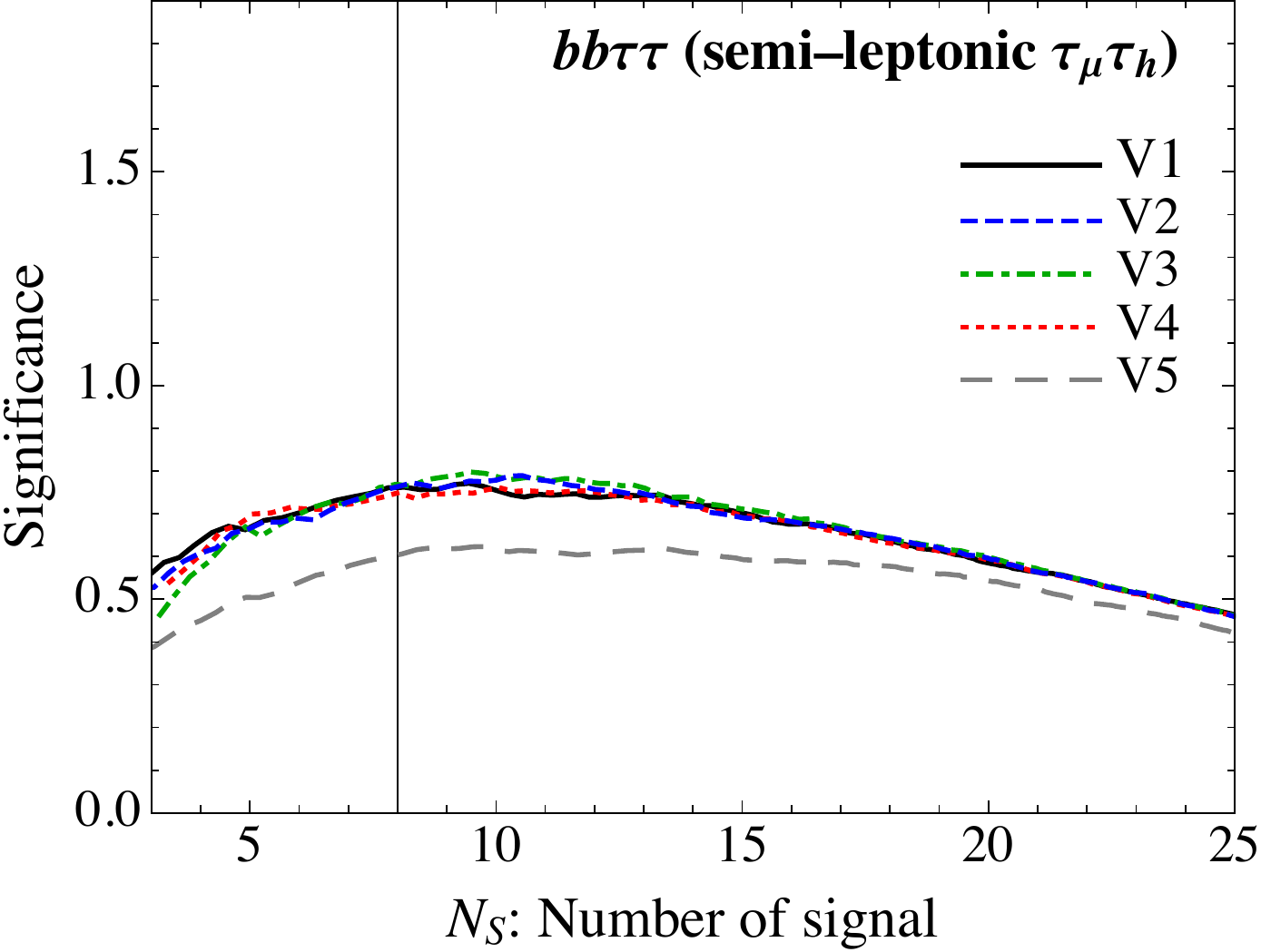}
	\caption{Similar plots to Fig.~\ref{BDT_signf_bbtt} for more BDT analyses wth the smaller set of variables. }
	\label{fig:app:bbtautau}
\end{figure}

%%%%%%%%%%%%%%%%%%%
As are shown in Figs.~\ref{fig:app:bbaa} and~\ref{fig:app:bbtautau} (similar plots to Figs.~\ref{fig:BDT_bbaa} and~\ref{BDT_signf_bbtt}), we have added more cases with a different subset of variables (cuts in Eq.~(\ref{train_condition}) were imposed in all cases):
\begin{itemize}
 \item V4  = $\{ m_{\gamma\gamma}^{\rm reco}, m_{bb}^{\rm reco}, 
 			\Delta R(b,b), \Delta R(\gamma,\gamma), \Delta R(b,\gamma)
		\}$, 
 \item V3 = V4 and $\{p_T(\gamma\gamma), p_T(bb), m_{hh}^{\rm reco}, p_T(hh) \}$, 
 \item V2 = V3 and $\{ N_{jet} \}$, 
 \item V1 = V2 and $\{ y(hh),  p_T(j_1), \Delta R(\gamma_i, b_j), E_T^{\rm miss}, p_T(\gamma), p_T(b), \eta(\gamma), \eta(b)  \}$ : all variables in our BDT analysis~,
\end{itemize}
where we start with the smallest set of variables that used to be part of the default set in literature, and then we kept adding more variables to see if they add a nonredundant discriminating power. While one could have taken into account all possible combinations of variables to figure out the effect of each individual variable, the above four cases reveals some detail of what is going on.

We have chosen the BDT cut minimum of 0.90 in our BDT analysis. As is evident in Fig.~\ref{fig:app:bbaa}, the significance gets improved as more variables are added. In V4 $\rightarrow$ V3, the enhanced significance is mainly due to $p_T(\gamma\gamma)$ and $p_T(bb)$. The jet multiplicity $N_{jet}$~\footnote{While the jet multiplicity is an efficient variable to suppress the backgrounds with the large multiplicity, the simulation of higher jet multiplicity is nontrivial in some processes including the signal: the signal simulation is at leading order (although it includes the virtual correction) in real emission and $\gamma\gamma +jets$ samples were matched up to three jets. Although we expect that the current result would be similar to the one using more rigorous simulation, its effect in the current analysis, in principle, should be taken with a grain of salt.} in V3 $\rightarrow$ V2 appears to be an efficient discriminator; for instance, it can efficiently suppress the backgrounds with the large jet multiplicity such as $t\bar{t}\gamma$ and $t\bar{t}H$. The additional variables in V1, which make the full set of variables used in our BDT analysis, only slightly improves the significance: they could be removed without affecting the result. It is interesting to notice that $p_T(\gamma)$ and $p_T(b)$ can be made redundant or can be traded for other variables.

%%%%%%%%%%%%%%%%%%%

Similarly for the $b\bar{b}\tau^+\tau^-$ channels, we have introduced the following intermediate analyses with smaller subset of variables (cuts in Eq.~(\ref{train_condition_tautau}) were imposed in all cases):
\begin{itemize}
 \item V5 = $\{ m_{bb}^{\rm reco}, m_{\tau\tau}^{\rm reco}, M_{T2} \}$, 
 \item V4 = V5 and $ \{ \Delta R(b,b), \Delta R(\tau,\tau), \Delta R(b,\tau), p_T(bb), p_T(\tau\tau) \} $, 
 \item V3 = V4 and $ \{ m_{hh}^{\rm reco}, p_T(hh), N_{jet} \} $, 
 \item V2 = V3 and $ \{ E_T^{\rm miss}, m_{\rm eff} \} $,
 \item V1  = V2 and $\{ \Delta R(\tau_i, b_j), p_T(\tau), p_T(b), \eta(\tau), \eta(b) \}$: all variables in our BDT analysis,
\end{itemize}
where the variables in V5 are similar to the subset of those used in the CMS analysis. The $\Delta R(\tau_i, b_j)$ in V1 includes three combinations except for the minimum one in V4. As is evident in Fig.~\ref{fig:app:bbtautau}, the $\Delta R$ and $p_T$ cuts of the $bb$  and $\tau^+\tau^-$ systems in V5 $\rightarrow$ V4 gives a nonredundant improvement of the significance. It is partly because $p_T(\tau)$ and $p_T(b)$ were not included in V5. While the variables in V4 are already the minimal set for the semileptonic $b\bar{b}\tau^+\tau^-$ channel, the hadronic $b\bar{b}\tau^+\tau^-$ channel continuously gains a series of improvements in V4 $\rightarrow$ V1 (for the given choice of the BDT cut that corresponds to 10 signal events). Similarly to the $b\bar{b}\gamma\gamma$ case, the $p_T(\tau)$ and $p_T(b)$ can be made redundant, or can be traded for other variables.

To make our BDT analysis stable against an overtraining issue, we have generated large enough Monte Carlo samples which guarantee smooth distributions of the variables used in the BDT analysis. As are seen in Figs.~\ref{fig:app:bbaa} and~\ref{fig:app:bbtautau}, the significance curves as a function of the BDT cut minimum, or the number of the signal events are smooth. We also made a following check to avoid overtraining: a data set is randomly split into two, namely $S_1$ and $S_2$.  A trained BDT function is obtained with $S_1$ ($S_2$), and it is applied to the other sample, $S_2$ ($S_1$).  Then, we combine two results if they look consistent.

\bibliography{lit}

\begin{thebibliography}{78}
\expandafter\ifx\csname natexlab\endcsname\relax\def\natexlab#1{#1}\fi
\expandafter\ifx\csname bibnamefont\endcsname\relax
  \def\bibnamefont#1{#1}\fi
\expandafter\ifx\csname bibfnamefont\endcsname\relax
  \def\bibfnamefont#1{#1}\fi
\expandafter\ifx\csname citenamefont\endcsname\relax
  \def\citenamefont#1{#1}\fi
\expandafter\ifx\csname url\endcsname\relax
  \def\url#1{\texttt{#1}}\fi
\expandafter\ifx\csname urlprefix\endcsname\relax\def\urlprefix{URL }\fi
\providecommand{\bibinfo}[2]{#2}
\providecommand{\eprint}[2][]{\url{#2}}

\bibitem[{\citenamefont{Aad et~al.}(2012)}]{Aad:2012tfa}
\bibinfo{author}{\bibfnamefont{G.}~\bibnamefont{Aad}} \bibnamefont{et~al.}
  (\bibinfo{collaboration}{ATLAS}), \bibinfo{journal}{Phys.Lett.}
  \textbf{\bibinfo{volume}{B716}}, \bibinfo{pages}{1} (\bibinfo{year}{2012}),
  \eprint{1207.7214}.

\bibitem[{\citenamefont{Chatrchyan et~al.}(2012)}]{Chatrchyan:2012ufa}
\bibinfo{author}{\bibfnamefont{S.}~\bibnamefont{Chatrchyan}}
  \bibnamefont{et~al.} (\bibinfo{collaboration}{CMS}),
  \bibinfo{journal}{Phys.Lett.} \textbf{\bibinfo{volume}{B716}},
  \bibinfo{pages}{30} (\bibinfo{year}{2012}), \eprint{1207.7235}.

\bibitem[{\citenamefont{Englert and Brout}(1964)}]{Englert:1964et}
\bibinfo{author}{\bibfnamefont{F.}~\bibnamefont{Englert}} \bibnamefont{and}
  \bibinfo{author}{\bibfnamefont{R.}~\bibnamefont{Brout}},
  \bibinfo{journal}{Phys. Rev. Lett.} \textbf{\bibinfo{volume}{13}},
  \bibinfo{pages}{321} (\bibinfo{year}{1964}).

\bibitem[{\citenamefont{Higgs}(1964)}]{Higgs:1964pj}
\bibinfo{author}{\bibfnamefont{P.~W.} \bibnamefont{Higgs}},
  \bibinfo{journal}{Phys. Rev. Lett.} \textbf{\bibinfo{volume}{13}},
  \bibinfo{pages}{508} (\bibinfo{year}{1964}).

\bibitem[{\citenamefont{Guralnik et~al.}(1964)\citenamefont{Guralnik, Hagen,
  and Kibble}}]{Guralnik:1964eu}
\bibinfo{author}{\bibfnamefont{G.~S.} \bibnamefont{Guralnik}},
  \bibinfo{author}{\bibfnamefont{C.~R.} \bibnamefont{Hagen}}, \bibnamefont{and}
  \bibinfo{author}{\bibfnamefont{T.~W.~B.} \bibnamefont{Kibble}},
  \bibinfo{journal}{Phys. Rev. Lett.} \textbf{\bibinfo{volume}{13}},
  \bibinfo{pages}{585} (\bibinfo{year}{1964}).

\bibitem[{\citenamefont{Weinberg}(1967)}]{Weinberg:1967tq}
\bibinfo{author}{\bibfnamefont{S.}~\bibnamefont{Weinberg}},
  \bibinfo{journal}{Phys. Rev. Lett.} \textbf{\bibinfo{volume}{19}},
  \bibinfo{pages}{1264} (\bibinfo{year}{1967}).

\bibitem[{\citenamefont{Khachatryan et~al.}(2016)}]{Khachatryan:2016sey}
\bibinfo{author}{\bibfnamefont{V.}~\bibnamefont{Khachatryan}}
  \bibnamefont{et~al.} (\bibinfo{collaboration}{CMS}), \bibinfo{journal}{Phys.
  Rev.} \textbf{\bibinfo{volume}{D94}}, \bibinfo{pages}{052012}
  (\bibinfo{year}{2016}), \eprint{1603.06896}.

\bibitem[{CMS(2017)}]{CMS-PAS-HIG-17-008}
\bibinfo{type}{Tech. Rep.} \bibinfo{number}{CMS-PAS-HIG-17-008},
  \bibinfo{institution}{CERN}, \bibinfo{address}{Geneva}
  (\bibinfo{year}{2017}), \urlprefix\url{http://cds.cern.ch/record/2273383}.

\bibitem[{\citenamefont{Baer et~al.}(2013)\citenamefont{Baer, Barklow, Fujii,
  Gao, Hoang, Kanemura, List, Logan, Nomerotski, Perelstein
  et~al.}}]{Baer:2013cma}
\bibinfo{author}{\bibfnamefont{H.}~\bibnamefont{Baer}},
  \bibinfo{author}{\bibfnamefont{T.}~\bibnamefont{Barklow}},
  \bibinfo{author}{\bibfnamefont{K.}~\bibnamefont{Fujii}},
  \bibinfo{author}{\bibfnamefont{Y.}~\bibnamefont{Gao}},
  \bibinfo{author}{\bibfnamefont{A.}~\bibnamefont{Hoang}},
  \bibinfo{author}{\bibfnamefont{S.}~\bibnamefont{Kanemura}},
  \bibinfo{author}{\bibfnamefont{J.}~\bibnamefont{List}},
  \bibinfo{author}{\bibfnamefont{H.~E.} \bibnamefont{Logan}},
  \bibinfo{author}{\bibfnamefont{A.}~\bibnamefont{Nomerotski}},
  \bibinfo{author}{\bibfnamefont{M.}~\bibnamefont{Perelstein}},
  \bibnamefont{et~al.} (\bibinfo{year}{2013}), \eprint{1306.6352}.

\bibitem[{\citenamefont{Barr et~al.}(2015)\citenamefont{Barr, Dolan, Englert,
  Ferreira~de Lima, and Spannowsky}}]{Barr:2014sga}
\bibinfo{author}{\bibfnamefont{A.~J.} \bibnamefont{Barr}},
  \bibinfo{author}{\bibfnamefont{M.~J.} \bibnamefont{Dolan}},
  \bibinfo{author}{\bibfnamefont{C.}~\bibnamefont{Englert}},
  \bibinfo{author}{\bibfnamefont{D.~E.} \bibnamefont{Ferreira~de Lima}},
  \bibnamefont{and}
  \bibinfo{author}{\bibfnamefont{M.}~\bibnamefont{Spannowsky}},
  \bibinfo{journal}{JHEP} \textbf{\bibinfo{volume}{02}}, \bibinfo{pages}{016}
  (\bibinfo{year}{2015}), \eprint{1412.7154}.

\bibitem[{\citenamefont{Azatov et~al.}(2015)\citenamefont{Azatov, Contino,
  Panico, and Son}}]{Azatov:2015oxa}
\bibinfo{author}{\bibfnamefont{A.}~\bibnamefont{Azatov}},
  \bibinfo{author}{\bibfnamefont{R.}~\bibnamefont{Contino}},
  \bibinfo{author}{\bibfnamefont{G.}~\bibnamefont{Panico}}, \bibnamefont{and}
  \bibinfo{author}{\bibfnamefont{M.}~\bibnamefont{Son}},
  \bibinfo{journal}{Phys. Rev.} \textbf{\bibinfo{volume}{D92}},
  \bibinfo{pages}{035001} (\bibinfo{year}{2015}), \eprint{1502.00539}.

\bibitem[{\citenamefont{He et~al.}(2016)\citenamefont{He, Ren, and
  Yao}}]{He:2015spf}
\bibinfo{author}{\bibfnamefont{H.-J.} \bibnamefont{He}},
  \bibinfo{author}{\bibfnamefont{J.}~\bibnamefont{Ren}}, \bibnamefont{and}
  \bibinfo{author}{\bibfnamefont{W.}~\bibnamefont{Yao}},
  \bibinfo{journal}{Phys. Rev.} \textbf{\bibinfo{volume}{D93}},
  \bibinfo{pages}{015003} (\bibinfo{year}{2016}), \eprint{1506.03302}.

\bibitem[{\citenamefont{Contino et~al.}(2017)}]{Contino:2016spe}
\bibinfo{author}{\bibfnamefont{R.}~\bibnamefont{Contino}} \bibnamefont{et~al.},
  \bibinfo{journal}{CERN Yellow Report} pp. \bibinfo{pages}{255--440}
  (\bibinfo{year}{2017}), \eprint{1606.09408}.

\bibitem[{\citenamefont{Trodden}(1999)}]{Trodden:1998ym}
\bibinfo{author}{\bibfnamefont{M.}~\bibnamefont{Trodden}},
  \bibinfo{journal}{Rev. Mod. Phys.} \textbf{\bibinfo{volume}{71}},
  \bibinfo{pages}{1463} (\bibinfo{year}{1999}), \eprint{hep-ph/9803479}.

\bibitem[{\citenamefont{Cline}(2006)}]{Cline:2006ts}
\bibinfo{author}{\bibfnamefont{J.~M.} \bibnamefont{Cline}}, in
  \emph{\bibinfo{booktitle}{{Les Houches Summer School - Session 86: Particle
  Physics and Cosmology: The Fabric of Spacetime Les Houches, France, July
  31-August 25, 2006}}} (\bibinfo{year}{2006}), \eprint{hep-ph/0609145}.

\bibitem[{\citenamefont{Morrissey and Ramsey-Musolf}(2012)}]{Morrissey:2012db}
\bibinfo{author}{\bibfnamefont{D.~E.} \bibnamefont{Morrissey}}
  \bibnamefont{and} \bibinfo{author}{\bibfnamefont{M.~J.}
  \bibnamefont{Ramsey-Musolf}}, \bibinfo{journal}{New J. Phys.}
  \textbf{\bibinfo{volume}{14}}, \bibinfo{pages}{125003}
  (\bibinfo{year}{2012}), \eprint{1206.2942}.

\bibitem[{\citenamefont{Degrassi et~al.}(2016)\citenamefont{Degrassi, Giardino,
  Maltoni, and Pagani}}]{Degrassi:2016wml}
\bibinfo{author}{\bibfnamefont{G.}~\bibnamefont{Degrassi}},
  \bibinfo{author}{\bibfnamefont{P.~P.} \bibnamefont{Giardino}},
  \bibinfo{author}{\bibfnamefont{F.}~\bibnamefont{Maltoni}}, \bibnamefont{and}
  \bibinfo{author}{\bibfnamefont{D.}~\bibnamefont{Pagani}},
  \bibinfo{journal}{JHEP} \textbf{\bibinfo{volume}{12}}, \bibinfo{pages}{080}
  (\bibinfo{year}{2016}), \eprint{1607.04251}.

\bibitem[{\citenamefont{Gorbahn and Haisch}(2016)}]{Gorbahn:2016uoy}
\bibinfo{author}{\bibfnamefont{M.}~\bibnamefont{Gorbahn}} \bibnamefont{and}
  \bibinfo{author}{\bibfnamefont{U.}~\bibnamefont{Haisch}},
  \bibinfo{journal}{JHEP} \textbf{\bibinfo{volume}{10}}, \bibinfo{pages}{094}
  (\bibinfo{year}{2016}), \eprint{1607.03773}.

\bibitem[{\citenamefont{Bizon et~al.}(2017)\citenamefont{Bizon, Gorbahn,
  Haisch, and Zanderighi}}]{Bizon:2016wgr}
\bibinfo{author}{\bibfnamefont{W.}~\bibnamefont{Bizon}},
  \bibinfo{author}{\bibfnamefont{M.}~\bibnamefont{Gorbahn}},
  \bibinfo{author}{\bibfnamefont{U.}~\bibnamefont{Haisch}}, \bibnamefont{and}
  \bibinfo{author}{\bibfnamefont{G.}~\bibnamefont{Zanderighi}},
  \bibinfo{journal}{JHEP} \textbf{\bibinfo{volume}{07}}, \bibinfo{pages}{083}
  (\bibinfo{year}{2017}), \eprint{1610.05771}.

\bibitem[{\citenamefont{Maltoni et~al.}(2017)\citenamefont{Maltoni, Pagani,
  Shivaji, and Zhao}}]{Maltoni:2017ims}
\bibinfo{author}{\bibfnamefont{F.}~\bibnamefont{Maltoni}},
  \bibinfo{author}{\bibfnamefont{D.}~\bibnamefont{Pagani}},
  \bibinfo{author}{\bibfnamefont{A.}~\bibnamefont{Shivaji}}, \bibnamefont{and}
  \bibinfo{author}{\bibfnamefont{X.}~\bibnamefont{Zhao}},
  \bibinfo{journal}{Eur. Phys. J.} \textbf{\bibinfo{volume}{C77}},
  \bibinfo{pages}{887} (\bibinfo{year}{2017}), \eprint{1709.08649}.

\bibitem[{\citenamefont{Di~Vita et~al.}(2017)\citenamefont{Di~Vita, Grojean,
  Panico, Riembau, and Vantalon}}]{DiVita:2017eyz}
\bibinfo{author}{\bibfnamefont{S.}~\bibnamefont{Di~Vita}},
  \bibinfo{author}{\bibfnamefont{C.}~\bibnamefont{Grojean}},
  \bibinfo{author}{\bibfnamefont{G.}~\bibnamefont{Panico}},
  \bibinfo{author}{\bibfnamefont{M.}~\bibnamefont{Riembau}}, \bibnamefont{and}
  \bibinfo{author}{\bibfnamefont{T.}~\bibnamefont{Vantalon}},
  \bibinfo{journal}{JHEP} \textbf{\bibinfo{volume}{09}}, \bibinfo{pages}{069}
  (\bibinfo{year}{2017}), \eprint{1704.01953}.

\bibitem[{\citenamefont{Cao et~al.}(2017{\natexlab{a}})\citenamefont{Cao, Liu,
  and Yan}}]{Cao:2015oxx}
\bibinfo{author}{\bibfnamefont{Q.-H.} \bibnamefont{Cao}},
  \bibinfo{author}{\bibfnamefont{Y.}~\bibnamefont{Liu}}, \bibnamefont{and}
  \bibinfo{author}{\bibfnamefont{B.}~\bibnamefont{Yan}},
  \bibinfo{journal}{Phys. Rev.} \textbf{\bibinfo{volume}{D95}},
  \bibinfo{pages}{073006} (\bibinfo{year}{2017}{\natexlab{a}}),
  \eprint{1511.03311}.

\bibitem[{\citenamefont{Bishara et~al.}(2017)\citenamefont{Bishara, Contino,
  and Rojo}}]{Bishara:2016kjn}
\bibinfo{author}{\bibfnamefont{F.}~\bibnamefont{Bishara}},
  \bibinfo{author}{\bibfnamefont{R.}~\bibnamefont{Contino}}, \bibnamefont{and}
  \bibinfo{author}{\bibfnamefont{J.}~\bibnamefont{Rojo}},
  \bibinfo{journal}{Eur. Phys. J.} \textbf{\bibinfo{volume}{C77}},
  \bibinfo{pages}{481} (\bibinfo{year}{2017}), \eprint{1611.03860}.

\bibitem[{\citenamefont{Kling et~al.}(2017)\citenamefont{Kling, Plehn, and
  Schichtel}}]{Kling:2016lay}
\bibinfo{author}{\bibfnamefont{F.}~\bibnamefont{Kling}},
  \bibinfo{author}{\bibfnamefont{T.}~\bibnamefont{Plehn}}, \bibnamefont{and}
  \bibinfo{author}{\bibfnamefont{P.}~\bibnamefont{Schichtel}},
  \bibinfo{journal}{Phys. Rev.} \textbf{\bibinfo{volume}{D95}},
  \bibinfo{pages}{035026} (\bibinfo{year}{2017}), \eprint{1607.07441}.

\bibitem[{\citenamefont{Baur et~al.}(2004)\citenamefont{Baur, Plehn, and
  Rainwater}}]{Baur:2003gp}
\bibinfo{author}{\bibfnamefont{U.}~\bibnamefont{Baur}},
  \bibinfo{author}{\bibfnamefont{T.}~\bibnamefont{Plehn}}, \bibnamefont{and}
  \bibinfo{author}{\bibfnamefont{D.~L.} \bibnamefont{Rainwater}},
  \bibinfo{journal}{Phys. Rev.} \textbf{\bibinfo{volume}{D69}},
  \bibinfo{pages}{053004} (\bibinfo{year}{2004}), \eprint{hep-ph/0310056}.

\bibitem[{\citenamefont{Baglio et~al.}(2013)\citenamefont{Baglio, Djouadi,
  Gröber, Mühlleitner, Quevillon, and Spira}}]{Baglio:2012np}
\bibinfo{author}{\bibfnamefont{J.}~\bibnamefont{Baglio}},
  \bibinfo{author}{\bibfnamefont{A.}~\bibnamefont{Djouadi}},
  \bibinfo{author}{\bibfnamefont{R.}~\bibnamefont{Gröber}},
  \bibinfo{author}{\bibfnamefont{M.~M.} \bibnamefont{Mühlleitner}},
  \bibinfo{author}{\bibfnamefont{J.}~\bibnamefont{Quevillon}},
  \bibnamefont{and} \bibinfo{author}{\bibfnamefont{M.}~\bibnamefont{Spira}},
  \bibinfo{journal}{JHEP} \textbf{\bibinfo{volume}{04}}, \bibinfo{pages}{151}
  (\bibinfo{year}{2013}), \eprint{1212.5581}.

\bibitem[{\citenamefont{Huang et~al.}(2016)\citenamefont{Huang, Joglekar, Li,
  and Wagner}}]{Huang:2015tdv}
\bibinfo{author}{\bibfnamefont{P.}~\bibnamefont{Huang}},
  \bibinfo{author}{\bibfnamefont{A.}~\bibnamefont{Joglekar}},
  \bibinfo{author}{\bibfnamefont{B.}~\bibnamefont{Li}}, \bibnamefont{and}
  \bibinfo{author}{\bibfnamefont{C.~E.~M.} \bibnamefont{Wagner}},
  \bibinfo{journal}{Phys. Rev.} \textbf{\bibinfo{volume}{D93}},
  \bibinfo{pages}{055049} (\bibinfo{year}{2016}), \eprint{1512.00068}.

\bibitem[{\citenamefont{Cao et~al.}(2016)\citenamefont{Cao, Yan, Zhang, and
  Zhang}}]{Cao:2015oaa}
\bibinfo{author}{\bibfnamefont{Q.-H.} \bibnamefont{Cao}},
  \bibinfo{author}{\bibfnamefont{B.}~\bibnamefont{Yan}},
  \bibinfo{author}{\bibfnamefont{D.-M.} \bibnamefont{Zhang}}, \bibnamefont{and}
  \bibinfo{author}{\bibfnamefont{H.}~\bibnamefont{Zhang}},
  \bibinfo{journal}{Phys. Lett.} \textbf{\bibinfo{volume}{B752}},
  \bibinfo{pages}{285} (\bibinfo{year}{2016}), \eprint{1508.06512}.

\bibitem[{\citenamefont{Cao et~al.}(2017{\natexlab{b}})\citenamefont{Cao, Li,
  Yan, Zhang, and Zhang}}]{Cao:2016zob}
\bibinfo{author}{\bibfnamefont{Q.-H.} \bibnamefont{Cao}},
  \bibinfo{author}{\bibfnamefont{G.}~\bibnamefont{Li}},
  \bibinfo{author}{\bibfnamefont{B.}~\bibnamefont{Yan}},
  \bibinfo{author}{\bibfnamefont{D.-M.} \bibnamefont{Zhang}}, \bibnamefont{and}
  \bibinfo{author}{\bibfnamefont{H.}~\bibnamefont{Zhang}},
  \bibinfo{journal}{Phys. Rev.} \textbf{\bibinfo{volume}{D96}},
  \bibinfo{pages}{095031} (\bibinfo{year}{2017}{\natexlab{b}}),
  \eprint{1611.09336}.

\bibitem[{\citenamefont{Alves et~al.}(2017)\citenamefont{Alves, Ghosh, and
  Sinha}}]{Alves:2017ued}
\bibinfo{author}{\bibfnamefont{A.}~\bibnamefont{Alves}},
  \bibinfo{author}{\bibfnamefont{T.}~\bibnamefont{Ghosh}}, \bibnamefont{and}
  \bibinfo{author}{\bibfnamefont{K.}~\bibnamefont{Sinha}},
  \bibinfo{journal}{Phys. Rev.} \textbf{\bibinfo{volume}{D96}},
  \bibinfo{pages}{035022} (\bibinfo{year}{2017}), \eprint{1704.07395}.

\bibitem[{CMS(2015)}]{CMS-PAS-FTR-15-002}
\bibinfo{type}{Tech. Rep.} \bibinfo{number}{CMS-PAS-FTR-15-002},
  \bibinfo{institution}{CERN}, \bibinfo{address}{Geneva}
  (\bibinfo{year}{2015}), \urlprefix\url{https://cds.cern.ch/record/2063038}.

\bibitem[{ATL(2014)}]{ATL-PHYS-PUB-2014-019}
\bibinfo{type}{Tech. Rep.} \bibinfo{number}{ATL-PHYS-PUB-2014-019},
  \bibinfo{institution}{CERN}, \bibinfo{address}{Geneva}
  (\bibinfo{year}{2014}), \urlprefix\url{http://cds.cern.ch/record/1956733}.

\bibitem[{\citenamefont{Barger et~al.}(2014)\citenamefont{Barger, Everett,
  Jackson, and Shaughnessy}}]{Barger:2013jfa}
\bibinfo{author}{\bibfnamefont{V.}~\bibnamefont{Barger}},
  \bibinfo{author}{\bibfnamefont{L.~L.} \bibnamefont{Everett}},
  \bibinfo{author}{\bibfnamefont{C.~B.} \bibnamefont{Jackson}},
  \bibnamefont{and}
  \bibinfo{author}{\bibfnamefont{G.}~\bibnamefont{Shaughnessy}},
  \bibinfo{journal}{Phys. Lett.} \textbf{\bibinfo{volume}{B728}},
  \bibinfo{pages}{433} (\bibinfo{year}{2014}), \eprint{1311.2931}.

\bibitem[{\citenamefont{Baur et~al.}(2003)\citenamefont{Baur, Plehn, and
  Rainwater}}]{Baur:2003gpa}
\bibinfo{author}{\bibfnamefont{U.}~\bibnamefont{Baur}},
  \bibinfo{author}{\bibfnamefont{T.}~\bibnamefont{Plehn}}, \bibnamefont{and}
  \bibinfo{author}{\bibfnamefont{D.~L.} \bibnamefont{Rainwater}},
  \bibinfo{journal}{Phys. Rev.} \textbf{\bibinfo{volume}{D68}},
  \bibinfo{pages}{033001} (\bibinfo{year}{2003}), \eprint{hep-ph/0304015}.

\bibitem[{\citenamefont{Dolan et~al.}(2012)\citenamefont{Dolan, Englert, and
  Spannowsky}}]{Dolan:2012rv}
\bibinfo{author}{\bibfnamefont{M.~J.} \bibnamefont{Dolan}},
  \bibinfo{author}{\bibfnamefont{C.}~\bibnamefont{Englert}}, \bibnamefont{and}
  \bibinfo{author}{\bibfnamefont{M.}~\bibnamefont{Spannowsky}},
  \bibinfo{journal}{JHEP} \textbf{\bibinfo{volume}{10}}, \bibinfo{pages}{112}
  (\bibinfo{year}{2012}), \eprint{1206.5001}.

\bibitem[{\citenamefont{Papaefstathiou
  et~al.}(2013)\citenamefont{Papaefstathiou, Yang, and
  Zurita}}]{Papaefstathiou:2012qe}
\bibinfo{author}{\bibfnamefont{A.}~\bibnamefont{Papaefstathiou}},
  \bibinfo{author}{\bibfnamefont{L.~L.} \bibnamefont{Yang}}, \bibnamefont{and}
  \bibinfo{author}{\bibfnamefont{J.}~\bibnamefont{Zurita}},
  \bibinfo{journal}{Phys. Rev.} \textbf{\bibinfo{volume}{D87}},
  \bibinfo{pages}{011301} (\bibinfo{year}{2013}), \eprint{1209.1489}.

\bibitem[{\citenamefont{Huang et~al.}(2017)\citenamefont{Huang, No, Pernié,
  Ramsey-Musolf, Safonov, Spannowsky, and Winslow}}]{Huang:2017jws}
\bibinfo{author}{\bibfnamefont{T.}~\bibnamefont{Huang}},
  \bibinfo{author}{\bibfnamefont{J.~M.} \bibnamefont{No}},
  \bibinfo{author}{\bibfnamefont{L.}~\bibnamefont{Pernié}},
  \bibinfo{author}{\bibfnamefont{M.}~\bibnamefont{Ramsey-Musolf}},
  \bibinfo{author}{\bibfnamefont{A.}~\bibnamefont{Safonov}},
  \bibinfo{author}{\bibfnamefont{M.}~\bibnamefont{Spannowsky}},
  \bibnamefont{and} \bibinfo{author}{\bibfnamefont{P.}~\bibnamefont{Winslow}},
  \bibinfo{journal}{Phys. Rev.} \textbf{\bibinfo{volume}{D96}},
  \bibinfo{pages}{035007} (\bibinfo{year}{2017}), \eprint{1701.04442}.

\bibitem[{\citenamefont{Ferreira~de Lima et~al.}(2014)\citenamefont{Ferreira~de
  Lima, Papaefstathiou, and Spannowsky}}]{deLima:2014dta}
\bibinfo{author}{\bibfnamefont{D.~E.} \bibnamefont{Ferreira~de Lima}},
  \bibinfo{author}{\bibfnamefont{A.}~\bibnamefont{Papaefstathiou}},
  \bibnamefont{and}
  \bibinfo{author}{\bibfnamefont{M.}~\bibnamefont{Spannowsky}},
  \bibinfo{journal}{JHEP} \textbf{\bibinfo{volume}{08}}, \bibinfo{pages}{030}
  (\bibinfo{year}{2014}), \eprint{1404.7139}.

\bibitem[{\citenamefont{Wardrope et~al.}(2015)\citenamefont{Wardrope, Jansen,
  Konstantinidis, Cooper, Falla, and Norjoharuddeen}}]{Wardrope:2014kya}
\bibinfo{author}{\bibfnamefont{D.}~\bibnamefont{Wardrope}},
  \bibinfo{author}{\bibfnamefont{E.}~\bibnamefont{Jansen}},
  \bibinfo{author}{\bibfnamefont{N.}~\bibnamefont{Konstantinidis}},
  \bibinfo{author}{\bibfnamefont{B.}~\bibnamefont{Cooper}},
  \bibinfo{author}{\bibfnamefont{R.}~\bibnamefont{Falla}}, \bibnamefont{and}
  \bibinfo{author}{\bibfnamefont{N.}~\bibnamefont{Norjoharuddeen}},
  \bibinfo{journal}{Eur. Phys. J.} \textbf{\bibinfo{volume}{C75}},
  \bibinfo{pages}{219} (\bibinfo{year}{2015}), \eprint{1410.2794}.

\bibitem[{\citenamefont{Behr et~al.}(2016)\citenamefont{Behr, Bortoletto,
  Frost, Hartland, Issever, and Rojo}}]{Behr:2015oqq}
\bibinfo{author}{\bibfnamefont{J.~K.} \bibnamefont{Behr}},
  \bibinfo{author}{\bibfnamefont{D.}~\bibnamefont{Bortoletto}},
  \bibinfo{author}{\bibfnamefont{J.~A.} \bibnamefont{Frost}},
  \bibinfo{author}{\bibfnamefont{N.~P.} \bibnamefont{Hartland}},
  \bibinfo{author}{\bibfnamefont{C.}~\bibnamefont{Issever}}, \bibnamefont{and}
  \bibinfo{author}{\bibfnamefont{J.}~\bibnamefont{Rojo}},
  \bibinfo{journal}{Eur. Phys. J.} \textbf{\bibinfo{volume}{C76}},
  \bibinfo{pages}{386} (\bibinfo{year}{2016}), \eprint{1512.08928}.

\bibitem[{\citenamefont{Barr et~al.}(2011)\citenamefont{Barr, French, Frost,
  and Lester}}]{Barr:2011he}
\bibinfo{author}{\bibfnamefont{A.~J.} \bibnamefont{Barr}},
  \bibinfo{author}{\bibfnamefont{S.~T.} \bibnamefont{French}},
  \bibinfo{author}{\bibfnamefont{J.~A.} \bibnamefont{Frost}}, \bibnamefont{and}
  \bibinfo{author}{\bibfnamefont{C.~G.} \bibnamefont{Lester}},
  \bibinfo{journal}{JHEP} \textbf{\bibinfo{volume}{10}}, \bibinfo{pages}{080}
  (\bibinfo{year}{2011}), \eprint{1106.2322}.

\bibitem[{\citenamefont{Contino et~al.}(2012)\citenamefont{Contino, Ghezzi,
  Moretti, Panico, Piccinini, and Wulzer}}]{Contino:2012xk}
\bibinfo{author}{\bibfnamefont{R.}~\bibnamefont{Contino}},
  \bibinfo{author}{\bibfnamefont{M.}~\bibnamefont{Ghezzi}},
  \bibinfo{author}{\bibfnamefont{M.}~\bibnamefont{Moretti}},
  \bibinfo{author}{\bibfnamefont{G.}~\bibnamefont{Panico}},
  \bibinfo{author}{\bibfnamefont{F.}~\bibnamefont{Piccinini}},
  \bibnamefont{and} \bibinfo{author}{\bibfnamefont{A.}~\bibnamefont{Wulzer}},
  \bibinfo{journal}{JHEP} \textbf{\bibinfo{volume}{08}}, \bibinfo{pages}{154}
  (\bibinfo{year}{2012}), \eprint{1205.5444}.

\bibitem[{\citenamefont{Goertz et~al.}(2015)\citenamefont{Goertz,
  Papaefstathiou, Yang, and Zurita}}]{Goertz:2014qta}
\bibinfo{author}{\bibfnamefont{F.}~\bibnamefont{Goertz}},
  \bibinfo{author}{\bibfnamefont{A.}~\bibnamefont{Papaefstathiou}},
  \bibinfo{author}{\bibfnamefont{L.~L.} \bibnamefont{Yang}}, \bibnamefont{and}
  \bibinfo{author}{\bibfnamefont{J.}~\bibnamefont{Zurita}},
  \bibinfo{journal}{JHEP} \textbf{\bibinfo{volume}{04}}, \bibinfo{pages}{167}
  (\bibinfo{year}{2015}), \eprint{1410.3471}.

\bibitem[{ATL(2017{\natexlab{a}})}]{ATL-PHYS-PUB-2017-001}
\bibinfo{type}{Tech. Rep.} \bibinfo{number}{ATL-PHYS-PUB-2017-001},
  \bibinfo{institution}{CERN}, \bibinfo{address}{Geneva}
  (\bibinfo{year}{2017}{\natexlab{a}}),
  \urlprefix\url{http://atlas.web.cern.ch/Atlas/GROUPS/PHYSICS/PUBNOTES/ATL-PHYS-PUB-2017-001/}.

\bibitem[{\citenamefont{Giudice et~al.}(2007)\citenamefont{Giudice, Grojean,
  Pomarol, and Rattazzi}}]{Giudice:2007fh}
\bibinfo{author}{\bibfnamefont{G.~F.} \bibnamefont{Giudice}},
  \bibinfo{author}{\bibfnamefont{C.}~\bibnamefont{Grojean}},
  \bibinfo{author}{\bibfnamefont{A.}~\bibnamefont{Pomarol}}, \bibnamefont{and}
  \bibinfo{author}{\bibfnamefont{R.}~\bibnamefont{Rattazzi}},
  \bibinfo{journal}{JHEP} \textbf{\bibinfo{volume}{06}}, \bibinfo{pages}{045}
  (\bibinfo{year}{2007}), \eprint{hep-ph/0703164}.

\bibitem[{\citenamefont{Falkowski et~al.}(2017)\citenamefont{Falkowski,
  Gonzalez-Alonso, Greljo, Marzocca, and Son}}]{Falkowski:2016cxu}
\bibinfo{author}{\bibfnamefont{A.}~\bibnamefont{Falkowski}},
  \bibinfo{author}{\bibfnamefont{M.}~\bibnamefont{Gonzalez-Alonso}},
  \bibinfo{author}{\bibfnamefont{A.}~\bibnamefont{Greljo}},
  \bibinfo{author}{\bibfnamefont{D.}~\bibnamefont{Marzocca}}, \bibnamefont{and}
  \bibinfo{author}{\bibfnamefont{M.}~\bibnamefont{Son}},
  \bibinfo{journal}{JHEP} \textbf{\bibinfo{volume}{02}}, \bibinfo{pages}{115}
  (\bibinfo{year}{2017}), \eprint{1609.06312}.

\bibitem[{\citenamefont{Ellis and Zanderighi}(2008)}]{Ellis:2007qk}
\bibinfo{author}{\bibfnamefont{R.~K.} \bibnamefont{Ellis}} \bibnamefont{and}
  \bibinfo{author}{\bibfnamefont{G.}~\bibnamefont{Zanderighi}},
  \bibinfo{journal}{JHEP} \textbf{\bibinfo{volume}{02}}, \bibinfo{pages}{002}
  (\bibinfo{year}{2008}), \eprint{0712.1851}.

\bibitem[{\citenamefont{de~Florian and Mazzitelli}(2013)}]{deFlorian:2013jea}
\bibinfo{author}{\bibfnamefont{D.}~\bibnamefont{de~Florian}} \bibnamefont{and}
  \bibinfo{author}{\bibfnamefont{J.}~\bibnamefont{Mazzitelli}},
  \bibinfo{journal}{Phys. Rev. Lett.} \textbf{\bibinfo{volume}{111}},
  \bibinfo{pages}{201801} (\bibinfo{year}{2013}), \eprint{1309.6594}.

\bibitem[{\citenamefont{Dawson et~al.}(1998)\citenamefont{Dawson, Dittmaier,
  and Spira}}]{Dawson:1998py}
\bibinfo{author}{\bibfnamefont{S.}~\bibnamefont{Dawson}},
  \bibinfo{author}{\bibfnamefont{S.}~\bibnamefont{Dittmaier}},
  \bibnamefont{and} \bibinfo{author}{\bibfnamefont{M.}~\bibnamefont{Spira}},
  \bibinfo{journal}{Phys. Rev.} \textbf{\bibinfo{volume}{D58}},
  \bibinfo{pages}{115012} (\bibinfo{year}{1998}), \eprint{hep-ph/9805244}.

\bibitem[{\citenamefont{de~Florian et~al.}(2016)}]{deFlorian:2016spz}
\bibinfo{author}{\bibfnamefont{D.}~\bibnamefont{de~Florian}}
  \bibnamefont{et~al.} (\bibinfo{collaboration}{LHC Higgs Cross Section Working
  Group}) (\bibinfo{year}{2016}), \eprint{1610.07922}.

\bibitem[{\citenamefont{de~Florian and Mazzitelli}(2015)}]{deFlorian:2015moa}
\bibinfo{author}{\bibfnamefont{D.}~\bibnamefont{de~Florian}} \bibnamefont{and}
  \bibinfo{author}{\bibfnamefont{J.}~\bibnamefont{Mazzitelli}},
  \bibinfo{journal}{JHEP} \textbf{\bibinfo{volume}{09}}, \bibinfo{pages}{053}
  (\bibinfo{year}{2015}), \eprint{1505.07122}.

\bibitem[{\citenamefont{Borowka
  et~al.}(2016{\natexlab{a}})\citenamefont{Borowka, Greiner, Heinrich, Jones,
  Kerner, Schlenk, Schubert, and Zirke}}]{Borowka:2016ehy}
\bibinfo{author}{\bibfnamefont{S.}~\bibnamefont{Borowka}},
  \bibinfo{author}{\bibfnamefont{N.}~\bibnamefont{Greiner}},
  \bibinfo{author}{\bibfnamefont{G.}~\bibnamefont{Heinrich}},
  \bibinfo{author}{\bibfnamefont{S.}~\bibnamefont{Jones}},
  \bibinfo{author}{\bibfnamefont{M.}~\bibnamefont{Kerner}},
  \bibinfo{author}{\bibfnamefont{J.}~\bibnamefont{Schlenk}},
  \bibinfo{author}{\bibfnamefont{U.}~\bibnamefont{Schubert}}, \bibnamefont{and}
  \bibinfo{author}{\bibfnamefont{T.}~\bibnamefont{Zirke}},
  \bibinfo{journal}{Phys. Rev. Lett.} \textbf{\bibinfo{volume}{117}},
  \bibinfo{pages}{012001} (\bibinfo{year}{2016}{\natexlab{a}}),
  \bibinfo{note}{[Erratum: Phys. Rev. Lett.117,no.7,079901(2016)]},
  \eprint{1604.06447}.

\bibitem[{\citenamefont{Grober et~al.}(2015)\citenamefont{Grober, Muhlleitner,
  Spira, and Streicher}}]{Grober:2015cwa}
\bibinfo{author}{\bibfnamefont{R.}~\bibnamefont{Grober}},
  \bibinfo{author}{\bibfnamefont{M.}~\bibnamefont{Muhlleitner}},
  \bibinfo{author}{\bibfnamefont{M.}~\bibnamefont{Spira}}, \bibnamefont{and}
  \bibinfo{author}{\bibfnamefont{J.}~\bibnamefont{Streicher}},
  \bibinfo{journal}{JHEP} \textbf{\bibinfo{volume}{09}}, \bibinfo{pages}{092}
  (\bibinfo{year}{2015}), \eprint{1504.06577}.

\bibitem[{\citenamefont{Borowka
  et~al.}(2016{\natexlab{b}})\citenamefont{Borowka, Greiner, Heinrich, Jones,
  Kerner, Schlenk, and Zirke}}]{Borowka:2016ypz}
\bibinfo{author}{\bibfnamefont{S.}~\bibnamefont{Borowka}},
  \bibinfo{author}{\bibfnamefont{N.}~\bibnamefont{Greiner}},
  \bibinfo{author}{\bibfnamefont{G.}~\bibnamefont{Heinrich}},
  \bibinfo{author}{\bibfnamefont{S.~P.} \bibnamefont{Jones}},
  \bibinfo{author}{\bibfnamefont{M.}~\bibnamefont{Kerner}},
  \bibinfo{author}{\bibfnamefont{J.}~\bibnamefont{Schlenk}}, \bibnamefont{and}
  \bibinfo{author}{\bibfnamefont{T.}~\bibnamefont{Zirke}},
  \bibinfo{journal}{JHEP} \textbf{\bibinfo{volume}{10}}, \bibinfo{pages}{107}
  (\bibinfo{year}{2016}{\natexlab{b}}), \eprint{1608.04798}.

\bibitem[{\citenamefont{de~Florian et~al.}(2017)\citenamefont{de~Florian,
  Fabre, and Mazzitelli}}]{deFlorian:2017qfk}
\bibinfo{author}{\bibfnamefont{D.}~\bibnamefont{de~Florian}},
  \bibinfo{author}{\bibfnamefont{I.}~\bibnamefont{Fabre}}, \bibnamefont{and}
  \bibinfo{author}{\bibfnamefont{J.}~\bibnamefont{Mazzitelli}},
  \bibinfo{journal}{JHEP} \textbf{\bibinfo{volume}{10}}, \bibinfo{pages}{215}
  (\bibinfo{year}{2017}), \eprint{1704.05700}.

\bibitem[{\citenamefont{Sjöstrand et~al.}(2015)\citenamefont{Sjöstrand, Ask,
  Christiansen, Corke, Desai, Ilten, Mrenna, Prestel, Rasmussen, and
  Skands}}]{Sjostrand:2014zea}
\bibinfo{author}{\bibfnamefont{T.}~\bibnamefont{Sjöstrand}},
  \bibinfo{author}{\bibfnamefont{S.}~\bibnamefont{Ask}},
  \bibinfo{author}{\bibfnamefont{J.~R.} \bibnamefont{Christiansen}},
  \bibinfo{author}{\bibfnamefont{R.}~\bibnamefont{Corke}},
  \bibinfo{author}{\bibfnamefont{N.}~\bibnamefont{Desai}},
  \bibinfo{author}{\bibfnamefont{P.}~\bibnamefont{Ilten}},
  \bibinfo{author}{\bibfnamefont{S.}~\bibnamefont{Mrenna}},
  \bibinfo{author}{\bibfnamefont{S.}~\bibnamefont{Prestel}},
  \bibinfo{author}{\bibfnamefont{C.~O.} \bibnamefont{Rasmussen}},
  \bibnamefont{and} \bibinfo{author}{\bibfnamefont{P.~Z.}
  \bibnamefont{Skands}}, \bibinfo{journal}{Comput. Phys. Commun.}
  \textbf{\bibinfo{volume}{191}}, \bibinfo{pages}{159} (\bibinfo{year}{2015}),
  \eprint{1410.3012}.

\bibitem[{\citenamefont{Alwall et~al.}(2014)\citenamefont{Alwall, Frederix,
  Frixione, Hirschi, Maltoni, Mattelaer, Shao, Stelzer, Torrielli, and
  Zaro}}]{Alwall:2014hca}
\bibinfo{author}{\bibfnamefont{J.}~\bibnamefont{Alwall}},
  \bibinfo{author}{\bibfnamefont{R.}~\bibnamefont{Frederix}},
  \bibinfo{author}{\bibfnamefont{S.}~\bibnamefont{Frixione}},
  \bibinfo{author}{\bibfnamefont{V.}~\bibnamefont{Hirschi}},
  \bibinfo{author}{\bibfnamefont{F.}~\bibnamefont{Maltoni}},
  \bibinfo{author}{\bibfnamefont{O.}~\bibnamefont{Mattelaer}},
  \bibinfo{author}{\bibfnamefont{H.~S.} \bibnamefont{Shao}},
  \bibinfo{author}{\bibfnamefont{T.}~\bibnamefont{Stelzer}},
  \bibinfo{author}{\bibfnamefont{P.}~\bibnamefont{Torrielli}},
  \bibnamefont{and} \bibinfo{author}{\bibfnamefont{M.}~\bibnamefont{Zaro}},
  \bibinfo{journal}{JHEP} \textbf{\bibinfo{volume}{07}}, \bibinfo{pages}{079}
  (\bibinfo{year}{2014}), \eprint{1405.0301}.

\bibitem[{\citenamefont{Hirschi and Mattelaer}(2015)}]{Hirschi:2015iia}
\bibinfo{author}{\bibfnamefont{V.}~\bibnamefont{Hirschi}} \bibnamefont{and}
  \bibinfo{author}{\bibfnamefont{O.}~\bibnamefont{Mattelaer}},
  \bibinfo{journal}{JHEP} \textbf{\bibinfo{volume}{10}}, \bibinfo{pages}{146}
  (\bibinfo{year}{2015}), \eprint{1507.00020}.

\bibitem[{ATL(2013{\natexlab{a}})}]{ATLAS:2013004}
\bibinfo{journal}{The ATLAS Collaboration, ATL-PHYS-PUB-2013-004}
  (\bibinfo{year}{2013}{\natexlab{a}}).

\bibitem[{\citenamefont{Cacciari et~al.}(2012)\citenamefont{Cacciari, Salam,
  and Soyez}}]{Cacciari:2011ma}
\bibinfo{author}{\bibfnamefont{M.}~\bibnamefont{Cacciari}},
  \bibinfo{author}{\bibfnamefont{G.~P.} \bibnamefont{Salam}}, \bibnamefont{and}
  \bibinfo{author}{\bibfnamefont{G.}~\bibnamefont{Soyez}},
  \bibinfo{journal}{Eur.Phys.J.} \textbf{\bibinfo{volume}{C72}},
  \bibinfo{pages}{1896} (\bibinfo{year}{2012}), \eprint{1111.6097}.

\bibitem[{\citenamefont{Cacciari et~al.}(2008)\citenamefont{Cacciari, Salam,
  and Soyez}}]{Cacciari:2008gp}
\bibinfo{author}{\bibfnamefont{M.}~\bibnamefont{Cacciari}},
  \bibinfo{author}{\bibfnamefont{G.~P.} \bibnamefont{Salam}}, \bibnamefont{and}
  \bibinfo{author}{\bibfnamefont{G.}~\bibnamefont{Soyez}},
  \bibinfo{journal}{JHEP} \textbf{\bibinfo{volume}{04}}, \bibinfo{pages}{063}
  (\bibinfo{year}{2008}), \eprint{0802.1189}.

\bibitem[{ATL(2013{\natexlab{b}})}]{ATL-PHYS-PUB-2013-009}
\bibinfo{type}{Tech. Rep.} \bibinfo{number}{ATL-PHYS-PUB-2013-009},
  \bibinfo{institution}{CERN}, \bibinfo{address}{Geneva}
  (\bibinfo{year}{2013}{\natexlab{b}}),
  \urlprefix\url{http://cds.cern.ch/record/1604420}.

\bibitem[{\citenamefont{Speckmayer et~al.}(2010)\citenamefont{Speckmayer,
  Hocker, Stelzer, and Voss}}]{Speckmayer:2010zz}
\bibinfo{author}{\bibfnamefont{P.}~\bibnamefont{Speckmayer}},
  \bibinfo{author}{\bibfnamefont{A.}~\bibnamefont{Hocker}},
  \bibinfo{author}{\bibfnamefont{J.}~\bibnamefont{Stelzer}}, \bibnamefont{and}
  \bibinfo{author}{\bibfnamefont{H.}~\bibnamefont{Voss}}, \bibinfo{journal}{J.
  Phys. Conf. Ser.} \textbf{\bibinfo{volume}{219}}, \bibinfo{pages}{032057}
  (\bibinfo{year}{2010}).

\bibitem[{\citenamefont{Brun and Rademakers}(1997)}]{Brun:1997pa}
\bibinfo{author}{\bibfnamefont{R.}~\bibnamefont{Brun}} \bibnamefont{and}
  \bibinfo{author}{\bibfnamefont{F.}~\bibnamefont{Rademakers}},
  \bibinfo{journal}{Nucl. Instrum. Meth.} \textbf{\bibinfo{volume}{A389}},
  \bibinfo{pages}{81} (\bibinfo{year}{1997}).

\bibitem[{CMS(2016)}]{CMS-PAS-TAU-16-002}
\bibinfo{type}{Tech. Rep.} \bibinfo{number}{CMS-PAS-TAU-16-002},
  \bibinfo{institution}{CERN}, \bibinfo{address}{Geneva}
  (\bibinfo{year}{2016}), \urlprefix\url{https://cds.cern.ch/record/2196972}.

\bibitem[{\citenamefont{Lester and Summers}(1999)}]{Lester:1999tx}
\bibinfo{author}{\bibfnamefont{C.~G.} \bibnamefont{Lester}} \bibnamefont{and}
  \bibinfo{author}{\bibfnamefont{D.~J.} \bibnamefont{Summers}},
  \bibinfo{journal}{Phys. Lett.} \textbf{\bibinfo{volume}{B463}},
  \bibinfo{pages}{99} (\bibinfo{year}{1999}), \eprint{hep-ph/9906349}.

\bibitem[{\citenamefont{Barr et~al.}(2003)\citenamefont{Barr, Lester, and
  Stephens}}]{Barr:2003rg}
\bibinfo{author}{\bibfnamefont{A.}~\bibnamefont{Barr}},
  \bibinfo{author}{\bibfnamefont{C.}~\bibnamefont{Lester}}, \bibnamefont{and}
  \bibinfo{author}{\bibfnamefont{P.}~\bibnamefont{Stephens}},
  \bibinfo{journal}{J. Phys.} \textbf{\bibinfo{volume}{G29}},
  \bibinfo{pages}{2343} (\bibinfo{year}{2003}), \eprint{hep-ph/0304226}.

\bibitem[{\citenamefont{Cheng and Han}(2008)}]{Cheng:2008hk}
\bibinfo{author}{\bibfnamefont{H.-C.} \bibnamefont{Cheng}} \bibnamefont{and}
  \bibinfo{author}{\bibfnamefont{Z.}~\bibnamefont{Han}},
  \bibinfo{journal}{JHEP} \textbf{\bibinfo{volume}{12}}, \bibinfo{pages}{063}
  (\bibinfo{year}{2008}), \eprint{0810.5178}.

\bibitem[{\citenamefont{Chatrchyan et~al.}(2014)}]{Chatrchyan:2014nva}
\bibinfo{author}{\bibfnamefont{S.}~\bibnamefont{Chatrchyan}}
  \bibnamefont{et~al.} (\bibinfo{collaboration}{CMS}), \bibinfo{journal}{JHEP}
  \textbf{\bibinfo{volume}{05}}, \bibinfo{pages}{104} (\bibinfo{year}{2014}),
  \eprint{1401.5041}.

\bibitem[{\citenamefont{Barr et~al.}(2014)\citenamefont{Barr, Dolan, Englert,
  and Spannowsky}}]{Barr:2013tda}
\bibinfo{author}{\bibfnamefont{A.~J.} \bibnamefont{Barr}},
  \bibinfo{author}{\bibfnamefont{M.~J.} \bibnamefont{Dolan}},
  \bibinfo{author}{\bibfnamefont{C.}~\bibnamefont{Englert}}, \bibnamefont{and}
  \bibinfo{author}{\bibfnamefont{M.}~\bibnamefont{Spannowsky}},
  \bibinfo{journal}{Phys. Lett.} \textbf{\bibinfo{volume}{B728}},
  \bibinfo{pages}{308} (\bibinfo{year}{2014}), \eprint{1309.6318}.

\bibitem[{ATL(2013{\natexlab{c}})}]{ATL-PHYS-PUB-2013-014}
\bibinfo{type}{Tech. Rep.} \bibinfo{number}{ATL-PHYS-PUB-2013-014},
  \bibinfo{institution}{CERN}, \bibinfo{address}{Geneva}
  (\bibinfo{year}{2013}{\natexlab{c}}),
  \urlprefix\url{https://cds.cern.ch/record/1611186}.

\bibitem[{ATL(2017{\natexlab{b}})}]{ATL-PHYS-PUB-2017-013}
\bibinfo{type}{Tech. Rep.} \bibinfo{number}{ATL-PHYS-PUB-2017-013},
  \bibinfo{institution}{CERN}, \bibinfo{address}{Geneva}
  (\bibinfo{year}{2017}{\natexlab{b}}),
  \urlprefix\url{https://cds.cern.ch/record/2273281}.

\bibitem[{ATL(2013{\natexlab{d}})}]{ATL-PHYS-PUB-2013-004}
\bibinfo{type}{Tech. Rep.} \bibinfo{number}{ATL-PHYS-PUB-2013-004},
  \bibinfo{institution}{CERN}, \bibinfo{address}{Geneva}
  (\bibinfo{year}{2013}{\natexlab{d}}),
  \urlprefix\url{https://cds.cern.ch/record/1527529}.

\bibitem[{\citenamefont{Adhikary et~al.}(2017)\citenamefont{Adhikary, Banerjee,
  Barman, Bhattacherjee, and Niyogi}}]{Adhikary:2017jtu}
\bibinfo{author}{\bibfnamefont{A.}~\bibnamefont{Adhikary}},
  \bibinfo{author}{\bibfnamefont{S.}~\bibnamefont{Banerjee}},
  \bibinfo{author}{\bibfnamefont{R.~K.} \bibnamefont{Barman}},
  \bibinfo{author}{\bibfnamefont{B.}~\bibnamefont{Bhattacherjee}},
  \bibnamefont{and} \bibinfo{author}{\bibfnamefont{S.}~\bibnamefont{Niyogi}}
  (\bibinfo{year}{2017}), \eprint{1712.05346}.

\bibitem[{\citenamefont{Alwall et~al.}(2008)\citenamefont{Alwall, Hoche,
  Krauss, Lavesson, Lonnblad et~al.}}]{Alwall:2007fs}
\bibinfo{author}{\bibfnamefont{J.}~\bibnamefont{Alwall}},
  \bibinfo{author}{\bibfnamefont{S.}~\bibnamefont{Hoche}},
  \bibinfo{author}{\bibfnamefont{F.}~\bibnamefont{Krauss}},
  \bibinfo{author}{\bibfnamefont{N.}~\bibnamefont{Lavesson}},
  \bibinfo{author}{\bibfnamefont{L.}~\bibnamefont{Lonnblad}},
  \bibnamefont{et~al.}, \bibinfo{journal}{Eur.Phys.J.}
  \textbf{\bibinfo{volume}{C53}}, \bibinfo{pages}{473} (\bibinfo{year}{2008}),
  \eprint{0706.2569}.

\bibitem[{\citenamefont{Dittmaier et~al.}(2011)}]{Dittmaier:2011ti}
\bibinfo{author}{\bibfnamefont{S.}~\bibnamefont{Dittmaier}}
  \bibnamefont{et~al.} (\bibinfo{collaboration}{LHC Higgs Cross Section Working
  Group}) (\bibinfo{year}{2011}), \eprint{1101.0593}.

\bibitem[{\citenamefont{Skands}(2010)}]{Skands:2010ak}
\bibinfo{author}{\bibfnamefont{P.~Z.} \bibnamefont{Skands}},
  \bibinfo{journal}{Phys. Rev.} \textbf{\bibinfo{volume}{D82}},
  \bibinfo{pages}{074018} (\bibinfo{year}{2010}), \eprint{1005.3457}.

\bibitem[{CMS(2011)}]{CMS-PAS-TOP-11-022}
\bibinfo{type}{Tech. Rep.} \bibinfo{number}{CMS-PAS-TOP-11-022},
  \bibinfo{institution}{CERN}, \bibinfo{address}{Geneva}
  (\bibinfo{year}{2011}), \urlprefix\url{http://cds.cern.ch/record/1385552}.

\end{thebibliography}

\end{document}